\documentclass[twocolumn,longauth]{aa}  
\usepackage[T1]{fontenc}
\usepackage[utf8]{inputenc}
\usepackage{natbib}
\usepackage{multirow}
\usepackage{comment}
\usepackage{graphicx}	
\usepackage{amsmath}	
\usepackage{amssymb}	
\usepackage{xspace}     
\usepackage[dvipsnames]{xcolor}
\usepackage{subcaption}
\usepackage{txfonts}
\usepackage{times}
\usepackage{tikz}
\usetikzlibrary{shapes.geometric, arrows}
\usepackage[dvipsnames]{xcolor}
\tikzstyle{input} = [rectangle, rounded corners, minimum width=3cm, minimum height=1cm,text centered, draw=black, align=center, fill=SpringGreen!50]
\tikzstyle{classpy} = [rectangle,  rounded corners, minimum width=3cm, minimum height=1cm, text centered, draw=black, fill=Cerulean!50]
\usepackage{tikz}  
\newcommand{\tikzhdashline}{\tikz[baseline] \draw[dashed] (5,2.7ex)--(18.5,2.7ex);}

\newcommand{\revint}[1]{{\color{black} #1}}
\newcommand{\revcollab}[1]{{\color{black} #1}}

\graphicspath{{/}}

\usepackage[colorlinks, citecolor=blue, linkcolor = blue]{hyperref}

\usepackage{cleveref}
\Crefname{section}{Sect.}{Sects.}
\Crefname{equation}{Eq.}{Eqs.}
\Crefname{figure}{Fig.}{Figs.}
\usepackage{lineno}
\newcommand{\review}[1]{{\color{black} #1}}

\begin{document}

   \title{KiDS-Legacy: Constraining dark energy, neutrino mass, and curvature}

 \author{Robert Reischke\inst{1,2}\thanks{rreischke@astro.uni-bonn.de, reischke@posteo.net},  Benjamin St\"{o}lzner\inst{2}, Benjamin Joachimi, Angus H. Wright\inst{2}, Marika Asgari\inst{3,7,8}, Andrej Dvornik\inst{2}, Hendrik Hildebrandt\inst{2}, Lucas Porth\inst{1}, Maximilian von Wietersheim-Kramsta\inst{4,5,6}, Jan Luca van den Busch\inst{2}, Ziang Yan\inst{2}, Maciej Bilicki\inst{9}, Pierre Burger\inst{10,1}, Nora Elisa Chisari\inst{11,15}, Joachim Harnois-D\'{e}raps\inst{3}, Christos Georgiou\inst{11}, Catherine Heymans\inst{2,12}, Priyanka Jalan\inst{9}, Shahab Joudaki\inst{13,14}, Konrad Kuijken\inst{15}, Shun-Sheng Li\inst{15,16}, Laila Linke\inst{17}, Constance Mahony \inst{18,2}, Tilman Tr\"{o}ster\inst{20}, Mijin Yoon\inst{15}
          }
   \institute{
   Argelander-Institut für Astronomie, Universität Bonn, Auf dem Hügel 71, D-53121 Bonn, Germany\\
\email{rreischk@uni-bonn.de, reischke@posteo.net} \and
Ruhr University Bochum, Faculty of Physics and Astronomy, Astronomical Institute (AIRUB), German Centre for Cosmological Lensing, 44780 Bochum, Germany
\and School of Mathematics, Statistics and Physics, Newcastle University, Herschel Building, NE1 7RU, Newcastle-upon-Tyne, UK
\and Department of Physics and Astronomy, University College London, Gower Street, London WC1E 6BT, UK
\and
            Department of Physics, Institute for Computational  Cosmology, Durham University, South Road, Durham DH1 3LE, UK
        \and
             Department of Physics, Centre for Extragalactic Astronomy, Durham University, South Road, Durham DH1 3LE, UK
    \and  E. A. Milne Centre, University of Hull, Cottingham Road, Hull, HU6 7RX, UK
    \and  Centre of Excellence for Data Science, AI, and Modelling (DAIM), University of Hull, Cottingham Road, Hull, HU6 7RX, UK
    \and Centre for Theoretical Physics, Polish Academy of Sciences, al. Lotnik\'ow 32/46, 02-668 Warsaw, Poland
    \and Department of Physics and Astronomy, University of Waterloo, 200 University Ave W, Waterloo, ON N2L 3G1, Canada
    \and Institute for Theoretical Physics, Utrecht University, Princetonplein 5, 3584CC Utrecht, The Netherlands
    \and Institute for Astronomy, University of Edinburgh, Blackford Hill, Edinburgh, EH9 3HJ, UK
    \and Centro de Investigaciones Energ\'{e}ticas, Medioambientales y Tecnol\'{o}gicas (CIEMAT), Av. Complutense 40, E-28040 Madrid, Spain
    \and Institute of Cosmology \& Gravitation, Dennis Sciama Building, University of Portsmouth, Portsmouth, PO1 3FX, United Kingdom
    \and Leiden Observatory, Leiden University, P.O. Box 9513, 2300 RA Leiden, the Netherlands,
    \and Aix-Marseille Universit\'{e}, CNRS, CNES, LAM, Marseille, France
    \and Universität Innsbruck, Institut für Astro- und Teilchenphysik, Technikerstr. 25/8, 6020 Innsbruck, Austria
    \and Donostia International Physics Centre, Manuel Lardizabal Ibilbidea, 4, 20018 Donostia, Gipuzkoa, Spain
    \and INAF - Osservatorio Astronomico di Roma, via Frascati 33, 00040 Monteporzio Catone (Roma), Italy
    \and Institute for Particle Physics and Astrophysics, ETH Zürich, Wolfgang-Pauli-Strasse 27, 8093 Zürich, Switzerland
}
    
   \date{\today}

\abstract{We constrained minimally extended cosmological models with the cosmic shear analysis of the final data release from the Kilo-Degree Survey (KiDS-Legacy) in combination with external probes. Due to the consistency of the KiDS-Legacy analysis with the cosmic microwave background (CMB), we could combine these datasets reliably for the first time. Additionally, we used CMB lensing, galaxy redshift-space distortions, and baryon acoustic oscillations. We assessed, in turn, the effects of spatial curvature, varying neutrino masses, and an evolving dark energy component on cosmological constraints from KiDS-Legacy alone and from KiDS-Legacy combined with external probes. We find KiDS-Legacy to be consistent with the fiducial flat $\Lambda$-cold dark matter ($\Lambda$CDM) analysis with $c^2 \sum m_\nu\leq 1.5\,$eV, $w_0 = -1.0\pm 0.7$, and $w_a = -1.3^{+1.9}_{-2.0}$ while $\Omega_K = 0.08^{+0.16}_{-0.17}$ (1$\sigma$ bounds) with an almost equal goodness of fit. The $w_0w_a$CDM model is not a significant improvement over $\Lambda$CDM when cosmic shear and CMB lensing are combined, yielding a Bayes factor $B = 0.07$. If all probes are combined, however, $B$ increases to 2.73, corresponding to a $2.6\sigma$ suspiciousness tension. The constraint on $S_8 = \sigma_8\sqrt{\Omega_\mathrm{m}/0.3}$ is robust to opening up the parameter space for cosmic shear. Adding all external datasets to KiDS-Legacy, we find $S_8 = 0.816 \pm 0.006$ in $\Lambda$CDM and $S_8 = 0.837 \pm 0.008$ in $w_0 w_a$CDM for all probes combined.}


  \titlerunning{KiDS-Legacy: Extended Cosmologies}
\authorrunning{R. Reischke et al.}

\author{
Robert Reischke\inst{1,2}\thanks{rreischk@uni-bonn.de
, reischke@posteo.net} \and
Benjamin St\"olzner\inst{2}\and
Benjamin Joachimi\inst{3}\and 
        Angus H. Wright\inst{2}\and
        Marika Asgari\inst{4}\and
        Maciej Bilicki\inst{5}\and
        Nora Elisa Chisari\inst{6,7}\and
        Andrej Dvornik\inst{2}\and
        Christos Georgiou\inst{8}\and
        Benjamin Giblin\inst{9}\and
        Joachim Harnois-D\'eraps\inst{4}\and        
        Catherine Heymans\inst{2,9}\and
        Hendrik Hildebrandt\inst{2}\and
        Henk Hoekstra\inst{7}\and
        Shahab Joudaki\inst{10}\and
        Konrad Kuijken\inst{7}\and
        Shun-Sheng Li\inst{11,12}\and
        Laila Linke\inst{13}\and
        Arthur Loureiro\inst{14,15}\and
        Constance Mahony\inst{16,17,2}\and
        Lauro Moscardini\inst{18,19,20}\and
        Nicola R. Napolitano\inst{21}\and
        Lucas Porth\inst{1}\and
        Mario Radovich\inst{22}\and
        Tilman Tr\"oster\inst{23}\and
        Maximilian von Wietersheim-Kramsta\inst{24,25}\and
        Ziang Yan\inst{2,26} \and
        Mijin Yoon\inst{7}\and
        Yun-Hao Zhang\inst{7,9}
    	}

\institute{
 Argelander-Institut für Astronomie, Universität Bonn, Auf dem Hügel 71, D-53121 Bonn, Germany
 \and Ruhr University Bochum, Faculty of Physics and Astronomy, Astronomical Institute (AIRUB), German Centre for Cosmological Lensing, 44780 Bochum, Germany 
\and Department of Physics and Astronomy, University College London, Gower Street, London WC1E 6BT, United Kingdom
\and School of Mathematics, Statistics and Physics, Newcastle University, Herschel Building, NE1 7RU, Newcastle-upon-Tyne, United Kingdom
\and Center for Theoretical Physics, Polish Academy of Sciences, al. Lotników 32/46, 02-668 Warsaw, Poland
\and Institute for Theoretical Physics, Utrecht University, Princetonplein 5, 3584CC Utrecht, The Netherlands
\and Leiden Observatory, Leiden University, P.O.Box 9513, 2300RA Leiden, The Netherlands
\and Institut de Física d’Altes Energies (IFAE), The Barcelona Institute of Science and Technology, Campus UAB, 08193 Bellaterra (Barcelona), Spain
\and Institute for Astronomy, University of Edinburgh, Royal Observatory, Blackford Hill, Edinburgh, EH9 3HJ, United Kingdom
\and Centro de Investigaciones Energéticas, Medioambientales y Tecnológicas (CIEMAT), Av. Complutense 40, E-28040 Madrid, Spain
\and Kavli Institute for Particle Astrophysics and Cosmology, Stanford University, Stanford, CA 94305, USA 
\and SLAC National Accelerator Laboratory, Menlo Park, CA 94025, USA
\and Universität Innsbruck, Institut für Astro- und Teilchenphysik, Technikerstr. 25/8, 6020 Innsbruck, Austria
\and The Oskar Klein Centre, Department of Physics, Stockholm University, AlbaNova University Centre, SE-106 91 Stockholm, Sweden
\and Imperial Centre for Inference and Cosmology (ICIC), Blackett Laboratory, Imperial College London, Prince Consort Road, London SW7 2AZ, United Kingdom
\and Department of Physics, University of Oxford, Denys Wilkinson Building, Keble Road, Oxford OX1 3RH, United Kingdom
\and Donostia International Physics Center, Manuel Lardizabal Ibilbidea, 4, 20018 Donostia, Gipuzkoa, Spain
\and Dipartimento di Fisica e Astronomia "Augusto Righi" - Alma Mater Studiorum Università di Bologna, via Piero Gobetti 93/2, I-40129 Bologna, Italy
\and Istituto Nazionale di Astrofisica (INAF) - Osservatorio di Astrofisica e Scienza dello Spazio (OAS), via Piero Gobetti 93/3, I-40129 Bologna, Italy
\and Istituto Nazionale di Fisica Nucleare (INFN) - Sezione di Bologna, viale Berti Pichat 6/2, I-40127 Bologna, Italy
\and Department of Physics ``E. Pancini'' University of Naples Federico II C.U. di Monte Sant'Angelo Via Cintia, 21 ed. 6, 80126 Naples, Italy
\and INAF - Osservatorio Astronomico di Padova, via dell'Osservatorio 5, 35122 Padova, Italy
\and Institute for Particle Physics and Astrophysics, ETH Zürich, Wolfgang-Pauli-Strasse 27, 8093 Zürich, Switzerland
\and Institute for Computational Cosmology, Ogden Centre for Fundamental Physics - West, Department of Physics, Durham University, South Road, Durham DH1 3LE, United Kingdom
\and Centre for Extragalactic Astronomy, Ogden Centre for Fundamental Physics - West, Department of Physics, Durham University, South Road, Durham DH1 3LE, United Kingdom
\and Graduate School of Science, Nagoya University, Furocho, Chikusa-ku, Nagoya, Aichi, 464-8602, Japan
}

\date{Received ; accepted }

   \maketitle

\section{Introduction}
The model currently best capturing the dynamics of the Universe on large scales is the $\Lambda$-cold dark matter model with vanishing spatial curvature (flat $\Lambda$CDM). Despite some discrepancies, this simple model with only a few parameters successfully explains a wide range of observations.

The weak gravitational lensing effect of the large-scale structure (LSS), also known as cosmic shear, has become a staple cosmological observable since its first detection over two decades ago \citep{kaiser2000,wittman2000,vanWaerbeke2000,bacon2000}. Due to its sensitivity to the growth of structure as well as the background cosmology, it has the potential to probe the $\Lambda$CDM model and its extensions. The most recent analysis comes from the so-called Stage-III weak lensing surveys, the Kilo-Degree Survey 
\citep[KiDS, see][for the most recent analysis]{wright_kids_2025,2025arXiv250319442S}\footnote{\url{https://kids.strw.leidenuniv.nl/}}, the Dark Energy Survey \citep[DES, see e.g.][]{amon_dark_2022,Secco_DES_scalecut_2022,2023MNRAS.524.2195S}\footnote{\url{https://www.darkenergysurvey.org/}}, and the Subaru Hyper Suprime-Cam Subaru Strategic Project \revint{\citep[HSC, see e.g.][]{2023PhRvD.108l3518L, 2023PhRvD.108l3519D}}\footnote{\url{https://hsc.mtk.nao.ac.jp/ssp/}}.
Cosmic shear is joined by other probes such as the cosmic microwave background \citep[CMB;][]{PlanckCollab_CMB2020,2025arXiv250620707C,2025arXiv250314452L}, baryon acoustic oscillations \citep[BAOs;][]{2025JCAP...02..021A,2025arXiv250314738D,2021PhRvD.103h3533A}, redshift space distortions \citep[RSDs;][]{2021MNRAS.500.1201H}, expansion history measurements via Type Ia supernovae \citep[SNeIa;][]{Scolnic22,Brout22}, as well as the present-day expansion rate, that is the Hubble constant, through Cepheids \citep{2019ApJ...876...85R}, strongly lensed quasars \citep{2020MNRAS.498.1420W}, or the tip of the red giant branch and J-region asymptotic giant branch stars \citep[see][for an overview]{2025ApJ...985..203F}. \revint{These last measurements technically do not depend on cosmology, as they are local and thus cannot distinguish between different cosmological models beyond the overall normalisation of the expansion rate.}

The first direct evidence of the accelerated expansion of the Universe \revint{\citep{1998AJ....116.1009R,1998ApJ...507...46S,1999ApJ...517..565P}} established cosmological probes across cosmic time as a test bed for physics beyond the standard model. These measurements implied, in hindsight, an unsurprisingly non-vanishing second coupling constant, $\Lambda$, in the gravity sector. One can express $\Lambda$ itself as an ideal fluid with constant energy density and with negative pressure. More generally, one can relate the energy density to the pressure via an equation of state, \revcollab{$w\left[a(z)\right]$, which can depend on redshift $z$.} This dynamic scenario can itself originate from an additional field that acts solely on the background cosmology and does not form perturbations.
A key target of cosmology is therefore to use cosmological observables to infer $w(a)$ and test whether it assumes its $\Lambda$CDM value of $w =-1$. 
Many constraints on $w$ use an expansion up to linear order, which \revcollab{effectively approximates many different} dark energy models \citep{2001IJMPD..10..213C,2003PhRvL..90i1301L}. Recently, the Dark Energy Spectroscopic Instrument \citep[DESI][]{2025JCAP...02..021A,2025arXiv250314738D} found tentative evidence that $w\neq -1$ in this parametrisation using BAO measurements and external probes including the CMB. This modification of the expansion history becomes particularly interesting in light of the Hubble tension, the discrepancy between local and early time measurements of the expansion rate \revint{\citep[see e.g.][for reviews and summaries]{2022PhR...984....1S,2023ARNPS..73..153K,2025PDU....4901965D}}. 

While \citet{wright_kids_2025} and \citet{2025arXiv250319442S} recently put forward constraints on flat $\Lambda$CDM using the final data release of KiDS \citep[DR5;][]{wright23_dr5}, the goal of this paper is to expand this analysis to minimal extensions to the baseline cosmological model and include external datasets. Unlike in the previous KiDS analyses \revint{\citep[KiDS-1000 and its reanalyses;][]{asgari_kids_2021,2021A&A...649A..88T,2022A&A...664A.170V,2023A&A...679A.133L,2025arXiv251001122Y}}, which found a tension in the amount of matter clustering in the Universe compared to the CMB (the $S_8$ tension), KiDS-Legacy is consistent with the other cosmological observables. The driving contributing factors are updated redshift calibration and scale cuts \citep[see appendix I in][]{wright_kids_2025}.
This consistency allows the combination of different probes and opens the possibility of constraining extended parameter spaces. 

On the data side, we used CMB primary anisotropies from Planck \citep{2020A&A...641A...5P}, the Atacama Cosmology Telescope \citep[ACT;][]{2025arXiv250314452L}, and the South Pole Telescope \citep[SPT;][]{2025arXiv250620707C}, the signal from gravitational lensing of the CMB \citep{2020A&A...641A...5P,2024ApJ...962..113M,2025PhRvD.111h3534G}, BAO data from DESI \citep{2025arXiv250314738D}, RSDs from the Extended Baryon Oscillation Spectroscopic Survey \citep[eBOSS, e.g.][]{2021MNRAS.500.1201H}, SNeIa from the Pantheon+ sample \citep{Scolnic22}, and combined cosmic shear measurements from KiDS-Legacy and DES Y3 based on \citet{des/kids:2023} with the updated data from \citet{wright23_dr5}. For the modelling, we focused on three main extensions to the fiducial analysis: $(i)$ varying neutrino mass ($\nu\Lambda$CDM),
$(ii)$ dynamic dark energy ($w_0w_a$CDM), and $(iii)$ a spatially non-flat Universe ($\Omega_K$CDM). \review{Although the linear growth factor is modified in dynamic dark energy scenarios, the extensions studied here do not explicitly alter the gravity sector; that is, we assume general relativity throughout this paper and introduce no new gravitational fields. In a companion paper \citep{2025arXiv251211039S}, we investigate, in more detail, extensions that alter the growth of structures by modifying the gravitational law in the context of Horndeski theory.}

We structure this paper as follows: \Cref{sec:method} briefly summarises the methodology, the inference pipeline, data, and likelihood used in this analysis. In \Cref{sec:constraints}, we present the results of the three types of extended models for different probe combinations. Lastly, we summarise our findings and conclude in \Cref{sec:concl}.

\section{Methodology and datasets}
\label{sec:method}
In this section, we first briefly summarise the methodology used for the cosmic shear analysis, which follows the one used in \citet{wright_kids_2025,2025arXiv250319442S,2025A&A...699A.124R}.

\subsection{Cosmic shear and KiDS-Legacy}
 {The KiDS-Legacy dataset combines two European Southern Observatory Public Surveys: KiDS \citep{2013ExA....35...25D} on the VLT Survey Telescope (VST) and on the VISTA telescope, the VISTA Kilo-Degree Infrared Galaxy Survey \citep[VIKING;][]{Edge13}. The KiDS data span} $1,347\,\mathrm{deg}^2$ \revint{in two stripes on and south of the celestial equator}, with optical bands $u,g,r,i$ from VST and $Z,Y,J,H,K_\mathrm{s}$ from VIKING. Compared with the previous release, DR5 adds a second $i$‑band pass and increases the area by 34 per cent. The KiDS‑Legacy lensing catalogue contains \revint{41 million source} galaxies over $967\,\mathrm{deg}^2$, with an effective number density of $8.79\,\mathrm{arcmin}^{-2}$.
 Deeper $i$‑band imaging and expanded spectroscopic calibration allow a higher photometric redshift limit of $z_B\leq 2$ and an extra high redshift tomographic bin. Sources are split into six roughly equally-populated bins by photometric redshifts, and the corresponding redshift distributions are calibrated against deep spectroscopy using self‑organising maps and the multicolour SKiLLS simulations \citep[see][for details]{li_2023b,2025arXiv250319440W}.
Galaxy shapes were measured with an updated version of the {\sc lensfit} algorithm \citep{miller_2013,fenech_2017} and calibrated with SKiLLS.

\begingroup
\renewcommand*{\arraystretch}{1.2}
\begin{table}[]
    \vspace{.075cm}
\caption{Data combinations used in this analysis. The details are described in \Cref{sec:method}.}
    \centering
    \begin{tabular}{ll}
    \hline \hline 
        \textbf{name} & \textbf{dataset}\\  \hline 
        KiDS & KiDS DR5 cosmic shear catalogue \\
        CMB& SPT 3G D1 + ACT DR6 + Planck 2018 \\ & lensing and primary (also known as SPA) \\
        lensing & KiDS + \revint{DES Y3} + CMBlensing \\
        low-$z$ & KiDS + DES Y3 + DESI DR2 BAO \\ &+ eBOSS DR16 RSD \\ \hline \\
    \end{tabular}
    \label{tab:data_combiantions}
    \tablefoot{\revint{With low-$z$ we refer to all probes which are entirely based on observations at low redshift. Therefore, the lensing signal of the CMB is not included in the low-$z$ set.}}
\end{table}
\endgroup

\begingroup
\renewcommand{\arraystretch}{1.2}
\begin{table*}
\caption{Model parameters and their priors grouped by the concordance model, extensions, and nuisance.}
\centering
\begin{tabular}{clll}
\hline\hline
\textbf{type} & \textbf{parameter} & \textbf{prior} & \textbf{description}\\
\hline
\multirow{6}{*}{\rotatebox[origin=c]{90}{flat $\Lambda$CDM}} 
&$\omega_{\rm cdm}$ 		& $\mathcal{U}(0.051,0.255)$ 						   & cold dark matter density\\
&$\omega_{\rm b}$   		& $\mathcal{U}(0.019,0.026)$ 						   & baryon density\\
&$h$ 			   		    & $\mathcal{U}(0.64,0.82)$ 							   & reduced Hubble parameter\\
&$n_{\rm s}$        		& $\mathcal{U}(0.84,1.1)$ 							   & spectral index of the primordial power spectrum \vspace{.1cm}\\ \tikzhdashline\hspace{-15cm}
&$S_8$              		& $\mathcal{U}(0.5,1.0)$ 							   & structure growth parameter \\
& $\ln 10^{10}A_\mathrm{s}$ & $\mathcal{U}(2.9,3.2)$ &  matter power spectrum amplitude \\
\hline
\multirow{4}{*}{\rotatebox[origin=c]{90}{extended}} 
&$\sum m_\nu [\mathrm{eV}]$ 		& $\mathcal{U}(0,3)$ 						   & sum of the neutrino masses\\
&$w_0$   		& $\mathcal{U}(-3,0)$ 						   & dark energy equation of state today\\
&$w_a$   		& $\mathcal{U}(-5,5)$ 						   & linear slope of the dark energy equation of state\\
&$\Omega_K$ 			   		    & $\mathcal{U}(-0.3,0.3)$ 							   & spatial curvature density parameter\\
\hline
\multirow{5}{*}{\rotatebox[origin=c]{90}{Nuisance}} 
&$\log_{10}T_{\rm AGN}$     & $\mathcal{U}(7.3,8.3)$ 							 & baryon feedback parameter\\
&$A_{\rm IA}$ 				& $\mathcal{N}(5.74,\mathbf{C}_{A_{\rm IA}, \beta})$ & amplitude of intrinsic galaxy alignments for red galaxies\\
&$\beta$ 					& $\mathcal{N}(0.44,\mathbf{C}_{A_{\rm IA}, \beta})$  & slope of the mass scaling of intrinsic galaxy alignments\\
&$\log_{10}M_{\rm i}$ 		& $\mathcal{N}(\boldsymbol{\mu},\mathbf{C}_M)$ 		 		   & mean halo mass of early-type galaxies per tomographic bin $i$\\
&$\delta_{{\rm z},i}$ 			& $\mathcal{N}(\boldsymbol{\mu},\mathbf{C}_z)$ 				   & shift of the mean of the redshift distribution per tomographic bin $i$\\\hline
 \\
\end{tabular}
\label{tab:parameters}
\tablefoot{The first two columns specify the types of the sampling parameters and the parameter names, respectively. The third column gives the priors used, with uniform priors denoted by their interval $\mathcal{U}$, and Gaussian priors by $\mathcal{N}(\mu,\sigma)$. The fourth column offers a brief description of each parameter. For more details on the chosen priors, see \citet{wright_kids_2025,2025arXiv250319442S,2025arXiv250319440W}. For nuisance parameters of the external probes, we use the fiducial priors from the respective analyses. 
When including any CMB data, we sample over $\log A_\mathrm{s}$ instead of $S_8$ and derive $S_8$ instead. All constraints in \Cref{tab:constraints} are, however, given in terms of $S_8$ only. We tested that this choice does not introduce any information and yields the same cosmological constraints.}
\end{table*}
\endgroup

We use CosmoPipe\footnote{\url{https://github.com/AngusWright/CosmoPipe/tree/KiDSLegacy_CosmicShear}} for the cosmic shear measurement, calibration, and inference infrastructure. As the summary statistic, we use the first six modes of the complete orthogonal sets of E/B-integrals \citep[COSEBIs;][]{Schneider_2010,Asgari_2012}, measured from 2 to $300\,$arcmin. The COSEBI E-modes are integrals over the angular power spectrum of the observed ellipticities, $C^{(ij)}_{\epsilon\epsilon}(\ell)$ in tomographic bins $i$ and $j$, which in turn can be written as
\begin{equation}
	\label{eq:cell}
C^{(ij)}_{\epsilon\epsilon}(\ell)= \int_0^{\chi_{\rm H}} {\rm d}\chi\, \frac{W^{(i)}_{{\epsilon}}(\chi)W^{(j)}_{\epsilon}(\chi)}{f_{ K}^2(\chi)}P_{\rm m, nl}\left(\frac{\ell+1/2}{f_{ K}(\chi)},z(\chi)\right).
	\end{equation}
Here, $P_{\rm m, nl}$ is the non-linear matter power spectrum, which we model using \revint{{\sc HMCODE2020} \citep{2021MNRAS.502.1401M}}. This employs a single-parameter ($\log_{10}T_\mathrm{AGN}$) baryonic feedback model to include the uncertainty of the modelling of baryonic feedback on scales $k > 0.1h/\mathrm{Mpc}$. Furthermore, $f_{ K}$, $\chi$, and $\chi_{\rm H}$ are the comoving angular diameter distance, the comoving radial distance, and the comoving horizon distance, respectively. Lastly, the lensing kernel $W^{(i)}_{\epsilon}$ contains the distribution of galaxies and the lensing geometry \citep[see e.g.][]{wright_kids_2025}.

We adopt the fiducial mass-dependent intrinsic alignment (IA) model NLA‑M \citep{wright_kids_2025}, an extension of the non-linear alignment (NLA) model that includes alignments for red, early‑type galaxies and assumes zero alignment for blue, late‑type galaxies. \review{\citet{wright_kids_2025} investigated different alignment models and their impact on the inferred value for $S_8$, finding changes of not larger than $0.2\sigma$. Our choice is further supported by a red-blue split analysis conducted by \citet{2025arXiv250319442S}, which found vanishing alignment for blue galaxies.}
Early types are identified by spectral type\footnote{To determine fractions, we use the template-fitting code {\sc BPZ} \citep{benitez:2000}, which provides $T_B$ as an output.} $T_B < 1.9$, fitting six SED templates with interpolation; the cut selects galaxies with an elliptical spectral contribution. The IA alignment strength for red galaxies is modelled as a power law in the mean halo mass for each tomographic bin. 
We use \revint{two alignment parameters}, $A_\mathrm{IA}$ and $\beta$, which represent the amplitude of IAs and the slope of the IA mass scaling, respectively. In our analysis, we incorporate the joint posterior on $A_\mathrm{IA}$ and $\beta$ from \citet{fortuna_2024} as a prior, approximating it with a bivariate Gaussian distribution. Furthermore, we apply a multivariate Gaussian prior on the halo mass for each tomographic bin.

\revcollab{The cosmic shear measurements are combined with the DES Y3 data \revint{\citep{amon_dark_2022,Secco_DES_scalecut_2022}}. As in \citet{des/kids:2023}, roughly eight per cent of the DES data are removed to avoid overlap with KiDS. \revint{We do not use HSC Y3 data \citep{2023PhRvD.108l3519D,2023PhRvD.108l3518L}}}; while the likelihood is publicly available, there is significant overlap between the surveys. Since the overall constraining power of HSC Y3 is weaker than the combined power of KiDS and DES, we do not expect this omission to change the results.

\subsection{BAO and RSD data}
We use BAO measurements (BAO scale, not including the Alcock-Paczy\'{n}ski effect) from DESI DR2 \citep{2025arXiv250314738D}. \revint{DESI contains a bright galaxy survey \citep[BGS,][]{hahn_2023}, luminous red galaxies \citep[LRGs,][]{zhou_2023}, emission line galaxies \citep[ELGs,][]{raichoor_2023}, the Lyman-$\alpha$ forest \citep{bourboux2020}, and quasars \citep[QSOs,][]{chaussidon_2023}; cosmological BAO results are given in \citet{2025JCAP...02..021A}.} We employ the public DESI likelihood measurements for BGS ($z\in [0.1,0.4))$, two LRG bins ($z\in[0.4,0.6)$ and $z\in[0.6,0.8)$, respectively), ELG ($z\in[1.1,1.6))$, combined LRG+ELG ($z\in [0.8,1.1)$, QSO ($z\in[0.8,2.1))$ and Lyman‑$\alpha$ forest ($z\in[1.77,4.16)$) samples.
From the RSD measurements of the growth rate, we use the following data: the Sloan Digital Sky Survey DR7 main galaxy sample \citep{ross_2015,howlett_2015}, BOSS DR12 \citep{alam_2021}, eBOSS DR16 ELGs \citep{tamone_2020,raichoor_2023,deMattia_2021}, eBOSS DR16 LRGs \citep{bautista_2021,gilmarin_2020}, eBOSS DR16 QSOs \citep{neveux_2021,hou_2021}.

\subsection{CMB data}
We use CMB data from \citet{2020A&A...641A...5P}, employing the compressed Planck likelihood of \citet{prince_2019} where the $\ell<30$ temperature likelihood is approximated by two Gaussian points, and {\sc plik‑lite} TTTEEE is used for $\ell>30$. 
We impose a Gaussian prior on the reionisation optical depth $\tau$, derived from Planck base $\Lambda$CDM constraints, to mimic the constraints from low-multipole polarisation measurements. \revint{We note that the low-multipole polarisation measurements are not directly sensitive to the extensions studied in this paper. However, as discussed, for example, in \citet{2025arXiv250416932S}, a larger optical depth would reduce some of the discrepancy with low-redshift probes. The origin of this remains unclear.}
For the CMB lensing signal, we employ the Planck lensing signal from \citet{2022JCAP...09..039C}. We do not vary the nuisance parameter $A_\mathrm{lens}$ in the fiducial analysis. This is motivated by the addition of CMB data where $A_\mathrm{lens}$ was found to be consistent with unity \citep{2025arXiv250620707C}. For an extended discussion of the effect of $A_\mathrm{lens}$, we refer to our companion paper (Stölzner et al. 2025b).

Furthermore, we add ACT DR6 lensing and primary CMB \citep{2024ApJ...962..112Q,2024ApJ...962..113M,2025arXiv250314451N,2025arXiv250314452L,2025arXiv250314454C} 
combined with SPT-3G D1 temperature and polarisation anisotropies \citep{2025arXiv250620707C}. The SPT lensing signal is taken from SPT-3G 2018 \citep{2023PhRvD.108l2005P}. For these datasets, we use the 
{\sc candl} likelihood \citep{2024A&A...686A..10B}. To remove correlations between ACT and Planck, the Planck data are used only for $\ell < 1000$ and $\ell < 600$ in TT and TEEE, respectively. The correlations between Planck + ACT and SPT are negligible, as SPT-3G D1 only covers a small portion of the sky. Although some cross-correlation is present in the CMB lensing signals, \citet{2025arXiv250420038Q} showed that it has a negligible impact on the reported error bars. \revint{We therefore append the ACT and SPT data to our data vector at multipoles larger than the limits above.}

Lastly, we neglect any cross-correlation between the cosmic shear and CMB lensing signals in the covariance matrix. \revint{\citet{2021A&A...649A.146R,2023A&A...673A.111Y} measured} the cross-correlation of KiDS-1000 with the CMB lensing signal of the Atacama Cosmology Telescope, and \citet{2023PhRvD.107b3529O} did the same for DES Y3, Planck, and SPT. 
\revint{Thus, although there is clear cross-correlation between the signals, the cross-covariance between the CMB and cosmic shear auto-correlation function will still be relatively small. The reason is that the auto-correlation covariance contains reconstruction noise and shape noise, in addition to the auto-correlation signal. For the cross-covariance, only the signal component of the cross-correlation is relevant, and it is smaller than the auto-correlation.}

\subsection{Supernova data}
For the evolving dark energy models, we furthermore use measurements of Type Ia supernovae from the Pantheon+ compilation \citep{Scolnic22}, consisting of 1701 light curves of 1550 spectroscopically confirmed SNIa with redshifts $z\in (0.001,2.26)$. Inference of cosmological parameters from this dataset was presented in \citet{Brout22}.

\begingroup
\renewcommand*{\arraystretch}{1.2}
\begin{table}[]
    \vspace{.075cm}
    \caption{Bayes factors, $B$ as defined in \Cref{eq:bayes}, for the extended models for the different probe combinations.}
    \centering
    \begin{tabular}{c|ccc}
    \hline \hline 
  \textbf{probes}  & \multicolumn{3}{c}{\textbf{Bayes factor}} \\
         & $w_0/\Lambda$ & $w_0w_a/\Lambda $ & $\Omega_K/\mathrm{flat}$ \\ \hline
        KiDS & $  0.38 $ & $ 0.2 $ & $ 0.95  $\\ 

lensing & $  0.19 $ & $ 0.07 $ & $ 0.56  $\\ 

low-$z$ & $  0.09 $ & $ 0.02 $ & $ 0.17  $\\ 

low-$z+$CMB& $  0.22 $ & $ 2.73 $ & $ 0.38  $\\ \hline
    \end{tabular}
    \label{tab:bayes_factor}
    \tablefoot{\revint{The columns always compare the extended model to the fiducial flat $\Lambda$CDM. This means that $B>1$ prefers the extension, while $B<1$ prefers flat $\Lambda$CDM. 
    According to the Jeffreys' scale, there is no significant evidence for any of the extended models being preferred over the extensions tested here for any data combination. Even more so, extensions to flat $\Lambda$CDM  are disfavoured in all but one case.}}
\end{table}
\endgroup

\subsection{Data combinations used}
The data combinations used are summarised in \Cref{tab:data_combiantions}. We focus on constraints from KiDS-Legacy alone, which we refer to as KiDS in the discussion and plots. The combination of SPT, Planck, and ACT is labelled CMB \citep[this corresponds to the SPA likelihood in][]{2025arXiv250620707C}. All probes that use the weak gravitational lensing effect, that is, CMB lensing, KiDS, and DES Y3, are collectively referred to as `lensing'. We label the low-redshift probes as low-$z$, that is, those that do not depend on the CMB. Lastly, we combine all probes but the supernovae into the low-$z$+CMB likelihood. 
All external likelihoods used in this work are publicly available and implemented in the {\sc CosmoSIS} standard library\footnote{\url{https://github.com/cosmosis-developers/cosmosis-standard-library/tree/v4.0/likelihood}} \citep{zuntz_2015}.

\subsection{Inference}
We use {\sc CosmoPipe} and sample the parameter space via the {\sc Nautilus}\footnote{\url{https://github.com/johannesulf/nautilus}} sampler \citep{lange2023} to generate samples from the posterior distribution. As a nested sampling method \citep{Skilling_Nested_2007}, {\sc Nautilus} allows for direct computation of the evidence and hence easy estimation of the Bayes factor ${B}$ for model comparison:
\begin{equation}
\label{eq:bayes}
    B = \frac{p(D|M_1)}{p(D|M_2)}\;, 
\end{equation}
where $p(D|M_1)$ is the evidence of the model $M_1$. If $B>1$, $M_1$ is preferred over model $M_2$ and vice versa for $B <1$.
The amount of preference can be quantified using the Jeffreys' scale. It quantifies a model preference with \revint{$|\log_{10} B| > [0.5,1,2]$ as `substantial', `strong', and `decisive'. The Bayes ratio suffers from several shortcomings: it is prior-dependent, not a probability, and, as a result, the Jeffreys scale is somewhat arbitrary. 
To avoid prior effects, we also report the tension between the two models based on suspiciousness \citep{Handley19}, where the parameter posteriors indicate some level of tension. For more details, we refer to section 3.2.1 in \citet{2025arXiv250319442S}.
However, since most models will be very consistent with the fiducial $\Lambda$CDM analysis, we will use the Bayes factors.}
For {\sc Nautilus}, we use the following settings: $f_\mathrm{live} = 0.01$, $N_\mathrm{live} = 3000$, and $N_\mathrm{eff} = 10000$.

\revint{We use the cosmological parameter inference code {\sc Cosmosis} \citep{zuntz_2015} to facilitate all of our computations and use {\sc CAMB} \citep{2000ApJ...538..473L} for the corresponding linear power spectrum to be fed into {\sc HMCODE2020} for the non-linear power spectra.} 
The likelihood for each probe is assumed to be a multivariate Gaussian likelihood with the cosmic shear covariance matrix calculated via the {\sc OneCovariance}\footnote{\url{https://github.com/rreischke/OneCovariance}} \citep{2025A&A...699A.124R} at the $\Lambda$CDM best-fit cosmology.

\review{A list of model parameters is provided in \Cref{tab:parameters}. For brevity, we show only nuisance parameters relevant to cosmic shear.
The flat $\Lambda$CDM parameters are the reduced Hubble constant $h = H_0[\mathrm{km}\,\mathrm{s}^{-1}\mathrm{Mpc}^{-1}]/100$, the cold dark matter and baryon densities $\omega_\mathrm{cdm} = \Omega_\mathrm{cdm}h^2$ and $\omega_\mathrm{b} = \Omega_\mathrm{b}h^2$, the spectral index of the primordial power spectrum and its amplitude, which we parametrise by either the variance $\sigma^2_8$ in spheres of $8\,\mathrm{Mpc}/h$, or by the amplitude of the matter power spectrum $A_\mathrm{s}$ at a pivot scale $k = 0.05\,\mathrm{Mpc}^{-1}$. Since cosmic shear is most sensitive to $S_8 = \sigma_8\sqrt{\Omega_\mathrm{m}/0.3}$, we use $S_8$ by default as the sampling parameter and derive the others. For the extended models, we use the sum of the neutrino masses $\sum m_\nu$, the spatial curvature parameter defined in \Cref{eq:curvature}, and the $w_0,w_a$-parametrisation as in \Cref{eq:cpl}.}

\begin{figure}
    \centering
    \includegraphics[width=0.95\linewidth]{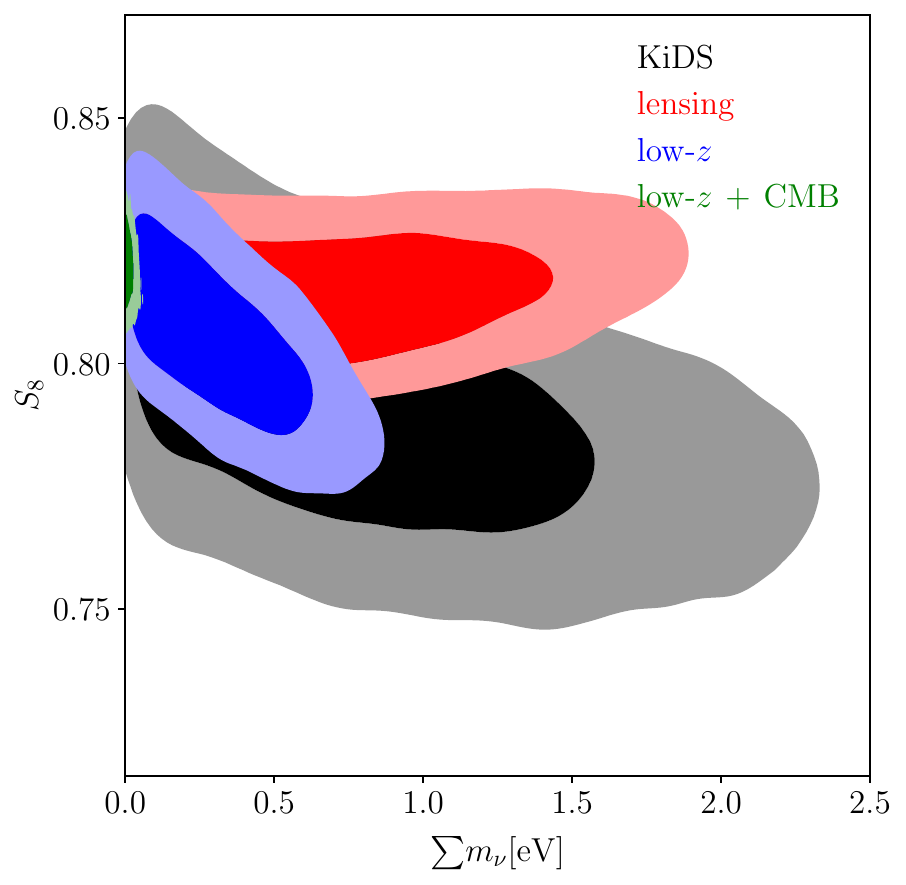}
    \caption{Marginalised two-dimensional $68\,\%$ and $95\,\%$ credible regions in the $S_8$,$\sum m_\nu$ plane. Different colours represent the probe combinations discussed in \Cref{tab:data_combiantions}.}
    \label{fig:mnu}
\end{figure}

\section{Constraints on extended models}
\label{sec:constraints}
In this section, we present the results for the extended cosmological models and for different combinations of datasets. For quick reference, the Bayes factors for model selection are shown in \Cref{tab:bayes_factor}. The marginal mean and its corresponding $68\,\%$ credible interval for all probe and model combinations for the main cosmological parameters are summarised in \Cref{tab:constraints}. We also show the $p$-values, calculated using the recipe discussed in \citet{2025arXiv250319442S} (see sect. 3.2.1), for the KiDS data alone for each case. This result differs slightly from \citet{wright_kids_2025}, where the maximum a posteriori (MAP) probability was explicitly calculated: here, instead, we use the MAP from the posterior samples. \revint{As a result, the $p$-values presented here are an upper limit.}

Here, we focus on the constraints on the parameters specific to the extended models studied and their interplay with $S_8$. For more details, we show the full contours for all cosmological parameters and the feedback parameter in \Cref{app:2}. We find that the nuisance parameters, in particular the intrinsic alignment parameters, as well as the redshift calibration parameters, are prior-dominated \review{and exhibit no cross-talk with extended models and probes. More precisely, the extended model parameters do not exhibit strong degeneracies with the nuisance parameters in the range allowed by the priors and the overall signal-to-noise of the cosmic shear measurement.}
Lastly, we note that constraints on $\Omega_K$ and $\sum m_\nu$ from lensing alone are not competitive with those from the CMB or BAO, and in particular with their combination. However, the current comparison is not on equal footing, as cosmic shear is still in its golden years. Hence, this situation will improve as we move to Stage IV surveys \citep[see][for forecasted constraints on curvature or neutrinos]{2020A&A...642A.191E,2025A&A...693A..58E}.

\subsection{Neutrinos}
The observation of neutrino oscillations necessitates that at least two of the three neutrino mass eigenstates have non-zero mass. Hence, their sum $\sum m_\nu >0$. In what follows, we express the neutrino mass in natural units, that is, $c=1$. In the fiducial KiDS analysis, ${\sum m_\nu = 0.06\,\mathrm{eV}}$ was assumed, which is in line with the lowest allowed mass range in the normal hierarchy from oscillation measurements \citep{deSalas:2020pgw, Esteban:2020cvm, 2024JHEP...12..216E}. Other laboratory experiments can instead give upper limits, $\leq 0.45\,\mathrm{eV}\;(\mathrm{at}\;90\,\%)$ via double-$\beta$ decay \citep[this is, though, a weighted sum of the neutrino masses with the weights given by the PMNS mixing matrix;][]{2025Sci...388..180K}. Due to their low mass and resulting high thermal velocities, neutrinos do not cluster on small scales, thus suppressing the growth of structures \citep[see e.g.][for overviews as well as first constraints with cosmic shear]{2006PhR...429..307L,2009A&A...500..657T,abazajian_neutrinos_2015,gerbino_status_2017}. Constraints from cosmology are mainly dominated by the CMB \citep{PlanckCollab_CMB2020,2025arXiv250620707C}, limiting $\sum m_\nu < 0.17\,\mathrm{eV}$. In combination with BOSS \citep{ivanov_cosmological_2020} and DESI \citep{2025JCAP...02..021A}, the cosmological measurement of $\sum m_\nu$ approaches the lower bound set by oscillation experiments, assuming the normal hierarchy.
Here, we investigate the influence of varying neutrino mass on the constraints on $S_8$ by KiDS as well as the upper limits on $\sum m_\nu$ themselves. In the analysis presented here, we assume the normal hierarchy.
For consistency with many other analyses, we run {\sc{camb}} \citep{2000ApJ...538..473L} with fixed $N_\mathrm{eff}$ and three neutrino species approximated as a single massive and two massless eigenstates \citep[as in][]{PlanckCollab_CMB2020} and pass their sum to {\sc{HMCODE2020}}, a combination which was tested against {\sc{halofit}} in \citet{2025arXiv250620707C} and found to give changes of $0.2\sigma$. However, this implies the existence of two ultra-relativistic species. As a result, the inferred total neutrino mass tends to be systematically lower compared with the fully degenerate mass scenario \citep[see e.g.][]{2019PhRvL.123h1301L,2020JCAP...07..037C}.

\Cref{fig:mnu} shows the marginalised constraints in the $S_8$,$\sum m_\nu$ plane for the different probe combinations listed in \Cref{tab:data_combiantions}. \revint{Since neutrinos affect clustering directly but also change the background by reducing $\Omega_\mathrm{m}$ when increasing $\sum m_\nu$, the interpretation is less straightforward than, for example, for feedback.}
Interestingly, we find a small anti-correlation between $S_8$ and $\sum m_\nu$ for KiDS alone. More probability mass at larger neutrino masses shifts the peak of the matter power spectrum towards lower wave numbers, $k$, thereby favouring slightly larger values of the spectral index, $n_\mathrm{s}$. This overall shift results in a slight amplitude gain for the cosmic shear signal, which, in turn, is accompanied by a slightly lower value of $S_8$. 

However, as this is not a significant shift in the parameters, the $p$-value at the MAP \revint{($S_8=0.793$ with neutrinos marginalised and $S_8=0.812$ for the fiducial analysis)} is equally good as the one for the $\Lambda$CDM case with a fixed neutrino mass.
The upper limit on the neutrino mass from KiDS alone is  
\begin{equation*}
    \sum m_\nu \leq   1.85\,\mathrm{eV} \quad (\mathrm{KiDS}\;\mathrm{at}\;95\,\%)\;,
\end{equation*}
\revint{that is, 95$\%$ of the marginal posterior probability in $\sum m_\nu$ lies below this value.}
Adding DES and, crucially, CMB lensing breaks the degeneracy with $S_8$, as the CMB lensing signal is sensitive to larger scales. The low-redshift probes, in turn, mainly fix the matter density, thus not breaking the degeneracy. The strongest constraints come from the CMB, particularly when combined with DESI data, as increasing $\sum m_\nu$ increases the expansion rate at the time of CMB release. To keep the sound horizon consistent with the data, $\Omega_\mathrm{m}$ must increase. However, DESI's lower $\Omega_\mathrm{m}$ thus gives very tight constraints on the neutrino masses for which we find $\sum m_\nu \leq 0.048\,\mathrm{eV}$ (at $95\,\%$), consistent with \citet{2025arXiv250620707C} and in tension with the neutrino oscillation measurements, $\sum m_\nu \geq 0.058$eV for the normal hierarchy. \revint{In particular, these values appear to reject the normal hierarchy at $2.1\sigma$ and over $3\sigma$ for the inverted hierarchy.} 
\revint{It is noteworthy that this value is also lower than the fiducial value in the KiDS-Legacy analysis. However, since this is a small change relative to the overall sensitivity of the cosmic shear data, it will not affect the results of that analysis.}

\begin{figure}
    \centering
    \includegraphics[width=0.95\linewidth]{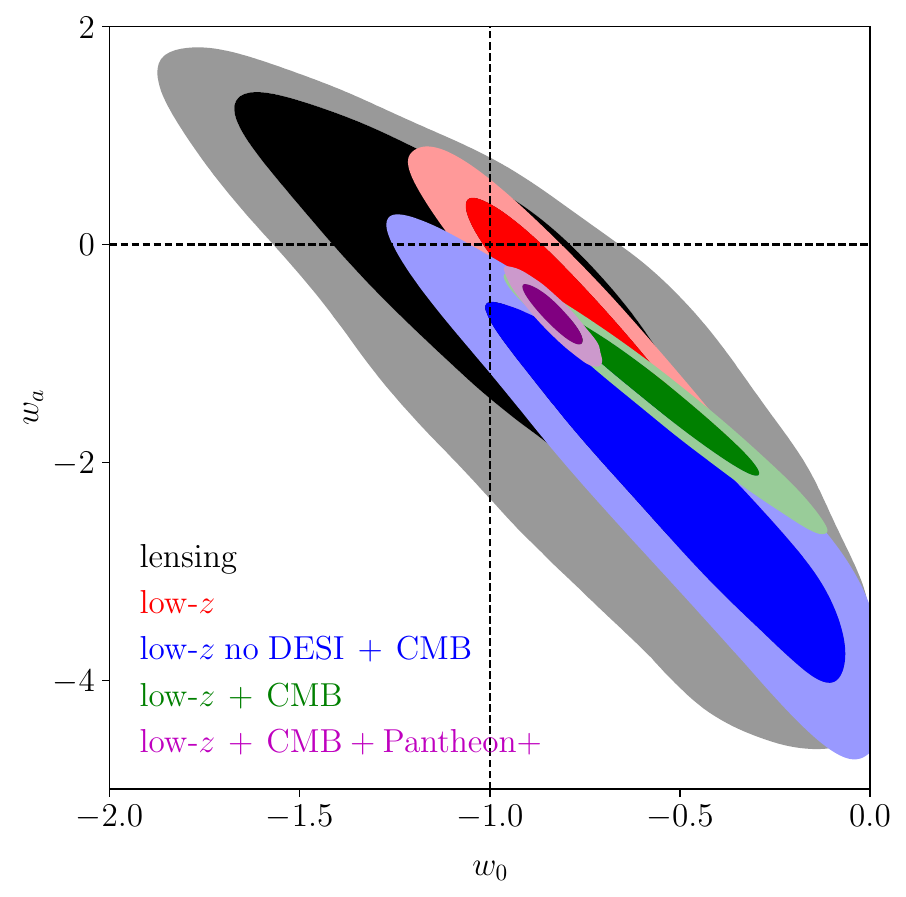}
    \caption{Marginalised constraints on dynamical dark energy in the \revint{$w_0,w_a$ parametrisation}, \Cref{eq:cpl}. The black contour contains all cosmic shear probes considered in this work, and the red contour is independent of the CMB. 
    The blue and green contours differ only in that the blue one excludes the DESI DR2 BAO. In purple, the Pantheon+ SN sample has been added. The dashed lines mark the $\Lambda$CDM values.}
    \label{fig:w0wa}
\end{figure}

As expected, the neutrino sector is not very well constrained with KiDS data alone, but constraints from cosmic shear alone are of similar quality compared to the previous KiDS-1000's {$3\times 2$pt} analysis \citep{2021A&A...649A..88T}.

\subsection{Dark energy equation of state}
We vary the dark energy equation of state parameter with the
standard \citep[CPL;][]{2001IJMPD..10..213C,2003PhRvL..90i1301L} parametrisation:
\begin{equation}
\label{eq:cpl}
    w(a) = w_0 +w_a(1-a),
\end{equation}
with the uniform priors given in \Cref{tab:parameters}. We do not explicitly impose $w < -1/3$, which is required for accelerated expansion, in the prior.
In this model, $w_0 = -1$ and $w_a=0$ corresponds to the cosmological constant $\Lambda$. The cosmological background is modified so that:
\begin{equation}
    E^2(a)  = \Omega_\mathrm{m}a^{-3} + (1-\Omega_\mathrm{m})\,\mathrm{exp}\left({-3\int_1^a\mathrm{d}a^\prime\frac{1+w(a^\prime)}{a^\prime}}\right),
\end{equation}
where the spatial curvature still vanishes.
This parametrisation has attracted considerable attention recently, owing to the DESI results, which have measured $w_0$ and $w_a$ to be discrepant with $\Lambda$.
This, in turn, has sparked a debate. For example, that the BAO data measures $w(z) = -1$ at the redshifts where the BAO has been measured \citep[see e.g.][]{2024JCAP...12..007C,2025MNRAS.540.2844E}, or whether prior volume effects play a role \citep[e.g.][]{2025arXiv250612004H,2025arXiv250403829S,2025JCAP...08..065B}, if theoretical priors \citep[e.g][]{2025JCAP...05..065L,2025arXiv250913318T,2025arXiv251107526C} can reconcile this tension, or if this hints towards non-minimally coupled gravity theories \citep{2025PhRvL.134r1002Y}.

\begin{figure}
    \centering
    \includegraphics[width=0.95\linewidth]{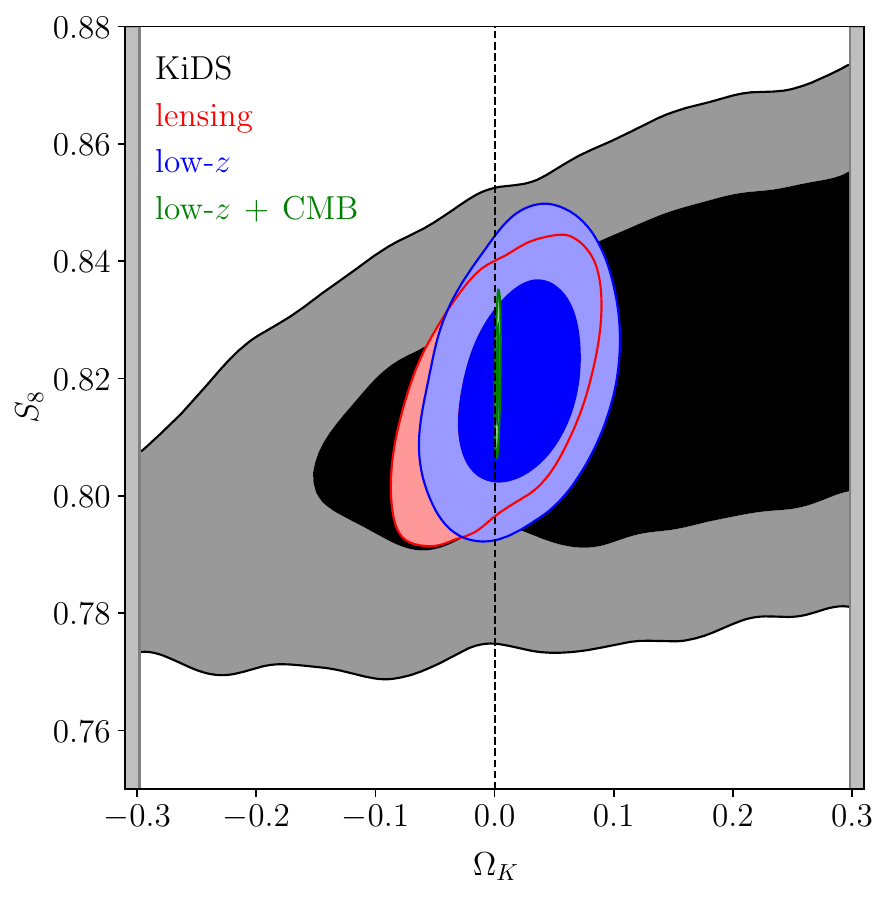}
    \caption{Constraints on $\Omega_K$CDM in the $S_8,\Omega_K$ plane. The colour scheme is the same as in \Cref{fig:mnu}. The low-$z+$CMB constraint (green) lies right next to the dashed line. The prior range is indicated by the grey bands.}
    \label{fig:OK}
\end{figure}

\begingroup
\renewcommand{\arraystretch}{1.8}
\begin{table*}[]
    \vspace{.075cm}
\setlength{\tabcolsep}{4.5pt}
\caption{Marginal constraints for all probe combinations (compare \Cref{tab:data_combiantions}) and models tested in this work.}
    \label{tab:constraints}
    \centering
    \scriptsize
\begin{tabular}{ccccccccccccc}    
    \hline\hline
model & probes &  $S_8$ & $\Omega_\mathrm{m}$ & $n_\mathrm{s}$  & $\sigma_8$  & $\log_{10} T_\mathrm{AGN}$&$\sum m_\nu\, [\mathrm{eV}]$ & $w_0$ & $w_a$ & $\Omega_K$ & $p_\mathrm{KiDS}$ \\ \hline

\multirow{4}{*}{\rotatebox[origin=c]{90}{$\Lambda$CDM}}  &
 KiDS & $\mathbf{0.813^{\mathstrut+\,0.018}_{\mathstrut-\,0.018}}$ &   $0.329^{\mathstrut+\,0.055}_{\mathstrut-\,0.054}$ &   $1.001^{\mathstrut+\,0.067}_{\mathstrut-\,0.070}$ &   $0.786^{\mathstrut+\,0.078}_{\mathstrut-\,0.079}$ &   $7.69^{\mathstrut+\,0.31}_{\mathstrut-\,0.29}$ & 0.06 & -1 & 0 & 0 & 0.39\\

 & lensing & $\mathbf{0.8181^{\mathstrut+\,0.0094}_{\mathstrut-\,0.0098}}$ &   $\mathbf{0.281^{\mathstrut+\,0.016}_{\mathstrut-\,0.016}}$ &  ${0.974^{\mathstrut+\,0.038}_{\mathstrut-\,0.041}}$ &   $\mathbf{0.846^{\mathstrut+\,0.023}_{\mathstrut-\,0.024}}$ &   $7.54^{\mathstrut+\,0.19}_{\mathstrut-\,0.18}$  & 0.06 & -1 & 0 & 0 & 0.43 \\
 
& low-$z$ & $\mathbf{0.817^{\mathstrut+\,0.011}_{\mathstrut-\,0.010}}$ &   $\mathbf{0.2965^{\mathstrut+\,0.0081}_{\mathstrut-\,0.0079}}$ &  $1.035^{\mathstrut+\,0.043}_{\mathstrut-\,0.041}$ &   $\mathbf{0.824^{\mathstrut+\,0.016}_{\mathstrut-\,0.016}}$ &   $7.60^{\mathstrut+\,0.23}_{\mathstrut-\,0.22}$ & 0.06 & -1 & 0 & 0 & 0.35 \\

& low-$z+$CMB& $\mathbf{0.8158^{\mathstrut+\,0.0061}_{\mathstrut-\,0.0056}}$ &   $\mathbf{0.3038^{\mathstrut+\,0.0031}_{\mathstrut-\,0.0029}}$ &   $\mathbf{0.9736^{\mathstrut+\,0.0032}_{\mathstrut-\,0.0031}}$ &   $\mathbf{0.8108^{\mathstrut+\,0.0033}_{\mathstrut-\,0.0041}}$ &   $7.50^{\mathstrut+\,0.15}_{\mathstrut-\,0.15}$  & 0.06 & -1 & 0 & 0 &  0.38 \\  \hline 

\multirow{4}{*}{\rotatebox[origin=c]{90}{$\nu\Lambda$CDM}}  & KiDS &   $\mathbf{0.79^{\mathstrut+\,0.02}_{\mathstrut-\,0.02}}$ &  $0.356^{\mathstrut+\,0.060}_{\mathstrut-\,0.061}$ &  $1.017^{\mathstrut+\,0.062}_{\mathstrut-\,0.065}$ &  $0.752^{\mathstrut+\,0.075}_{\mathstrut-\,0.075}$ &  $7.69^{\mathstrut+\,0.31}_{\mathstrut-\,0.29}$ &  $\mathbf{\leq 1.97}$ &  -1 & 0& 0 & 0.49\\

& lensing &   $\mathbf{0.8149^{\mathstrut+\,0.0092}_{\mathstrut-\,0.0095}}$ &  $\mathbf{0.296^{\mathstrut+\,0.017}_{\mathstrut-\,0.018}}$ &  $0.981^{\mathstrut+\,0.043}_{\mathstrut-\,0.045}$ &  $\mathbf{0.821^{\mathstrut+\,0.024}_{\mathstrut-\,0.024}}$ &  $7.58^{\mathstrut+\,0.22}_{\mathstrut-\,0.21}$ &  $\mathbf{\leq 1.61}$ & -1 & 0& 0 & 0.39 \\

& low-$z$ &  $\mathbf{0.805^{\mathstrut+\,0.014}_{\mathstrut-\,0.014}}$ &  $\mathbf{0.2980^{\mathstrut+\,0.0077}_{\mathstrut-\,0.0083}}$ &  $1.050^{\mathstrut+\,0.037}_{\mathstrut-\,0.038}$ &  $\mathbf{0.818^{\mathstrut+\,0.017}_{\mathstrut-\,0.016}}$ &  $7.59^{\mathstrut+\,0.22}_{\mathstrut-\,0.22}$ &  $\mathbf{\leq 0.74}$ &-1 & 0& 0 & 0.39 \\

& low-$z+$CMB&  $\mathbf{0.820^{\mathstrut+\,0.0059}_{\mathstrut-\,0.0057}}$ &  $\mathbf{0.3004^{\mathstrut+\,0.0029}_{\mathstrut-\,0.0030}}$ &  $\mathbf{0.9731^{\mathstrut+\,0.0031}_{\mathstrut-\,0.0031}}$ &  $\mathbf{0.8195^{\mathstrut+\,0.0048}_{\mathstrut-\,0.0047}}$ &  $7.51^{\mathstrut+\,0.16}_{\mathstrut-\,0.16}$ &  $\mathbf{\leq 0.048}$ &-1 & 0& 0 & 0.40 \\ \hline

\multirow{4}{*}{\rotatebox[origin=c]{90}{$w$CDM}}  & KiDS
&   $\mathbf{0.809^{\mathstrut+\,0.040}_{\mathstrut-\,0.041}}$ &  $0.332^{\mathstrut+\,0.058}_{\mathstrut-\,0.059}$ &  $0.998^{\mathstrut+\,0.075}_{\mathstrut-\,0.080}$ &  $0.780^{\mathstrut+\,0.079}_{\mathstrut-\,0.078}$ &  $7.70^{\mathstrut+\,0.33}_{\mathstrut-\,0.30}$ & 0.06 & $\mathbf{-1.10}^{\mathstrut+\,0.45}_{\mathbf{\mathstrut-\,0.47}}$ & 0 & 0 & 0.39\\ 

& lensing &  $\mathbf{0.811^{\mathstrut+\,0.017}_{\mathstrut-\,0.016}}$ &  $\mathbf{0.268^{\mathstrut+\,0.039}_{\mathstrut-\,0.038}}$ &  $0.980^{\mathstrut+\,0.040}_{\mathstrut-\,0.042}$ &  $\mathbf{0.864^{\mathstrut+\,0.049}_{\mathstrut-\,0.048}}$ &  $7.54^{\mathstrut+\,0.19}_{\mathstrut-\,0.18}$ & 0.06 &$\mathbf{-1.09^{\mathstrut+\,0.21}_{\mathstrut-\,0.21}}$ & 0 & 0 & 0.35\\ 

& low-$z$ &   $\mathbf{0.824^{\mathstrut+\,0.013}_{\mathstrut-\,0.013}}$ &  $\mathbf{0.2970^{\mathstrut+\,0.0079}_{\mathstrut-\,0.0081}}$ &  $1.04^{\mathstrut+\,0.04}_{\mathstrut-\,0.04}$ &  $\mathbf{0.830^{\mathstrut+\,0.017}_{\mathstrut-\,0.017}}$ &  $7.60^{\mathstrut+\,0.23}_{\mathstrut-\,0.23}$ & 0.06  & $\mathbf{-0.943^{\mathstrut+\,0.062}_{\mathstrut-\,0.064}}$ & 0 & 0 & 0.36\\ 

& low-$z+$CMB &   $\mathbf{0.8150^{\mathstrut+\,0.0052}_{\mathstrut-\,0.0063}}$ &  $\mathbf{0.2905^{\mathstrut+\,0.0069}_{\mathstrut-\,0.0067}}$ &  $\mathbf{0.972^{\mathstrut+\,0.003}_{\mathstrut-\,0.003}}$ &  $\mathbf{0.8284^{\mathstrut+\,0.0099}_{\mathstrut-\,0.0089}}$ &  $7.51^{\mathstrut+\,0.16}_{\mathstrut-\,0.16}$ & 0.06 & $\mathbf{-1.06^{\mathstrut+\,0.03}_{\mathstrut-\,0.03}}$ & 0 & 0 & 0.32 \\ \hline

\multirow{6}{*}{\rotatebox[origin=c]{90}{$w_0w_a$CDM}} & KiDS &   $\mathbf{0.799^{\mathstrut+\,0.041}_{\mathstrut-\,0.042}}$ &  $0.341^{\mathstrut+\,0.063}_{\mathstrut-\,0.063}$ &  $1.007^{\mathstrut+\,0.069}_{\mathstrut-\,0.075}$ &  $0.760^{\mathstrut+\,0.078}_{\mathstrut-\,0.078}$ &  $7.67^{\mathstrut+\,0.31}_{\mathstrut-\,0.27}$ & 0.06  & $-0.98^{\mathstrut+\,0.65}_{\mathstrut-\,0.65}$ &  
$\mathbf{-1.3}^{\mathbf{\mathstrut+\,1.93}}_{\mathstrut-\,2.01}$ & 0 & 0.41 \\

& lensing  & $\mathbf{0.815^{\mathstrut+\,0.020}_{\mathstrut-\,0.021}}$ &  $\mathbf{0.270}^{\mathstrut+\,\mathbf{0.041}}_{\mathstrut-\,0.040}$ &  $0.984^{\mathstrut+\,0.041}_{\mathstrut-\,0.042}$ &  $\mathbf{0.864^{\mathstrut+\,0.047}_{\mathstrut-\,0.047}}$ &  $7.54^{\mathstrut+\,0.19}_{\mathstrut-\,0.18}$ & 0.06 & $\mathbf{-0.92}^{\mathstrut+\,0.45}_{\mathbf{\mathstrut-\,0.42}}$ &  $\mathbf{-0.88}^{\mathbf{\mathstrut+\,1.45}}_{\mathstrut-\,1.56}$ & 0 & 0.41 \\

& low-$z$  &  $\mathbf{0.827^{\mathstrut+\,0.014}_{\mathstrut-\,0.013}}$ &  $\mathbf{0.322^{\mathstrut+\,0.026}_{\mathstrut-\,0.026}}$ &  $1.032^{\mathstrut+\,0.046}_{\mathstrut-\,0.044}$ &  $\mathbf{0.803^{\mathstrut+\,0.032}_{\mathstrut-\,0.032}}$ &  $7.61^{\mathstrut+\,0.24}_{\mathstrut-\,0.23}$ & 0.06  & $\mathbf{-0.74^{\mathstrut+\,0.22}_{\mathstrut-\,0.21}}$ &  $\mathbf{-0.80^{\mathstrut+\,0.80}_{\mathstrut-\,0.83}}$ & 0 &0.38 \\

& low-$z+$CMB &  $\mathbf{0.8365^{\mathstrut+\,0.0084}_{\mathstrut-\,0.0076}}$ &  $\mathbf{0.329^{\mathstrut+\,0.013}_{\mathstrut-\,0.013}}$ &  $\mathbf{0.9709^{\mathstrut+\,0.0031}_{\mathstrut-\,0.0032}}$ &  $\mathbf{0.799^{\mathstrut+\,0.011}_{\mathstrut-\,0.011}}$ &  $7.53^{\mathstrut+\,0.18}_{\mathstrut-\,0.17}$ &  0.06 & $\mathbf{-0.63^{\mathstrut+\,0.13}_{\mathstrut-\,0.13}}$ &  $\mathbf{-1.20^{\mathstrut+\,0.39}_{\mathstrut-\,0.38}}$ & 0 & 0.28 \\

& low-$z+$SPA$+$Pantheon+ &   $\mathbf{0.8270^{\mathstrut+\,0.0063}_{\mathstrut-\,0.0068}}$ &  $\mathbf{0.3101^{\mathstrut+\,0.0052}_{\mathstrut-\,0.0056}}$ &  $\mathbf{0.9713^{\mathstrut+\,0.0032}_{\mathstrut-\,0.0031}}$ &  $\mathbf{0.8135^{\mathstrut+\,0.0073}_{\mathstrut-\,0.0073}}$ &  $7.52^{\mathstrut+\,0.17}_{\mathstrut-\,0.17}$ & 0.06 &  $\mathbf{-0.831^{\mathstrut+\,0.052}_{\mathstrut-\,0.053}}$ &  $\mathbf{-0.65^{\mathstrut+\,0.19}_{\mathstrut-\,0.18}}$ & 0 & 0.32\\

& low-$z$ no DESI$+$CMB &  $\mathbf{0.829^{\mathstrut+\,0.013}_{\mathstrut-\,0.012}}$ &  $\mathbf{0.302^{\mathstrut+\,0.027}_{\mathstrut-\,0.027}}$ &  $\mathbf{0.9709^{\mathstrut+\,0.0032}_{\mathstrut-\,0.0033}}$ &  $\mathbf{0.828^{\mathstrut+\,0.025}_{\mathstrut-\,0.025}}$ &  $7.51^{\mathstrut+\,0.16}_{\mathstrut-\,0.16}$ &  0.06 & $\mathbf{-0.51}^{\mathstrut+\,0.32}_{\mathbf{\mathstrut-\,0.31}}$ &  $\mathbf{-2.26^{\mathstrut+\,1.15}_{\mathstrut-\,1.12}}$ & 0 & 0.38\\ \hline

\multirow{4}{*}{\rotatebox[origin=c]{90}{$\Omega_K\Lambda$CDM}} & KiDS &   $\mathbf{0.816^{\mathstrut+\,0.021}_{\mathstrut-\,0.021}}$ &  $0.336^{\mathstrut+\,0.058}_{\mathstrut-\,0.057}$ &  $0.994^{\mathstrut+\,0.073}_{\mathstrut-\,0.075}$ &  $0.782^{\mathstrut+\,0.080}_{\mathstrut-\,0.079}$ &  $7.72^{\mathstrut+\,0.33}_{\mathstrut-\,0.30}$ &  0.06 &-1 & 0& $0.08^{\mathstrut+\,0.16}_{\mathstrut-\,0.17}$ & 0.48\\

& lensing &  $\mathbf{0.818^{\mathstrut+\,0.010}_{\mathstrut-\,0.011}}$ &  $\mathbf{0.282^{\mathstrut+\,0.018}_{\mathstrut-\,0.018}}$ &  $0.97^{\mathstrut+\,0.07}_{\mathstrut-\,0.07}$ &  $\mathbf{0.85^{\mathstrut+\,0.03}_{\mathstrut-\,0.03}}$ &  $7.55^{\mathstrut+\,0.20}_{\mathstrut-\,0.19}$ &  0.06 &-1 & 0 & $\mathbf{0.001^{\mathstrut+\,0.040}_{\mathstrut-\,0.041}}$  & 0.34 \\ 

& low-$z$ &  $\mathbf{0.820^{\mathstrut+\,0.011}_{\mathstrut-\,0.011}}$ &  $\mathbf{0.2932^{\mathstrut+\,0.0096}_{\mathstrut-\,0.0097}}$ &  $1.040^{\mathstrut+\,0.041}_{\mathstrut-\,0.041}$ &  $\mathbf{0.832^{\mathstrut+\,0.019}_{\mathstrut-\,0.019}}$ &  $7.60^{\mathstrut+\,0.23}_{\mathstrut-\,0.22}$ &  0.06 &-1 & 0 & $\mathbf{0.022^{\mathstrut+\,0.034}_{\mathstrut-\,0.034}}$   & 0.43 \\

& low-$z+$CMB&   $\mathbf{0.8208^{\mathstrut+\,0.0061}_{\mathstrut-\,0.0052}}$ &  $\mathbf{0.3032^{\mathstrut+\,0.0027}_{\mathstrut-\,0.0028}}$ &  $\mathbf{0.9695^{\mathstrut+\,0.0035}_{\mathstrut-\,0.0036}}$ &  $\mathbf{0.8165^{\mathstrut+\,0.0042}_{\mathstrut-\,0.0045}}$ &  $7.52^{\mathstrut+\,0.17}_{\mathstrut-\,0.17}$ & 0.06 &-1 & 0 &  $\mathbf{0.0026^{\mathstrut+\,0.001}_{\mathstrut-\,0.001}}$   & 0.32 \\ \hline
    \end{tabular}
    \tablefoot{The constraints are provided as the marginal mean and the $68\,\%$ confidence interval relative to it. For neutrinos, we quote the $95\,\%$ upper limit.
    The last column shows the $p$-values (calculated as the probability-to-exceed) at the maximum a posteriori (MAP) for KiDS in the specific data and model combination. \revint{Numbers shown in bold are parameters which are constrained by the given model and probe combination \citep[see Appendix A in][]{asgari_kids_2021}.} 
We note that, because parameters such as $\log_{10}T_\mathrm{AGN}$ have highly asymmetric posterior distributions, we do not report a mode. Lastly, the constraints on $h$ are not explicitly given, as we focus on those parameters potentially constrained by current lensing data.}
\end{table*}
\endgroup

Here, we test the full $w_0,w_a$ parametrisation and the subclass where $w_a = 0$, these are called $w_0w_a$CDM and $w$CDM respectively. 
For KiDS alone, we find the following constraints:
\begin{equation*}
\begin{split}
       & w_0 =-1.1^{\mathstrut+\,0.5}_{\mathstrut-\,0.5} \quad(\mathrm{KiDS\; for\;}w_0\mathrm{CDM\;at}\;68\%),\\
       & w_0 = -1.0^{\mathstrut+\,0.7}_{\mathstrut-\,0.7} \quad(\mathrm{KiDS\; for\;}w_0w_a\mathrm{CDM\;at}\;68\%),\\
       & w_a = -1.3^{\mathstrut+\,1.9}_{\mathstrut-\,2.0} \quad(\mathrm{KiDS\; for\;}w_0w_a\mathrm{CDM\;at}\;68\%),
\end{split}
\end{equation*}
consistent with $\Lambda$CDM and thus the fiducial analysis of KiDS. The dark energy equation of state alters the expansion function and, consequently, the growth of structure. The constraints obtained from weak lensing on the equation of state are determined by the effective equation of state, $w_\mathrm{eff}$, \revint{at the mean redshift of the source distribution for each tomographic bin}. Different models are then distinguished by the amount of structure growth from that point in time to today, and by what that point in time represents, that is, by the distance-redshift relationship. 
Increasing $w_\mathrm{eff}$ moves the source closer to the observer at fixed $z$, decreasing the lensing signal, requiring an increase in $S_8$.
At the same time, however, the amount of structure growth at a given $z>0$ is increased by increasing $w_\mathrm{eff}$ if the amplitude of the matter power spectrum is kept fixed.
The exact degeneracy between $S_8$ and $w_\mathrm{eff}$ therefore depends on the redshift baseline of the survey in question and is difficult to disentangle since $S_8$ depends on $\Omega_\mathrm{m}$, which itself modifies the distance-redshift relation. Generally, however, one does find a strong positive correlation between $S_8$ and $w_\mathrm{eff}$. This is why weak lensing benefits significantly from the addition of external probes that fix either $\sigma_8$ or $\Omega_\mathrm{m}$.

In \Cref{fig:w0wa}, we present the results for the $w_0,w_a$ parametrisation with external probes, showing that the lensing and low-$z$ probe combinations are consistent with $\Lambda$CDM. The addition of CMB data to the low-redshift probes (green contours) moves the best fit away from $\Lambda$CDM, since the CMB is able to break degeneracies in the DESI data with $\Omega_\mathrm{m}$, making the low-$z+$CMB combination particularly strong. Crucially, as the green contour, although shifted away slightly from a cosmological constant, is consistent with $\Lambda$CDM, the addition of BAO data as measured by DESI is extremely important to move the contours away from $\Lambda$CDM. {We remark that the prior range on $h$ (see \Cref{tab:parameters}) is fairly tight. This choice is based on the $\Lambda$CDM analysis of KiDS-Legacy \citep{wright_kids_2025}}. It does not restrict the analysis in $\Lambda$CDM when combining with the CMB \citep{2025arXiv250319442S}. However, opening the parameter space for dynamical dark energy leads to a large posterior probability at low values of $h$ and high values of $\Omega_\mathrm{m}$, as can be seen in \Cref{fig:all_full_post}. This occurs since the CMB is mostly sensitive to the physical density $\rho_\mathrm{m} \sim h^2\Omega_\mathrm{m}$. Hence, when performing an analysis that includes SPA, we increase the prior range for $h$.
\review{In turn, this changes the Bayes factor as it depends on the prior volume, again motivating reporting the suspiciousness as well.}

Since the lensing probes are consistent with $\Lambda$CDM and provide an excellent fit with a $p$-value of 0.3, the Bayes factor, $B$ in \Cref{eq:bayes}, that is the evidence of the $w_0w_a$CDM model over the evidence of $\Lambda$CDM given the same data,
clearly prefers a cosmological constant over dynamical dark energy with $B = 0.06$. It is noteworthy that this is a stronger preference for $\Lambda$CDM than the Bayes factor suggests for the low-$z+$CMB combination, for which we find $B=2.73$ in favour of $w_0w_a$CDM. Especially adding SN from Pantheon+ removes the lower part of the posterior and moves it closer to $\Lambda$CDM; the preference for dynamical dark energy becomes smaller in this case. In terms of suspiciousness, we find tensions of $2.6\sigma$ and $0.4\sigma$ with respect to $\Lambda$CDM for the low-$z+$CMB and lensing, respectively. \review{The $2.6\sigma$ suspiciousness tension is not the same as the $2.6\sigma$ difference of $\Omega_K$ from zero. The former is related to the Bayes factor, while the latter is a significance in the posterior space. In fact, those significances for $w_0$ and $w_a$ are $3.25\sigma$ and $3.4\sigma$, respectively.}
Furthermore, we investigate the tension between our lensing dataset and the CMB$+$BAO. We use the Monte Carlo parameter-shift method introduced by \citet{2020PhRvD.101j3527R}
and find a tension of $0.5\sigma$ in the $w_0,w_a$ plane, driven by the larger uncertainties for dynamical dark energy when using the lensing data. In general, we do not find any shifts larger than $2\sigma$ with this method.

\subsection{Spatial curvature}
In a universe with spatial curvature, the (comoving) angular diameter distance changes from being equal to the comoving distance, $\chi$, to
\begin{equation}
\label{eq:curvature}
    f_K(\chi) =\left\{ \begin{array}{lc}
        K^{-1/2}\sin(K^{1/2}\chi)   & \mathrm{for}\;K> 0 \\ 
        \chi &\mathrm{for}\; K = 0\\
         |K|^{-1/2}\sinh(|K|^{1/2}\chi)   & \mathrm{for}\;K< 0 
    \end{array} \right..
\end{equation}
Due to the closure equation, the Hubble function is modified as
\begin{equation}
    E(a) \coloneqq \frac{H(a)}{H_0} = \sqrt{\Omega_\mathrm{m}a^{-3} + (1-\Omega_\mathrm{m}-\Omega_K)+\Omega_Ka^{-2}},
\end{equation}
where $\Omega_K = -(c/H_0)^2K$ is the spatial curvature parameter. Lastly, we assume that the modified structure formation in a spatially curved universe is entirely dictated by the linear power spectrum, while non-linear corrections are implicitly accounted for via the background. \revint{\citet{2022PhRvD.106h3504T} discuss the effect of $\Omega_K\neq0$ on the non-linear power spectrum using a separate universe approach and the halo model. The leading-order terms are captured via the linear growth in our approach, while higher-order terms are neglected. Hence, the results presented here can be seen as conservative.}

The CMB \citep[e.g.][]{PlanckCollab_CMB2020,2025arXiv250620707C} yields strong constraints on $\Omega_K$ via the position of the first acoustic peak in the CMB angular power spectrum. Due to its degeneracy with the matter density, a combination with BAO data tightens constraints significantly, providing ${\Omega_K = (0.26\pm 0.11)\times 10^{-2}}$ roughly  $2\sigma$ away from spatial flatness. 

\Cref{fig:OK} displays the constraints on $\Omega_K$ together with $S_8$ from KiDS, lensing, low-$z$, and all considered probes in black, red, blue, and green, respectively. The dashed line indicates a spatially flat universe, and $\Omega_K$ is shown over the whole prior range. One can see that KiDS alone provides only mild constraints on $\Omega_K$, as it lies within the prior edge for $\Omega_K > 0$ and within the prior edge for $\Omega_K < 0$. This is expected, as measuring spatial curvature requires a meaningful observation of absolute physical scales, which lensing struggles to provide. Alternatively, adding CMB lensing already increases the constraining power substantially, as it provides measurements on larger scales and can therefore resolve the peak of the matter power spectrum, providing constraints on $\Omega_\mathrm{m}$ and hence $\Omega_K$, yielding
\begin{equation*}
    \Omega_K = 0.00\pm 0.04\quad(\mathrm{lensing\;for\;}\Omega_K\Lambda\mathrm{CDM\;at\; 68\%}).
\end{equation*}
Adding BAO to cosmic shear provides similar constraints, again via breaking the $\Omega_\mathrm{m}$ degeneracy. The influence of the constraints on $S_8$ is negligible. 
Adding the CMB recovers the constraints provided in \citet{2025arXiv250620707C} due to the high signal-to-noise ratio of the acoustic peak position.

\begin{figure}
    \centering
    \includegraphics[width=0.49\textwidth]{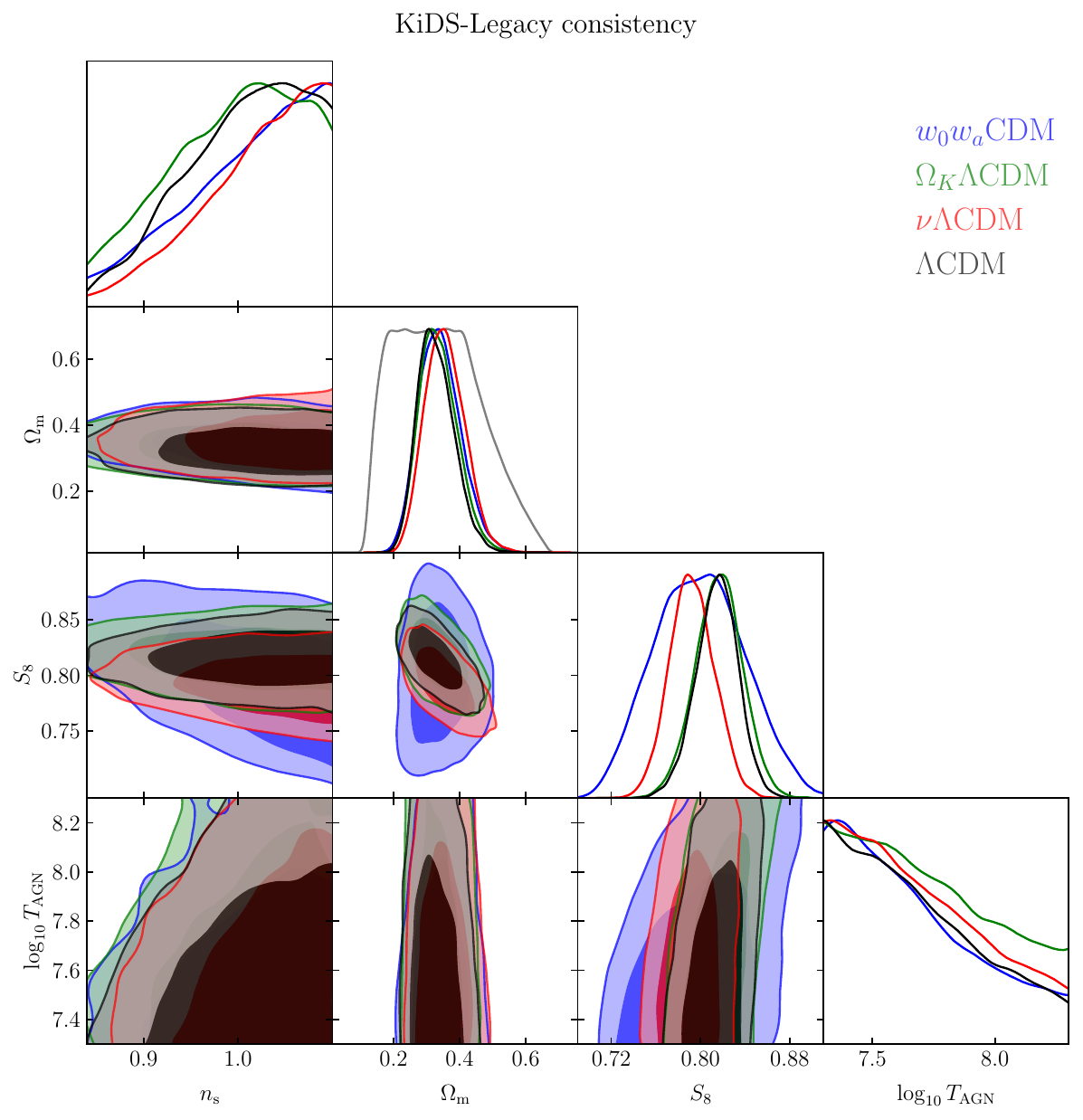}
    \caption{Constraints on $\Lambda$CDM parameters from KiDS-Legacy alone when opening up the parameter space. The grey line in the marginalised $\Omega_\mathrm{m}$ indicates the prior range. All other parameters have hard priors as shown.}
    \label{fig:kids-legacy-fiducial}
\end{figure}

\subsection{Consistency of KiDS-Legacy}
\citet{2025arXiv250319442S} provided several internal consistency checks of the KiDS-Legacy dataset. Here, we also summarise the consistency when adding more datasets and expanding the parameter space. 

First, \Cref{fig:kids-legacy-fiducial} shows the constraints for $\Omega_\mathrm{m}$, the spectral index $n_\mathrm{s}$ and $S_8$ for the three extended models studied here in comparison to the fiducial analysis from \citet{wright_kids_2025}. It can be seen that the modes of the posterior marginals are very stable, with the largest shift occurring in $S_8$ when varying the neutrino mass, as discussed already before. Furthermore, the constraints on $S_8$ degrade when dynamical dark energy is included. This is because a more negative $w_\mathrm{eff}$ moves sources further away at a given redshift, requiring less matter clustering, leading to a positive correlation between $w_\mathrm{eff}$ and $S_8$.
We also measure the Kullback-Leibler divergence, $D_\mathrm{KL}$, between marginal posterior distributions $p_1$ and $p_2$, that is their relative entropy:
\begin{equation}
D_\mathrm{KL}(p_1, p_2) = \int \mathrm{d}x\,p_1(x)\log\frac{p_1(x)}{p_2(x)}\;.    
\end{equation}
The largest Kullback-Leibler divergence in $S_8$ is $D_\mathrm{KL}=0.45$ in the case of the neutrinos, a shift of $1.2\sigma$ of the marginal mean relative to the fiducial analysis.

\begin{figure}
    \centering
    \includegraphics[width=0.48\textwidth]{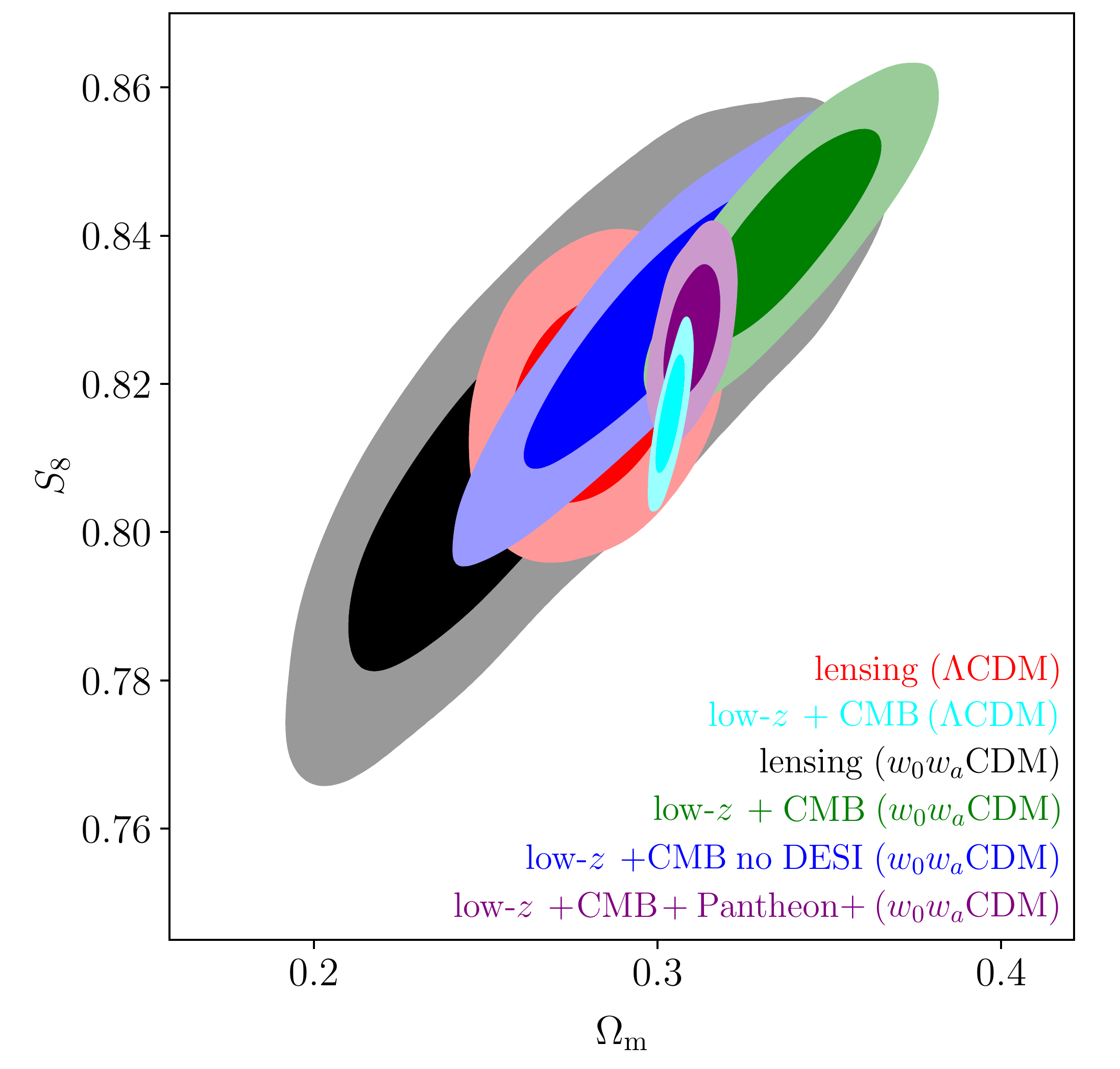}
    \caption{\revint{Constraints on $\Omega_\mathrm{m}$ and $S_8$ for different data combinations in the $w_0,w_a$ parametrisation, \Cref{eq:cpl}. The red and the cyan contours assume a $\Lambda$CDM prior.}}
    \label{fig:kids-w0was8}
\end{figure}

In \Cref{fig:kids-w0was8}, we show the consistency of different probe combinations in the $\Omega_\mathrm{m},S_8$ plane and how dynamical dark energy moves these specific contours. One can see the shift to higher $\Omega_\mathrm{m}$ and hence $S_8$ when adding the DESI results to the low-redshift and the CMB data (from the blue to the green contour). \revint{This shift, however, is reduced when adding in the SN from Pantheon+, bringing $\Omega_\mathrm{m}$ down again. This does not mean that the tension with $\Lambda$ is reduced, as only the negative $w_a$ tail of the posterior is removed. While this moves the peak of the posterior closer to $\Lambda$CDM, the overall discrepancy with it stays the same (as was shown in \Cref{fig:w0wa}).}
We observe the same shift for this data combination when moving from $\Lambda$CDM to $w_0w_a$CDM (from the cyan to the purple contour). It is only in these scenarios where SN observations add constraining power. \revint{While we do not show it explicitly, low-$z+$SPA$+$Pantheon$+$ in $\Lambda$CDM agrees with low-$z+$SPA}. Lastly, we show the effect on the lensing probes (from the black contour to the red contour) when using the $w_0,w_a$ parametrisation, which is consistent with the fiducial analysis.

\begin{figure}
    \centering
    \includegraphics[width=0.48\textwidth]{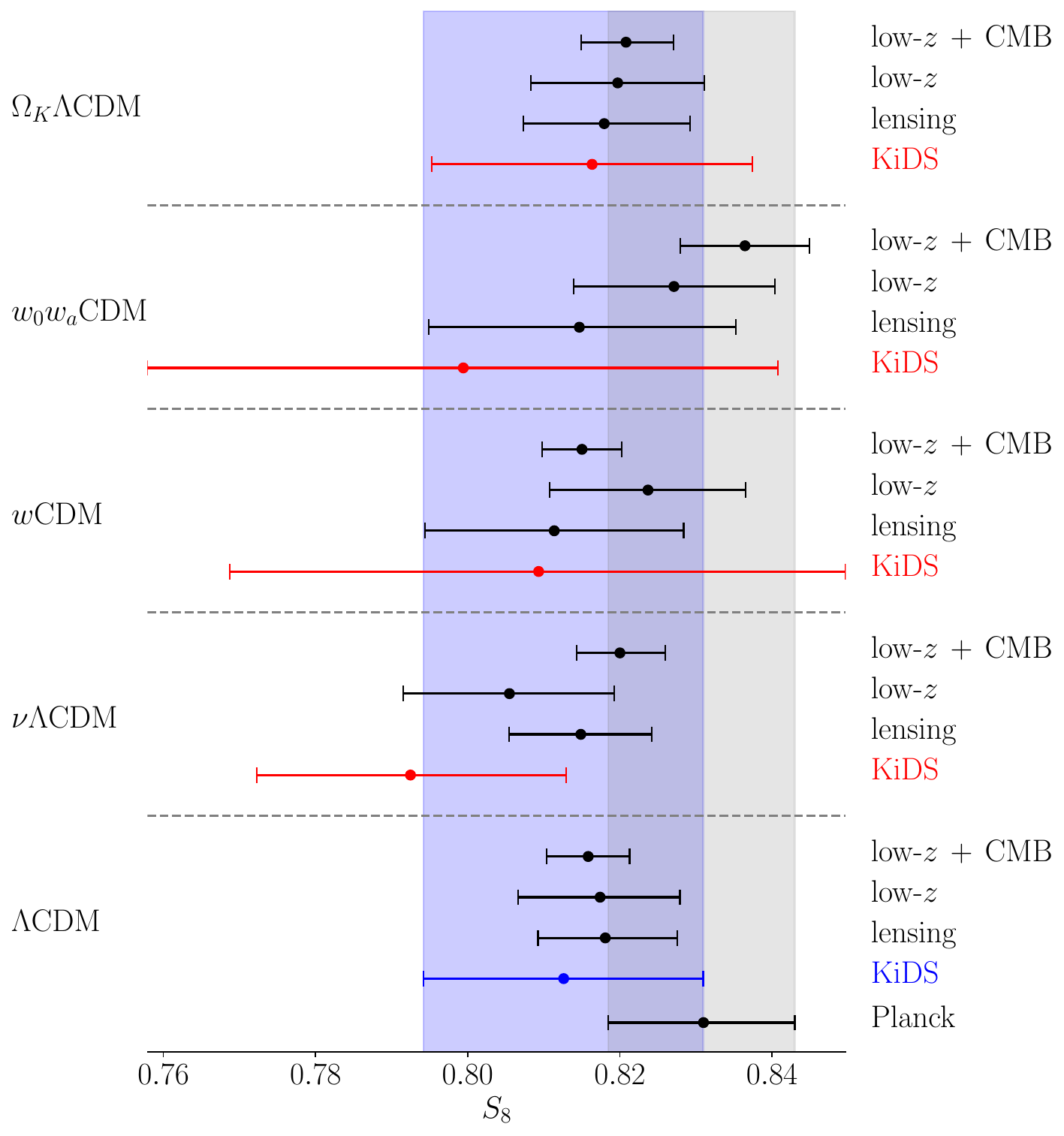}
    \caption{Marginal mean of $S_8$ and the $68\,\%$ credible interval for all extended models and probe combinations tested in this work. The blue band shows the $\Lambda$CDM constraints from KiDS \citep{wright_kids_2025}, while the grey band shows the constraints from Planck \citep{PlanckCollab_CMB2020}.}
    \label{fig:all_consistency}
\end{figure}

\Cref{fig:all_consistency} displays a summary of the main probe and model combinations tested in this work. We show the marginal mean, together with the $68\,\%$ confidence interval, of the main parameter constrained by cosmic shear, $S_8$, for each combination, and highlight the fiducial $\Lambda$CDM analyses by KiDS-Legacy and Planck in blue and grey, respectively. Additionally, we present the case in which we keep $w_a$ fixed at zero and only vary $w_0$, referred to as $w$CDM. We observe that the marginal constraints on $S_8$ remain stable as the parameter space is expanded or additional probes are included. In $\Lambda$CDM, we find $S_8$ to be unchanged and a per cent measurement from lensing alone, which turns into a sub-per cent measurement when combining all probes considered here.
The shift in $S_8$ when varying the neutrino mass reduces when including external data, as expected, as the combination of CMB and BAO pushes the neutrino mass very close to the value used in the fiducial analysis.

For dynamical dark energy, $w$CDM does not affect the marginal mean of $S_8$ strongly and simply increases the error bars. This is consistent with the fact that BAO and the CMB agree at $w_0 = -1$ when $w_a$ is held fixed. However, when considering the full $w_0,w_a$ parametrisation, the combination of all probes actually prefers high $S_8$. The shift is mainly driven by an increasing $\Omega_\mathrm{m}$, while $\sigma_8$ does change less. This can, again, be quantified by the Kullback-Leibler divergence for which we find \revint{$D_\mathrm{KL} = 3$ (2.3$\sigma$) for $S_8$ but $D_\mathrm{KL} = 1.4$ (1.1$\sigma$) for $\sigma_8$}. Interestingly, the value of $\sigma_8$ is lowered in the dynamical dark energy scenario.  
However, as previously discussed, the preference for this extended model is relatively small. 

\revint{As indicated in \Cref{tab:constraints}, some of the parameters are not constrained by the likelihood but rather via our prior choice; this is most pronounced for $\log_{10} T_\mathrm{AGN}$. Consequently, those posteriors have very non-Gaussian distributions, which can lead to projection effects. The main parameter of interest in the previous discussion, however, is $S_8$ whose interplay with $\log_{10} T_\mathrm{AGN}$ is studied in \Cref{app:1}. Since we do not find significant changes in $S_8$ due to $\log_{10} T_\mathrm{AGN}$, we conclude that projection effects are not dominant.}

\section{Summary}
\label{sec:concl}
We present the extended cosmology analysis of the final cosmic shear catalogue of the Kilo Degree Survey (KiDS-Legacy) in combination with external cosmic shear data from the Dark Energy Survey (DES), as well as external probes from the cosmic microwave background (CMB), baryon acoustic oscillations (BAOs), and redshift space distortion (RSDs). 
Due to the consistency of KiDS-Legacy with other cosmological probes, especially in $S_8 = \sigma_8\sqrt{\Omega_\mathrm{m}/0.3}$, we can, for the first time, reliably combine with all external datasets. We derive constraints on the sum of neutrino masses, the spatial curvature of the Universe, and the equation of state of dark energy. Furthermore, we investigate the robustness of the KiDS-Legacy data to different combinations of data and parameter space extensions.

We find that KiDS-Legacy prefers $\Lambda$CDM over extended models, with almost identical goodness of fit across all extended models considered, and that the Bayes factors are all below unity, disfavouring extended models.  
By combining KiDS-Legacy with DES Y3 and CMB lensing data from Planck, the Atacama Cosmology Telescope (ACT) and the South Pole Telescope (SPT), we define a lensing probe which enhances the consistency with $\Lambda$CDM, \revint{rejecting extended models according to the Jeffreys scale, especially time-evolving dark energy models, with a Bayes factor $B < 1$}. The equation of state of these models is constrained to {$w_0 = -0.92^{+0.45}_{-0.42}$} and $w_a = -0.88^{+1.45}_{-1.56}$ from weak gravitational lensing alone.
Curvature and neutrinos are only poorly constrained by weak gravitational lensing with ${\Omega_K = 0.001^{+0.040}_{-0.041}}$ and $\sum m_\nu \leq  1.61\,\mathrm{eV}$.

Adding clustering data from BAOs and RSDs significantly improves constraints compared to those from weak gravitational lensing alone, as $\Omega_\mathrm{m}$ becomes tightly constrained, thereby breaking one of the main lensing degeneracies and allowing it to measure the amplitude of the linear matter power spectrum, $\sigma_8$. This low-redshift, low-$z$ probe collection is independent of the CMB and is, to a substantial degree, consistent with $\Lambda$CDM, with all constraints consistent with $\Lambda$CDM within $1\sigma$, and hence all Bayes factors disfavour the extended models.

When combining low-redshift data with CMB primary anisotropies and CMB lensing, we recover constraints on the neutrino mass and curvature from the CMB alone, which are much more sensitive to these parameters. For dynamical dark energy, however, the different preferences for $\Omega_\mathrm{m}$ between the CMB and low-$z$ probes shift $w(a)$ away from $\Lambda$CDM by roughly $3\sigma$ if allowed to vary with redshift. \revint{This, however, is less strong in terms of model selection than the preference of lensing probes for $\Lambda$CDM. More quantitatively, the Bayes factor, $B$ is 0.07 (0.02) using our lensing (low-$z$) probes in preference of $w_0w_a$CDM over $\Lambda$CDM, thus corresponding to strong (very strong) evidence in favour of $\Lambda$CDM. In contrast, when combining all probes, we only find $B=2.73$.}

The cosmic shear analysis of KiDS-Legacy is robust to minimal extensions of $\Lambda$CDM, preferring it over all other models while maintaining statistically stable inferred values of $S_8$ when the parameter space is expanded, or external probes are added. This extends the fiducial analysis of KiDS-Legacy naturally, while arriving at a similar conclusion: at the end of cosmic shear Stage-III surveys, cosmological probes, spanning 13 billion years, are strongly in favour of $\Lambda$CDM. With the tentative evidence for dynamical dark energy in DESI, cosmic shear is set to move to Stage-IV as well, providing much tighter constraints and decisively measuring $w(z)$.

\begin{acknowledgements}
The authors would like to thank L. Balkenhol, E. Camphuis, and S. Galli for their help and for pointing us to the right places for constructing the CMB likelihood. The authors also thank an anonymous referee for helpful comments on the manuscript.
RR, AD, HHi, and CM are supported by a European Research Council Consolidator Grant (No. 770935). AHW is supported by the Deutsches Zentrum für Luft- und Raumfahrt (DLR), under project 50QE2305, made possible by the Bundesministerium für Wirtschaft und Klimaschutz, and acknowledges funding from the German Science Foundation DFG, via the Collaborative Research Center SFB1491 “Cosmic Interacting Matters - From Source to Signal”.
BS, and ZY acknowledge support from the Max Planck Society and the Alexander von Humboldt Foundation in the framework of the Max Planck-Humboldt Research Award endowed by the Federal Ministry of Education and Research.
HHi is supported by a DFG Heisenberg grant (Hi 1495/5-1), the DFG Collaborative Research Centre SFB149.1, and the DLR project 50QE2305.
BJ acknowledges support from the ERC-selected UKRI Frontier Research Grant EP/Y03015X/1 and by STFC Consolidated Grant ST/V000780/1.
MA is supported by the UK Science and Technology Facilities Council (STFC) under grant number ST/Y002652/1 and the Royal Society under grant numbers RGSR2222268 and ICAR1231094.
LM acknowledges the financial contribution from the PRIN-MUR 2022 20227RNLY3 grant “The concordance cosmological model: stress-tests with galaxy clusters” supported by Next Generation EU and from the grant ASI n. 2024-10-HH.0 “Attività scientifiche per la missione Euclid – fase E”.
AL acknowledges support from the Swedish National Space Agency (Rymdstyrelsen) under Career Grant Project Dnr 2024-00171.
MvWK acknowledges the support by UK STFC (grant no. ST/X001075/1), the UK Space Agency (grant no. ST/X001997/1), and InnovateUK (grant no. TS/Y014693/1). 
MB is supported by the Polish National Science Center through grant no. 2020/38/E/ST9/00395.
JHD acknowledges support from an STFC
Ernest Rutherford Fellowship (project reference ST/S004858/1). BG acknowledges support from the UKRI Stephen Hawking Fellowship (grant reference EP/Y017137/1).
CH acknowledges support from the Max Planck Society and the Alexander von Humboldt Foundation in the framework of the Max Planck-Humboldt Research Award endowed by the Federal Ministry of Education and Research, and the UK Science and Technology Facilities Council (STFC) under grant ST/V000594/1.
CG is funded by the MICINN project PID2022-141079NB-C32. IFAE is partially funded by the CERCA program of the Generalitat de Catalunya.
SJ acknowledges the Dennis Sciama Fellowship at the University of Portsmouth and the Ram\'{o}n y Cajal Fellowship from the Spanish Ministry of Science.
KK acknowledges support from the Royal Society and Imperial College.
LL is supported by the Austrian Science Fund (FWF) [ESP 357-N].
CM acknowledges support under the grant number PID2021-128338NB-I00 from the Spanish Ministry of Science.
NRN acknowledges financial support from the National Science Foundation of China, Research Fund for Excellent International Scholars (grant n. 12150710511), and from the research grant from China Manned Space Project n. CMS-CSST-2021-A01.
TT acknowledges funding from the Swiss National Science Foundation under the Ambizione project PZ00P2\_193352 
MY, SSL and HHo acknowledge funding from the European Research Council (ERC) under the European Union’s Horizon 2020 research and innovation program (Grant agreement No. 101053992). \newline

{\it Software:} The figures in this work were created with {\sc matplotlib} \citep{Hunter_matplotlib_2007}, making use of the {\sc NumPy} \citep{harris2020array}, {\sc SciPy} \citep{2020SciPy-NMeth}, {\sc CosmoPower} \citep{2022MNRAS.511.1771S} and {\sc GetDist} \citep{2019arXiv191013970L} software packages. \newline
{\it Kilo-Degree Survey:} Based on observations made with ESO Telescopes at the La Silla Paranal Observatory under programme IDs 179.A-2004, 177.A-3016, 177.A-3017, 177.A-3018, 298.A-5015.
\\
{\it Dark Energy Spectroscopic Instrument:} This research used data obtained with the Dark Energy Spectroscopic Instrument (DESI). DESI construction and operations are managed by the Lawrence Berkeley National Laboratory. This material is based upon work supported by the U.S. Department of Energy, Office of Science, Office of High-Energy Physics, under Contract No. DE–AC02–05CH11231, and by the National Energy Research Scientific Computing Centre, a DOE Office of Science User Facility under the same contract. Additional support for DESI was provided by the U.S. National Science Foundation (NSF), Division of Astronomical Sciences under Contract No. AST-0950945 to the NSF’s National Optical-Infrared Astronomy Research Laboratory; the Science and Technology Facilities Council of the United Kingdom; the Gordon and Betty Moore Foundation; the Heising-Simons Foundation; the French Alternative Energies and Atomic Energy Commission (CEA); the National Council of Science and Technology of Mexico (CONACYT); the Ministry of Science and Innovation of Spain (MICINN), and by the DESI Member Institutions: www.desi.lbl.gov/collaborating-institutions. The DESI collaboration is honoured to be permitted to conduct scientific research on Iolkam Du’ag (Kitt Peak), a mountain with particular significance to the Tohono O’odham Nation. Any opinions, findings, and conclusions or recommendations expressed in this material are those of the author(s) and do not necessarily reflect the views of the U.S. National Science Foundation, the U.S. Department of Energy, or any of the listed funding agencies.
\\
{\it SDSS-IV:} Funding for the Sloan Digital Sky 
Survey IV has been provided by the 
Alfred P. Sloan Foundation, the U.S. 
Department of Energy Office of 
Science, and the Participating 
Institutions. 

SDSS-IV acknowledges support and 
resources from the Center for High 
Performance Computing  at the 
University of Utah. The SDSS 
website is www.sdss4.org.

SDSS-IV is managed by the 
Astrophysical Research Consortium 
for the Participating Institutions 
of the SDSS Collaboration, including 
the Brazilian Participation Group, 
the Carnegie Institution for Science, 
Carnegie Mellon University, Centre for 
Astrophysics | Harvard \& 
Smithsonian, the Chilean Participation 
Group, the French Participation Group, 
Instituto de Astrof\'isica de 
Canarias, The Johns Hopkins 
University, Kavli Institute for the 
Physics and Mathematics of the 
Universe (IPMU) / University of 
Tokyo, the Korean Participation Group, 
Lawrence Berkeley National Laboratory, 
Leibniz Institut f\"ur Astrophysik 
Potsdam (AIP),  Max-Planck-Institut 
f\"ur Astronomie (MPIA Heidelberg), 
Max-Planck-Institut f\"ur 
Astrophysik (MPA Garching), 
Max-Planck-Institut f\"ur 
Extraterrestrische Physik (MPE), 
National Astronomical Observatories of 
China, New Mexico State University, 
New York University, University of 
Notre Dame, Observat\'ario 
Nacional / MCTI, The Ohio State 
University, Pennsylvania State 
University, Shanghai 
Astronomical Observatory, United 
Kingdom Participation Group, 
Universidad Nacional Aut\'onoma 
de M\'exico, University of Arizona, 
University of Colorado Boulder, 
University of Oxford, University of 
Portsmouth, University of Utah, 
University of Virginia, University 
of Washington, University of 
Wisconsin, Vanderbilt University, 
and Yale University.
\\
{\it Dark Energy Survey:} This project used public archival data from the Dark Energy Survey (DES). Funding for the DES Projects has been provided by the U.S. Department of Energy, the U.S. National Science Foundation, the Ministry of Science and Education of Spain, the Science and Technology FacilitiesCouncil of the United Kingdom, the Higher Education Funding Council for England, the National Center for Supercomputing Applications at the University of Illinois at Urbana-Champaign, the Kavli Institute of Cosmological Physics at the University of Chicago, the Center for Cosmology and Astro-Particle Physics at the Ohio State University, the Mitchell Institute for Fundamental Physics and Astronomy at Texas A\&M University, Financiadora de Estudos e Projetos, Funda{\c c}{\~a}o Carlos Chagas Filho de Amparo {\`a} Pesquisa do Estado do Rio de Janeiro, Conselho Nacional de Desenvolvimento Cient{\'i}fico e Tecnol{\'o}gico and the Minist{\'e}rio da Ci{\^e}ncia, Tecnologia e Inova{\c c}{\~a}o, the Deutsche Forschungsgemeinschaft, and the Collaborating Institutions in the Dark Energy Survey.
The Collaborating Institutions are Argonne National Laboratory, the University of California at Santa Cruz, the University of Cambridge, Centro de Investigaciones Energ{\'e}ticas, Medioambientales y Tecnol{\'o}gicas-Madrid, the University of Chicago, University College London, the DES-Brazil Consortium, the University of Edinburgh, the Eidgen{\"o}ssische Technische Hochschule (ETH) Z{\"u}rich,  Fermi National Accelerator Laboratory, the University of Illinois at Urbana-Champaign, the Institut de Ci{\`e}ncies de l'Espai (IEEC/CSIC), the Institut de F{\'i}sica d'Altes Energies, Lawrence Berkeley National Laboratory, the Ludwig-Maximilians Universit{\"a}t M{\"u}nchen and the associated Excellence Cluster Universe, the University of Michigan, the National Optical Astronomy Observatory, the University of Nottingham, The Ohio State University, the OzDES Membership Consortium, the University of Pennsylvania, the University of Portsmouth, SLAC National Accelerator Laboratory, Stanford University, the University of Sussex, and Texas A\&M University.
Based in part on observations at Cerro Tololo Inter-American Observatory, National Optical Astronomy Observatory, which is operated by the Association of Universities for Research in Astronomy (AURA) under a cooperative agreement with the National Science Foundation. 
\\
{\it Planck:} Based on observations obtained with Planck (http://www.esa.int/Planck), an ESA science mission with instruments and contributions directly funded by ESA Member States, NASA, and Canada.
\\
{\it SPT:} Supported by the National Science Foundation (NSF) through awards OPP-1852617 and OPP-2332483. Partial support is also provided by the Kavli Institute of Cosmological Physics at the University of Chicago. Argonne National Laboratory’s work was supported by the U.S. Department of Energy, Office of High Energy Physics, under contract DE-AC02-06CH11357
\\
{\it ACT}: Support for ACT was through the U.S. National Science Foundation through awards AST-0408698, AST- 0965625, and AST-1440226 for the ACT project, as well as awards PHY-0355328, PHY-0855887 and PHY- 1214379. Funding was also provided by Princeton University, the University of Pennsylvania, and a Canada Foundation for Innovation (CFI) award to UBC. ACT operated in the Parque Astron\'{o}mico Atacama in northern Chile under the auspices of the Agencia Nacional de Investigaci\'{o}n y Desarrollo (ANID).
\\
\textit{Author contributions}: All authors contributed to the development and writing of this paper. The authorship list is given in three groups: the lead authors (RR, BS, BJ) followed by two alphabetical groups. The first alphabetical group includes those who are key contributors to both the scientific analysis and the data products. The second group covers those who have either made a significant contribution to the data products or to the scientific analysis.
\end{acknowledgements}

\bibliographystyle{aa}
\bibliography{library}

@ARTICLE{2025MNRAS.540.2844E,
       author = {{Efstathiou}, George},
        title = "{Baryon acoustic oscillations from a different angle}",
      journal = {\mnras},
         year = 2025,
        month = jul,
       volume = {540},
       number = {3},
        pages = {2844-2852},
          doi = {10.1093/mnras/staf906},
archivePrefix = {arXiv},
       eprint = {2505.02658},
 primaryClass = {astro-ph.CO},
       adsurl = {https://ui.adsabs.harvard.edu/abs/2025MNRAS.540.2844E}
}

@ARTICLE{2025Sci...388..180K,
       author = {{KATRIN Collaboration} and {Aker}, Max and {Batzler}, Dominic and {Beglarian}, Armen and {Behrens}, Jan and {Beisenk{\"o}tter}, Justus and {Biassoni}, Matteo and {Bieringer}, Benedikt and {Biondi}, Yanina and {Block}, Fabian and {Bobien}, Steffen and {B{\"o}ttcher}, Matthias and {Bornschein}, Beate and {Bornschein}, Lutz and {Caldwell}, Tom S. and {Carminati}, Marco and {Chatrabhuti}, Auttakit and {Chilingaryan}, Suren and {Daniel}, Byron A. and {Debowski}, Karol and {Descher}, Martin and {Barrero}, Deseada D{\'\i}az and {Doe}, Peter J. and {Dragoun}, Otokar and {Drexlin}, Guido and {Edzards}, Frank and {Eitel}, Klaus and {Ellinger}, Enrico and {Engel}, Ralph and {Enomoto}, Sanshiro and {Felden}, Arne and {Fengler}, Caroline and {Fiorini}, Carlo and {Formaggio}, Joseph A. and {Forstner}, Christian and {Fr{\"a}nkle}, Florian M. and {Gauda}, Kevin and {Gavin}, Andrew S. and {Gil}, Woosik and {Gl{\"u}ck}, Ferenc and {Grohmann}, Steffen and {Gr{\"o}ssle}, Robin and {Gumbsheimer}, Rainer and {Gutknecht}, Nathanael and {Hannen}, Volker and {Hasselmann}, Leonard and {Hau{\ss}mann}, Norman and {Helbing}, Klaus and {Henke}, Hanna and {Heyns}, Svenja and {Hickford}, Stephanie and {Hiller}, Roman and {Hillesheimer}, David and {Hinz}, Dominic and {H{\"o}hn}, Thomas and {Huber}, Anton and {Jansen}, Alexander and {Karl}, Christian and {Kellerer}, Jonas and {Khosonthongkee}, Khanchai and {Kleifges}, Matthias and {Klein}, Manuel and {Kohpei{\ss}}, Joshua and {K{\"o}hler}, Christoph and {K{\"o}llenberger}, Leonard and {Kopmann}, Andreas and {Kova{\v{c}}}, Neven and {Koval{\'\i}k}, Alojz and {Krause}, Holger and {La Cascio}, Luisa and {Lasserre}, Thierry and {Lauer}, Joscha and {Le}, Thanh-Long and {Lebeda}, Ond{\v{r}}ej and {Lehnert}, Bjoern and {Li}, Gen and {Lokhov}, Alexey and {Machatschek}, Moritz and {Mark}, Martin and {Marsteller}, Alexander and {Martin}, Eric L. and {Melzer}, Christin and {Mertens}, Susanne and {Mohanty}, Shailaja and {Mostafa}, Jalal and {M{\"u}ller}, Klaus and {Nava}, Andrea and {Neumann}, Holger and {Niemes}, Simon and {Onillon}, Anthony and {Parno}, Diana S. and {Pavan}, Maura and {Pinsook}, Udomsilp and {Poon}, Alan W.~P. and {Lopez Poyato}, Jose Manuel and {Pozzi}, Stefano and {Priester}, Florian and {R{\'a}li{\v{s}}}, Jan and {Ramachandran}, Shivani and {Robertson}, R.~G. Hamish and {Rodenbeck}, Caroline and {R{\"o}llig}, Marco and {R{\"o}ttele}, Carsten and {Ry{\v{s}}av{\'y}}, Milos and {Sack}, Rudolf and {Saenz}, Alejandro and {Salomon}, Richard and {Sch{\"a}fer}, Peter and {Schl{\"o}sser}, Magnus and {Schl{\"o}sser}, Klaus and {Schl{\"u}ter}, Lisa and {Schneidewind}, Sonja and {Schnurr}, Ulrich and {Schrank}, Michael and {Sch{\"u}rmann}, Jannis and {Sch{\"u}tz}, Ann-Kathrin and {Schwemmer}, Alessandro and {Schwenck}, Adrian and {{\v{S}}ef{\v{c}}{\'\i}k}, Michal and {Siegmann}, Daniel and {Simon}, Frank and {Spanier}, Felix and {Spreng}, Daniela and {Sreethawong}, Warintorn and {Steidl}, Markus and {{\v{S}}torek}, Jaroslav and {Stribl}, Xaver and {Sturm}, Michael and {Suwonjandee}, Narumon and {Jerome}, Nicholas Tan and {Telle}, Helmut H. and {Thorne}, Larisa A. and {Th{\"u}mmler}, Thomas and {Tirolf}, Simon and {Titov}, Nikita and {Tkachev}, Igor and {Urban}, Korbinian and {Valerius}, Kathrin and {V{\'e}nos}, Drahoslav and {Weinheimer}, Christian and {Welte}, Stefan and {Wendel}, J{\"u}rgen and {Wiesinger}, Christoph and {Wilkerson}, John F. and {Wolf}, Joachim and {W{\"u}stling}, Sascha and {Wydra}, Johanna and {Xu}, Weiran and {Zadorozhny}, Sergey and {Zeller}, Genrich},
        title = "{Direct neutrino-mass measurement based on 259 days of KATRIN data}",
      journal = {Science},
         year = 2025,
        month = apr,
       volume = {388},
       number = {6743},
        pages = {180-185},
          doi = {10.1126/science.adq9592},
archivePrefix = {arXiv},
       eprint = {2406.13516},
 primaryClass = {nucl-ex},
       adsurl = {https://ui.adsabs.harvard.edu/abs/2025Sci...388..180K}
}

@ARTICLE{2024JHEP...12..216E,
       author = {{Esteban}, Ivan and {Gonzalez-Garcia}, M.~C. and {Maltoni}, Michele and {Martinez-Soler}, Ivan and {Pinheiro}, Jo{\~a}o Paulo and {Schwetz}, Thomas},
        title = "{NuFit-6.0: updated global analysis of three-flavor neutrino oscillations}",
      journal = {Journal of High Energy Physics},
         year = 2024,
        month = dec,
       volume = {2024},
       number = {12},
          eid = {216},
        pages = {216},
          doi = {10.1007/JHEP12(2024)216},
archivePrefix = {arXiv},
       eprint = {2410.05380},
 primaryClass = {hep-ph},
       adsurl = {https://ui.adsabs.harvard.edu/abs/2024JHEP...12..216E}
}

@ARTICLE{2023MNRAS.524.2195S,
       author = {{Samuroff}, S. and {Mandelbaum}, R. and {Blazek}, J. and {Campos}, A. and {MacCrann}, N. and {Zacharegkas}, G. and {Amon}, A. and {Prat}, J. and {Singh}, S. and {Elvin-Poole}, J. and {Ross}, A.~J. and {Alarcon}, A. and {Baxter}, E. and {Bechtol}, K. and {Becker}, M.~R. and {Bernstein}, G.~M. and {Rosell}, A. Carnero and {Kind}, M. Carrasco and {Cawthon}, R. and {Chang}, C. and {Chen}, R. and {Choi}, A. and {Crocce}, M. and {Davis}, C. and {DeRose}, J. and {Dodelson}, S. and {Doux}, C. and {Drlica-Wagner}, A. and {Eckert}, K. and {Everett}, S. and {Fert{\'e}}, A. and {Gatti}, M. and {Giannini}, G. and {Gruen}, D. and {Gruendl}, R.~A. and {Harrison}, I. and {Herner}, K. and {Huff}, E.~M. and {Jarvis}, M. and {Kuropatkin}, N. and {Leget}, P.-F. and {Lemos}, P. and {McCullough}, J. and {Myles}, J. and {Navarro-Alsina}, A. and {Pandey}, S. and {Porredon}, A. and {Raveri}, M. and {Rodriguez-Monroy}, M. and {Rollins}, R.~P. and {Roodman}, A. and {Rossi}, G. and {Rykoff}, E.~S. and {S{\'a}nchez}, C. and {Secco}, L.~F. and {Sevilla-Noarbe}, I. and {Sheldon}, E. and {Shin}, T. and {Troxel}, M.~A. and {Tutusaus}, I. and {Weaverdyck}, N. and {Yanny}, B. and {Yin}, B. and {Zhang}, Y. and {Zuntz}, J. and {Aguena}, M. and {Alves}, O. and {Annis}, J. and {Bacon}, D. and {Bertin}, E. and {Bocquet}, S. and {Brooks}, D. and {Burke}, D.~L. and {Carretero}, J. and {Costanzi}, M. and {da Costa}, L.~N. and {Pereira}, M.~E.~S. and {De Vicente}, J. and {Desai}, S. and {Diehl}, H.~T. and {Dietrich}, J.~P. and {Doel}, P. and {Ferrero}, I. and {Flaugher}, B. and {Frieman}, J. and {Garc{\'\i}a-Bellido}, J. and {Hinton}, S.~R. and {Hollowood}, D.~L. and {Honscheid}, K. and {James}, D.~J. and {Kuehn}, K. and {Lahav}, O. and {Marshall}, J.~L. and {Melchior}, P. and {Mena-Fern{\'a}ndez}, J. and {Menanteau}, F. and {Miquel}, R. and {Newman}, J. and {Palmese}, A. and {Pieres}, A. and {Malag{\'o}n}, A.~A. Plazas and {Sanchez}, E. and {Scarpine}, V. and {Smith}, M. and {Suchyta}, E. and {Swanson}, M.~E.~C. and {Tarle}, G. and {To}, C. and {DES Collaboration}},
        title = "{The Dark Energy Survey Year 3 and eBOSS: constraining galaxy intrinsic alignments across luminosity and colour space}",
      journal = {\mnras},
         year = 2023,
        month = sep,
       volume = {524},
       number = {2},
        pages = {2195-2223},
          doi = {10.1093/mnras/stad2013},
archivePrefix = {arXiv},
       eprint = {2212.11319},
 primaryClass = {astro-ph.CO},
       adsurl = {https://ui.adsabs.harvard.edu/abs/2023MNRAS.524.2195S}
}

@ARTICLE{2025arXiv250416932S,
       author = {{Sailer}, Noah and {Farren}, Gerrit S. and {Ferraro}, Simone and {White}, Martin},
        title = "{Addressing Tensions in {\ensuremath{\Lambda}}CDM Cosmology by an Increase in the Optical Depth to Reionization}",
      journal = {\prl},
         year = 2026,
        month = feb,
       volume = {136},
       number = {8},
          eid = {081002},
        pages = {081002},
          doi = {10.1103/6r54-8lv4},
archivePrefix = {arXiv},
       eprint = {2504.16932},
 primaryClass = {astro-ph.CO},
       adsurl = {https://ui.adsabs.harvard.edu/abs/2026PhRvL.136h1002S}
}

@ARTICLE{2022PhRvD.106h3504T,
       author = {{Terasawa}, Ryo and {Takahashi}, Ryuichi and {Nishimichi}, Takahiro and {Takada}, Masahiro},
        title = "{Separate universe approach to evaluate nonlinear matter power spectrum for nonflat {\ensuremath{\Lambda}} CDM model}",
      journal = {\prd},
         year = 2022,
        month = oct,
       volume = {106},
       number = {8},
          eid = {083504},
        pages = {083504},
          doi = {10.1103/PhysRevD.106.083504},
archivePrefix = {arXiv},
       eprint = {2205.10339},
 primaryClass = {astro-ph.CO},
       adsurl = {https://ui.adsabs.harvard.edu/abs/2022PhRvD.106h3504T}
}

@ARTICLE{2025arXiv251001122Y,
       author = {{Yoon}, Mijin and {Hoekstra}, Henk and {Li}, Shun-Sheng and {Kuijken}, Konrad and {Miller}, Lance and {Hildebrandt}, Hendrik and {Heymans}, Catherine and {Joachimi}, Benjamin and {Wright}, Angus H. and {Asgari}, Marika and {van den Busch}, Jan Luca and {Reischke}, Robert and {St{\"o}lzner}, Benjamin},
        title = "{KiDS-1000 cosmic shear reanalysis using MetaCalibration}",
      journal = {arXiv e-prints, A\&A submitted},
         year = 2025,
        month = oct,
          eid = {arXiv:2510.01122},
        pages = {arXiv:2510.01122},
          doi = {10.48550/arXiv.2510.01122},
archivePrefix = {arXiv},
       eprint = {2510.01122},
 primaryClass = {astro-ph.CO},
       adsurl = {https://ui.adsabs.harvard.edu/abs/2025arXiv251001122Y}
}

@ARTICLE{2013ExA....35...25D,
       author = {{de Jong}, Jelte T.~A. and {Verdoes Kleijn}, Gijs A. and {Kuijken}, Konrad H. and {Valentijn}, Edwin A.},
        title = "{The Kilo-Degree Survey}",
      journal = {Experimental Astronomy},
         year = 2013,
        month = jan,
       volume = {35},
       number = {1-2},
        pages = {25-44},
          doi = {10.1007/s10686-012-9306-1},
archivePrefix = {arXiv},
       eprint = {1206.1254},
 primaryClass = {astro-ph.CO},
       adsurl = {https://ui.adsabs.harvard.edu/abs/2013ExA....35...25D}
}

@ARTICLE{Handley19,
       author = {{Handley}, Will and {Lemos}, Pablo},
        title = "{Quantifying dimensionality: Bayesian cosmological model complexities}",
      journal = {\prd},
         year = 2019,
        month = jul,
       volume = {100},
       number = {2},
          eid = {023512},
        pages = {023512},
          doi = {10.1103/PhysRevD.100.023512},
archivePrefix = {arXiv},
       eprint = {1903.06682},
 primaryClass = {astro-ph.CO},
       adsurl = {https://ui.adsabs.harvard.edu/abs/2019PhRvD.100b3512H}
}

@ARTICLE{2020PhRvD.101j3527R,
       author = {{Raveri}, Marco and {Zacharegkas}, Georgios and {Hu}, Wayne},
        title = "{Quantifying concordance of correlated cosmological data sets}",
      journal = {\prd},
         year = 2020,
        month = may,
       volume = {101},
       number = {10},
          eid = {103527},
        pages = {103527},
          doi = {10.1103/PhysRevD.101.103527},
archivePrefix = {arXiv},
       eprint = {1912.04880},
 primaryClass = {astro-ph.CO},
       adsurl = {https://ui.adsabs.harvard.edu/abs/2020PhRvD.101j3527R}
}

@ARTICLE{2021A&A...649A.146R,
       author = {{Robertson}, Naomi Clare and {Alonso}, David and {Harnois-D{\'e}raps}, Joachim and {Darwish}, Omar and {Kannawadi}, Arun and {Amon}, Alexandra and {Asgari}, Marika and {Bilicki}, Maciej and {Calabrese}, Erminia and {Choi}, Steve K. and {Devlin}, Mark J. and {Dunkley}, Jo and {Dvornik}, Andrej and {Erben}, Thomas and {Ferraro}, Simone and {Fortuna}, Maria Cristina and {Giblin}, Benjamin and {Han}, Dongwon and {Heymans}, Catherine and {Hildebrandt}, Hendrik and {Hill}, J. Colin and {Hilton}, Matt and {Ho}, Shuay-Pwu P. and {Hoekstra}, Henk and {Hubmayr}, Johannes and {Hughes}, John P. and {Joachimi}, Benjamin and {Joudaki}, Shahab and {Knowles}, Kenda and {Kuijken}, Konrad and {Madhavacheril}, Mathew S. and {Moodley}, Kavilan and {Miller}, Lance and {Namikawa}, Toshiya and {Nati}, Federico and {Niemack}, Michael D. and {Page}, Lyman A. and {Partridge}, Bruce and {Schaan}, Emmanuel and {Schillaci}, Alessandro and {Schneider}, Peter and {Sehgal}, Neelima and {Sherwin}, Blake D. and {Sif{\'o}n}, Crist{\'o}bal and {Staggs}, Suzanne T. and {Tr{\"o}ster}, Tilman and {van Engelen}, Alexander and {Valentijn}, Edwin and {Wollack}, Edward J. and {Wright}, Angus H. and {Xu}, Zhilei},
        title = "{Strong detection of the CMB lensing and galaxy weak lensing cross-correlation from ACT-DR4, Planck Legacy, and KiDS-1000}",
      journal = {\aap},
         year = 2021,
        month = may,
       volume = {649},
          eid = {A146},
        pages = {A146},
          doi = {10.1051/0004-6361/202039975},
archivePrefix = {arXiv},
       eprint = {2011.11613},
 primaryClass = {astro-ph.CO},
       adsurl = {https://ui.adsabs.harvard.edu/abs/2021A&A...649A.146R}
}

@ARTICLE{2025PDU....4901965D,
       author = {{Di Valentino}, Eleonora and {Said}, Jackson Levi and {Riess}, Adam and {Pollo}, Agnieszka and {Poulin}, Vivian and {G{\'o}mez-Valent}, Adri{\`a} and {Weltman}, Amanda and {Palmese}, Antonella and {Huang}, Caroline D. and {van de Bruck}, Carsten and {Saraf}, Chandra Shekhar and {Kuo}, Cheng-Yu and {Uhlemann}, Cora and {Grand{\'o}n}, Daniela and {Paz}, Dante and {Eckert}, Dominique and {Teixeira}, Elsa M. and {Saridakis}, Emmanuel N. and {Colg{\'a}in}, Eoin {\'O}. and {Beutler}, Florian and {Niedermann}, Florian and {Bajardi}, Francesco and {Barenboim}, Gabriela and {Gubitosi}, Giulia and {Musella}, Ilaria and {Banik}, Indranil and {Szapudi}, Istvan and {Singal}, Jack and {Cases}, Jaume Haro and {Chluba}, Jens and {Torrado}, Jes{\'u}s and {Mifsud}, Jurgen and {Jedamzik}, Karsten and {Said}, Khaled and {Dialektopoulos}, Konstantinos and {Herold}, Laura and {Perivolaropoulos}, Leandros and {Zu}, Lei and {Galbany}, Llu{\'\i}s and {Breuval}, Louise and {Visinelli}, Luca and {Escamilla}, Luis A. and {Anchordoqui}, Luis A. and {Sheikh-Jabbari}, M.~M. and {Lembo}, Margherita and {Dainotti}, Maria Giovanna and {Vincenzi}, Maria and {Asgari}, Marika and {Gerbino}, Martina and {Forconi}, Matteo and {Cantiello}, Michele and {Moresco}, Michele and {Benetti}, Micol and {Sch{\"o}neberg}, Nils and {Akarsu}, {\"O}zg{\"u}r and {Nunes}, Rafael C. and {Bernardo}, Reginald Christian and {Ch{\'a}vez}, Ricardo and {Anderson}, Richard I. and {Watkins}, Richard and {Capozziello}, Salvatore and {Li}, Siyang and {Vagnozzi}, Sunny and {Pan}, Supriya and {Treu}, Tommaso and {Irsic}, Vid and {Handley}, Will and {Giar{\`e}}, William and {Murakami}, Yukei and {Banihashemi}, Abdolali and {Poudou}, Ad{\`e}le and {Heavens}, Alan and {Kogut}, Alan and {Domi}, Alba and {Lenart}, Aleksander {\L}ukasz and {Melchiorri}, Alessandro and {Vadal{\`a}}, Alessandro and {Amon}, Alexandra and {Rivera}, Alexander Bonilla and {Reeves}, Alexander and {Zhuk}, Alexander and {Bonanno}, Alfio and {{\"O}vg{\"u}n}, Ali and {Pisani}, Alice and {Talebian}, Alireza and {Abebe}, Amare and {Aboubrahim}, Amin and {Gonz{\'a}lez Mor{\'a}n}, Ana Luisa and {Kov{\'a}cs}, Andr{\'a}s and {Lymperis}, Andreas and {Papatriantafyllou}, Andreas and {Liddle}, Andrew R. and {Paliathanasis}, Andronikos and {Borowiec}, Andrzej and {Yadav}, Anil Kumar and {Yadav}, Anita and {Sen}, Anjan Ananda and {William}, Anjitha John and {Davis}, Anne Christine and {Shajib}, Anowar J. and {Walters}, Anthony and {Lonappan}, Anto Idicherian and {Chudaykin}, Anton and {Capodagli}, Antonio and {da Silva}, Antonio and {De Felice}, Antonio and {Racioppi}, Antonio and {Oficial}, Araceli Soler and {Montiel}, Ariadna and {Favale}, Arianna and {Bernui}, Armando and {Velasco}, Arrianne Crystal and {Heinesen}, Asta and {Bakopoulos}, Athanasios and {Chatzistavrakidis}, Athanasios and {Khanpour}, Bahman and {Sathyaprakash}, Bangalore S. and {Zgirski}, Bartek and {L'Huillier}, Benjamin and {Famaey}, Benoit and {Jain}, Bhuvnesh and {Zhang}, Bing and {Karmakar}, Biswajit and {Dragovich}, Branko and {Thomas}, Brooks and {Correa}, Carlos and {Boiza}, Carlos G. and {Marques}, Catarina and {Escamilla-Rivera}, Celia and {Tzerefos}, Charalampos and {Zhang}, Chi and {De Leo}, Chiara and {Pfeifer}, Christian and {Lee}, Christine and {Venter}, Christo and {Gomes}, Cl{\'a}udio and {Roque De bom}, Clecio and {Moreno-Pulido}, Cristian and {Iosifidis}, Damianos and {Grin}, Dan and {Blixt}, Daniel and {Scolnic}, Dan and {Oriti}, Daniele and {Dobrycheva}, Daria and {Bettoni}, Dario and {Benisty}, David and {Fern{\'a}ndez-Arenas}, David and {Wiltshire}, David L. and {Sanchez Cid}, David and {Tamayo}, David and {Valls-Gabaud}, David and {Pedrotti}, Davide and {Wang}, Deng and {Staicova}, Denitsa and {Totolou}, Despoina and {Rubiera-Garcia}, Diego and {Milakovi{\'c}}, Dinko and {Pesce}, Dominic W. and {Sluse}, Dominique and {Borka}, Du{\v{s}}ko and {Yusofi}, Ebrahim and {Giusarma}, Elena and {Terlevich}, Elena and {Tomasetti}, Elena and {Vagenas}, Elias C. and {Fazzari}, Elisa and {Ferreira}, Elisa G.~M. and {Barakovic}, Elvis and {Dimastrogiovanni}, Emanuela and {Holm}, Emil Brinch and {Mottola}, Emil and {{\"O}z{\"u}lker}, Emre and {Specogna}, Enrico and {Brocato}, Enzo and {Jensko}, Erik and {Enriquez}, Erika Antonette and {Bhatia}, Esha and {Bresolin}, Fabio and {Avila}, Felipe and {Bouch{\`e}}, Filippo and {Bombacigno}, Flavio and {Anagnostopoulos}, Fotios K. and {Pace}, Francesco and {Sorrenti}, Francesco and {Lobo}, Francisco S.~N. and {Courbin}, Fr{\'e}d{\'e}ric and {Hansen}, Frode K. and {Sloan}, Greg and {Farrugia}, Gabriel and {Lynch}, Gabriel and {Garcia-Arroyo}, Gabriela and {Raimondo}, Gabriella and {Lambiase}, Gaetano and {Anand}, Gagandeep S. and {Poulot}, Gaspard and {Leon}, Genly and {Kouniatalis}, Gerasimos and {Nardini}, Germano and {Cs{\"o}rnyei}, G{\'e}za and {Galloni}, Giacomo},
        title = "{The CosmoVerse White Paper: Addressing observational tensions in cosmology with systematics and fundamental physics}",
      journal = {Physics of the Dark Universe},
         year = 2025,
        month = sep,
       volume = {49},
          eid = {101965},
        pages = {101965},
          doi = {10.1016/j.dark.2025.101965},
archivePrefix = {arXiv},
       eprint = {2504.01669},
 primaryClass = {astro-ph.CO},
       adsurl = {https://ui.adsabs.harvard.edu/abs/2025PDU....4901965D}
}

@ARTICLE{2023A&A...679A.133L,
       author = {{Li}, Shun-Sheng and {Hoekstra}, Henk and {Kuijken}, Konrad and {Asgari}, Marika and {Bilicki}, Maciej and {Giblin}, Benjamin and {Heymans}, Catherine and {Hildebrandt}, Hendrik and {Joachimi}, Benjamin and {Miller}, Lance and {van den Busch}, Jan Luca and {Wright}, Angus H. and {Kannawadi}, Arun and {Reischke}, Robert and {Shan}, HuanYuan},
        title = "{KiDS-1000: Cosmology with improved cosmic shear measurements}",
      journal = {\aap},
         year = 2023,
        month = nov,
       volume = {679},
          eid = {A133},
        pages = {A133},
          doi = {10.1051/0004-6361/202347236},
archivePrefix = {arXiv},
       eprint = {2306.11124},
 primaryClass = {astro-ph.CO},
       adsurl = {https://ui.adsabs.harvard.edu/abs/2023A&A...679A.133L}
}

@ARTICLE{2023PhRvD.108l3518L,
       author = {{Li}, Xiangchong and {Zhang}, Tianqing and {Sugiyama}, Sunao and {Dalal}, Roohi and {Terasawa}, Ryo and {Rau}, Markus M. and {Mandelbaum}, Rachel and {Takada}, Masahiro and {More}, Surhud and {Strauss}, Michael A. and {Miyatake}, Hironao and {Shirasaki}, Masato and {Hamana}, Takashi and {Oguri}, Masamune and {Luo}, Wentao and {Nishizawa}, Atsushi J. and {Takahashi}, Ryuichi and {Nicola}, Andrina and {Osato}, Ken and {Kannawadi}, Arun and {Sunayama}, Tomomi and {Armstrong}, Robert and {Bosch}, James and {Komiyama}, Yutaka and {Lupton}, Robert H. and {Lust}, Nate B. and {MacArthur}, Lauren A. and {Miyazaki}, Satoshi and {Murayama}, Hitoshi and {Nishimichi}, Takahiro and {Okura}, Yuki and {Price}, Paul A. and {Tait}, Philip J. and {Tanaka}, Masayuki and {Wang}, Shiang-Yu},
        title = "{Hyper Suprime-Cam Year 3 results: Cosmology from cosmic shear two-point correlation functions}",
      journal = {\prd},
         year = 2023,
        month = dec,
       volume = {108},
       number = {12},
          eid = {123518},
        pages = {123518},
          doi = {10.1103/PhysRevD.108.123518},
archivePrefix = {arXiv},
       eprint = {2304.00702},
 primaryClass = {astro-ph.CO},
       adsurl = {https://ui.adsabs.harvard.edu/abs/2023PhRvD.108l3518L}
}

@ARTICLE{1998ApJ...507...46S,
       author = {{Schmidt}, Brian P. and {Suntzeff}, Nicholas B. and {Phillips}, M.~M. and {Schommer}, Robert A. and {Clocchiatti}, Alejandro and {Kirshner}, Robert P. and {Garnavich}, Peter and {Challis}, Peter and {Leibundgut}, B. and {Spyromilio}, J. and {Riess}, Adam G. and {Filippenko}, Alexei V. and {Hamuy}, Mario and {Smith}, R. Chris and {Hogan}, Craig and {Stubbs}, Christopher and {Diercks}, Alan and {Reiss}, David and {Gilliland}, Ron and {Tonry}, John and {Maza}, Jos{\'e} and {Dressler}, A. and {Walsh}, J. and {Ciardullo}, R.},
        title = "{The High-Z Supernova Search: Measuring Cosmic Deceleration and Global Curvature of the Universe Using Type IA Supernovae}",
      journal = {\apj},
         year = 1998,
        month = nov,
       volume = {507},
       number = {1},
        pages = {46-63},
          doi = {10.1086/306308},
archivePrefix = {arXiv},
       eprint = {astro-ph/9805200},
 primaryClass = {astro-ph},
       adsurl = {https://ui.adsabs.harvard.edu/abs/1998ApJ...507...46S}
}

@ARTICLE{amon_dark_2022,
       author = {{Amon}, A. and {Gruen}, D. and {Troxel}, M.~A. and {MacCrann}, N. and {Dodelson}, S. and {Choi}, A. and {Doux}, C. and {Secco}, L.~F. and {Samuroff}, S. and {Krause}, E. and {Cordero}, J. and {Myles}, J. and {DeRose}, J. and {Wechsler}, R.~H. and {Gatti}, M. and {Navarro-Alsina}, A. and {Bernstein}, G.~M. and {Jain}, B. and {Blazek}, J. and {Alarcon}, A. and {Fert{\'e}}, A. and {Lemos}, P. and {Raveri}, M. and {Campos}, A. and {Prat}, J. and {S{\'a}nchez}, C. and {Jarvis}, M. and {Alves}, O. and {Andrade-Oliveira}, F. and {Baxter}, E. and {Bechtol}, K. and {Becker}, M.~R. and {Bridle}, S.~L. and {Camacho}, H. and {Carnero Rosell}, A. and {Carrasco Kind}, M. and {Cawthon}, R. and {Chang}, C. and {Chen}, R. and {Chintalapati}, P. and {Crocce}, M. and {Davis}, C. and {Diehl}, H.~T. and {Drlica-Wagner}, A. and {Eckert}, K. and {Eifler}, T.~F. and {Elvin-Poole}, J. and {Everett}, S. and {Fang}, X. and {Fosalba}, P. and {Friedrich}, O. and {Gaztanaga}, E. and {Giannini}, G. and {Gruendl}, R.~A. and {Harrison}, I. and {Hartley}, W.~G. and {Herner}, K. and {Huang}, H. and {Huff}, E.~M. and {Huterer}, D. and {Kuropatkin}, N. and {Leget}, P. and {Liddle}, A.~R. and {McCullough}, J. and {Muir}, J. and {Pandey}, S. and {Park}, Y. and {Porredon}, A. and {Refregier}, A. and {Rollins}, R.~P. and {Roodman}, A. and {Rosenfeld}, R. and {Ross}, A.~J. and {Rykoff}, E.~S. and {Sanchez}, J. and {Sevilla-Noarbe}, I. and {Sheldon}, E. and {Shin}, T. and {Troja}, A. and {Tutusaus}, I. and {Tutusaus}, I. and {Varga}, T.~N. and {Weaverdyck}, N. and {Yanny}, B. and {Yin}, B. and {Zhang}, Y. and {Zuntz}, J. and {Aguena}, M. and {Allam}, S. and {Annis}, J. and {Bacon}, D. and {Bertin}, E. and {Bhargava}, S. and {Brooks}, D. and {Buckley-Geer}, E. and {Burke}, D.~L. and {Carretero}, J. and {Costanzi}, M. and {da Costa}, L.~N. and {Pereira}, M.~E.~S. and {De Vicente}, J. and {Desai}, S. and {Dietrich}, J.~P. and {Doel}, P. and {Ferrero}, I. and {Flaugher}, B. and {Frieman}, J. and {Garc{\'\i}a-Bellido}, J. and {Gaztanaga}, E. and {Gerdes}, D.~W. and {Giannantonio}, T. and {Gschwend}, J. and {Gutierrez}, G. and {Hinton}, S.~R. and {Hollowood}, D.~L. and {Honscheid}, K. and {Hoyle}, B. and {James}, D.~J. and {Kron}, R. and {Kuehn}, K. and {Lahav}, O. and {Lima}, M. and {Lin}, H. and {Maia}, M.~A.~G. and {Marshall}, J.~L. and {Martini}, P. and {Melchior}, P. and {Menanteau}, F. and {Miquel}, R. and {Mohr}, J.~J. and {Morgan}, R. and {Ogando}, R.~L.~C. and {Palmese}, A. and {Paz-Chinch{\'o}n}, F. and {Petravick}, D. and {Pieres}, A. and {Romer}, A.~K. and {Sanchez}, E. and {Scarpine}, V. and {Schubnell}, M. and {Serrano}, S. and {Smith}, M. and {Soares-Santos}, M. and {Tarle}, G. and {Thomas}, D. and {To}, C. and {Weller}, J. and {DES Collaboration}},
        title = "{Dark Energy Survey Year 3 results: Cosmology from cosmic shear and robustness to data calibration}",
      journal = {\prd},
         year = 2022,
        month = jan,
       volume = {105},
       number = {2},
          eid = {023514},
        pages = {023514},
          doi = {10.1103/PhysRevD.105.023514},
archivePrefix = {arXiv},
       eprint = {2105.13543},
 primaryClass = {astro-ph.CO},
       adsurl = {https://ui.adsabs.harvard.edu/abs/2022PhRvD.105b3514A}
}

@ARTICLE{Schaye2023_flamingo,
       author = {{Schaye}, Joop and {Kugel}, Roi and {Schaller}, Matthieu and {Helly}, John C. and {Braspenning}, Joey and {Elbers}, Willem and {McCarthy}, Ian G. and {van Daalen}, Marcel P. and {Vandenbroucke}, Bert and {Frenk}, Carlos S. and {Kwan}, Juliana and {Salcido}, Jaime and {Bah{\'e}}, Yannick M. and {Borrow}, Josh and {Chaikin}, Evgenii and {Hahn}, Oliver and {Hu{\v{s}}ko}, Filip and {Jenkins}, Adrian and {Lacey}, Cedric G. and {Nobels}, Folkert S.~J.},
        title = "{The FLAMINGO project: cosmological hydrodynamical simulations for large-scale structure and galaxy cluster surveys}",
      journal = {\mnras},
         year = 2023,
        month = dec,
       volume = {526},
       number = {4},
        pages = {4978-5020},
          doi = {10.1093/mnras/stad2419},
archivePrefix = {arXiv},
       eprint = {2306.04024},
 primaryClass = {astro-ph.CO},
       adsurl = {https://ui.adsabs.harvard.edu/abs/2023MNRAS.526.4978S}
}

@ARTICLE{Schaller_pkFlamingo_2024,
       author = {{Schaller}, Matthieu and {Schaye}, Joop and {Kugel}, Roi and {Broxterman}, Jeger C. and {van Daalen}, Marcel P.},
        title = "{The FLAMINGO project: baryon effects on the matter power spectrum}",
      journal = {\mnras},
         year = 2025,
        month = may,
       volume = {539},
       number = {2},
        pages = {1337-1351},
          doi = {10.1093/mnras/staf569},
archivePrefix = {arXiv},
       eprint = {2410.17109},
 primaryClass = {astro-ph.CO},
       adsurl = {https://ui.adsabs.harvard.edu/abs/2025MNRAS.539.1337S}
}

@ARTICLE{asgari_kids_2021,
       author = {{Asgari}, Marika and {Lin}, Chieh-An and {Joachimi}, Benjamin and {Giblin}, Benjamin and {Heymans}, Catherine and {Hildebrandt}, Hendrik and {Kannawadi}, Arun and {St{\"o}lzner}, Benjamin and {Tr{\"o}ster}, Tilman and {van den Busch}, Jan Luca and {Wright}, Angus H. and {Bilicki}, Maciej and {Blake}, Chris and {de Jong}, Jelte and {Dvornik}, Andrej and {Erben}, Thomas and {Getman}, Fedor and {Hoekstra}, Henk and {K{\"o}hlinger}, Fabian and {Kuijken}, Konrad and {Miller}, Lance and {Radovich}, Mario and {Schneider}, Peter and {Shan}, HuanYuan and {Valentijn}, Edwin},
        title = "{KiDS-1000 cosmology: Cosmic shear constraints and comparison between two point statistics}",
      journal = {\aap},
         year = 2021,
        month = jan,
       volume = {645},
          eid = {A104},
        pages = {A104},
          doi = {10.1051/0004-6361/202039070},
archivePrefix = {arXiv},
       eprint = {2007.15633},
 primaryClass = {astro-ph.CO},
       adsurl = {https://ui.adsabs.harvard.edu/abs/2021A&A...645A.104A}
}

@ARTICLE{1999ApJ...517..565P,
       author = {{Perlmutter}, S. and {Aldering}, G. and {Goldhaber}, G. and {Knop}, R.~A. and {Nugent}, P. and {Castro}, P.~G. and {Deustua}, S. and {Fabbro}, S. and {Goobar}, A. and {Groom}, D.~E. and {Hook}, I.~M. and {Kim}, A.~G. and {Kim}, M.~Y. and {Lee}, J.~C. and {Nunes}, N.~J. and {Pain}, R. and {Pennypacker}, C.~R. and {Quimby}, R. and {Lidman}, C. and {Ellis}, R.~S. and {Irwin}, M. and {McMahon}, R.~G. and {Ruiz-Lapuente}, P. and {Walton}, N. and {Schaefer}, B. and {Boyle}, B.~J. and {Filippenko}, A.~V. and {Matheson}, T. and {Fruchter}, A.~S. and {Panagia}, N. and {Newberg}, H.~J.~M. and {Couch}, W.~J. and {Project}, The Supernova Cosmology},
        title = "{Measurements of {\ensuremath{\Omega}} and {\ensuremath{\Lambda}} from 42 High-Redshift Supernovae}",
      journal = {\apj},
         year = 1999,
        month = jun,
       volume = {517},
       number = {2},
        pages = {565-586},
          doi = {10.1086/307221},
archivePrefix = {arXiv},
       eprint = {astro-ph/9812133},
 primaryClass = {astro-ph},
       adsurl = {https://ui.adsabs.harvard.edu/abs/1999ApJ...517..565P}
}

@ARTICLE{2021PhRvD.103h3533A,
       author = {{Alam}, Shadab and {Aubert}, Marie and {Avila}, Santiago and {Balland}, Christophe and {Bautista}, Julian E. and {Bershady}, Matthew A. and {Bizyaev}, Dmitry and {Blanton}, Michael R. and {Bolton}, Adam S. and {Bovy}, Jo and {Brinkmann}, Jonathan and {Brownstein}, Joel R. and {Burtin}, Etienne and {Chabanier}, Sol{\`e}ne and {Chapman}, Michael J. and {Choi}, Peter Doohyun and {Chuang}, Chia-Hsun and {Comparat}, Johan and {Cousinou}, Marie-Claude and {Cuceu}, Andrei and {Dawson}, Kyle S. and {de la Torre}, Sylvain and {de Mattia}, Arnaud and {Agathe}, Victoria de Sainte and {des Bourboux}, H{\'e}lion du Mas and {Escoffier}, Stephanie and {Etourneau}, Thomas and {Farr}, James and {Font-Ribera}, Andreu and {Frinchaboy}, Peter M. and {Fromenteau}, Sebastien and {Gil-Mar{\'\i}n}, H{\'e}ctor and {Le Goff}, Jean-Marc and {Gonzalez-Morales}, Alma X. and {Gonzalez-Perez}, Violeta and {Grabowski}, Kathleen and {Guy}, Julien and {Hawken}, Adam J. and {Hou}, Jiamin and {Kong}, Hui and {Parker}, James and {Klaene}, Mark and {Kneib}, Jean-Paul and {Lin}, Sicheng and {Long}, Daniel and {Lyke}, Brad W. and {de la Macorra}, Axel and {Martini}, Paul and {Masters}, Karen and {Mohammad}, Faizan G. and {Moon}, Jeongin and {Mueller}, Eva-Maria and {Mu{\~n}oz-Guti{\'e}rrez}, Andrea and {Myers}, Adam D. and {Nadathur}, Seshadri and {Neveux}, Richard and {Newman}, Jeffrey A. and {Noterdaeme}, Pasquier and {Oravetz}, Audrey and {Oravetz}, Daniel and {Palanque-Delabrouille}, Nathalie and {Pan}, Kaike and {Paviot}, Romain and {Percival}, Will J. and {P{\'e}rez-R{\`a}fols}, Ignasi and {Petitjean}, Patrick and {Pieri}, Matthew M. and {Prakash}, Abhishek and {Raichoor}, Anand and {Ravoux}, Corentin and {Rezaie}, Mehdi and {Rich}, James and {Ross}, Ashley J. and {Rossi}, Graziano and {Ruggeri}, Rossana and {Ruhlmann-Kleider}, Vanina and {S{\'a}nchez}, Ariel G. and {S{\'a}nchez}, F. Javier and {S{\'a}nchez-Gallego}, Jos{\'e} R. and {Sayres}, Conor and {Schneider}, Donald P. and {Seo}, Hee-Jong and {Shafieloo}, Arman and {Slosar}, An{\v{z}}e and {Smith}, Alex and {Stermer}, Julianna and {Tamone}, Amelie and {Tinker}, Jeremy L. and {Tojeiro}, Rita and {Vargas-Maga{\~n}a}, Mariana and {Variu}, Andrei and {Wang}, Yuting and {Weaver}, Benjamin A. and {Weijmans}, Anne-Marie and {Y{\`e}che}, Christophe and {Zarrouk}, Pauline and {Zhao}, Cheng and {Zhao}, Gong-Bo and {Zheng}, Zheng},
        title = "{Completed SDSS-IV extended Baryon Oscillation Spectroscopic Survey: Cosmological implications from two decades of spectroscopic surveys at the Apache Point Observatory}",
      journal = {\prd},
         year = 2021,
        month = apr,
       volume = {103},
       number = {8},
          eid = {083533},
        pages = {083533},
          doi = {10.1103/PhysRevD.103.083533},
archivePrefix = {arXiv},
       eprint = {2007.08991},
 primaryClass = {astro-ph.CO},
       adsurl = {https://ui.adsabs.harvard.edu/abs/2021PhRvD.103h3533A}
}

@ARTICLE{Scolnic22,
       author = {{Scolnic}, Dan and {Brout}, Dillon and {Carr}, Anthony and {Riess}, Adam G. and {Davis}, Tamara M. and {Dwomoh}, Arianna and {Jones}, David O. and {Ali}, Noor and {Charvu}, Pranav and {Chen}, Rebecca and {Peterson}, Erik R. and {Popovic}, Brodie and {Rose}, Benjamin M. and {Wood}, Charlotte M. and {Brown}, Peter J. and {Chambers}, Ken and {Coulter}, David A. and {Dettman}, Kyle G. and {Dimitriadis}, Georgios and {Filippenko}, Alexei V. and {Foley}, Ryan J. and {Jha}, Saurabh W. and {Kilpatrick}, Charles D. and {Kirshner}, Robert P. and {Pan}, Yen-Chen and {Rest}, Armin and {Rojas-Bravo}, Cesar and {Siebert}, Matthew R. and {Stahl}, Benjamin E. and {Zheng}, WeiKang},
        title = "{The Pantheon+ Analysis: The Full Data Set and Light-curve Release}",
      journal = {\apj},
         year = 2022,
        month = oct,
       volume = {938},
       number = {2},
          eid = {113},
        pages = {113},
          doi = {10.3847/1538-4357/ac8b7a},
archivePrefix = {arXiv},
       eprint = {2112.03863},
 primaryClass = {astro-ph.CO},
       adsurl = {https://ui.adsabs.harvard.edu/abs/2022ApJ...938..113S}
}

@ARTICLE{Brout22,
       author = {{Brout}, Dillon and {Scolnic}, Dan and {Popovic}, Brodie and {Riess}, Adam G. and {Carr}, Anthony and {Zuntz}, Joe and {Kessler}, Rick and {Davis}, Tamara M. and {Hinton}, Samuel and {Jones}, David and {Kenworthy}, W. D'Arcy and {Peterson}, Erik R. and {Said}, Khaled and {Taylor}, Georgie and {Ali}, Noor and {Armstrong}, Patrick and {Charvu}, Pranav and {Dwomoh}, Arianna and {Meldorf}, Cole and {Palmese}, Antonella and {Qu}, Helen and {Rose}, Benjamin M. and {Sanchez}, Bruno and {Stubbs}, Christopher W. and {Vincenzi}, Maria and {Wood}, Charlotte M. and {Brown}, Peter J. and {Chen}, Rebecca and {Chambers}, Ken and {Coulter}, David A. and {Dai}, Mi and {Dimitriadis}, Georgios and {Filippenko}, Alexei V. and {Foley}, Ryan J. and {Jha}, Saurabh W. and {Kelsey}, Lisa and {Kirshner}, Robert P. and {M{\"o}ller}, Anais and {Muir}, Jessie and {Nadathur}, Seshadri and {Pan}, Yen-Chen and {Rest}, Armin and {Rojas-Bravo}, Cesar and {Sako}, Masao and {Siebert}, Matthew R. and {Smith}, Mat and {Stahl}, Benjamin E. and {Wiseman}, Phil},
        title = "{The Pantheon+ Analysis: Cosmological Constraints}",
      journal = {\apj},
         year = 2022,
        month = oct,
       volume = {938},
       number = {2},
          eid = {110},
        pages = {110},
          doi = {10.3847/1538-4357/ac8e04},
archivePrefix = {arXiv},
       eprint = {2202.04077},
 primaryClass = {astro-ph.CO},
       adsurl = {https://ui.adsabs.harvard.edu/abs/2022ApJ...938..110B}
}

@ARTICLE{2025ApJ...985..203F,
       author = {{Freedman}, Wendy L. and {Madore}, Barry F. and {Hoyt}, Taylor J. and {Jang}, In Sung and {Lee}, Abigail J. and {Owens}, Kayla A.},
        title = "{Status Report on the Chicago-Carnegie Hubble Program (CCHP): Measurement of the Hubble Constant Using the Hubble and James Webb Space Telescopes}",
      journal = {\apj},
         year = 2025,
        month = jun,
       volume = {985},
       number = {2},
          eid = {203},
        pages = {203},
          doi = {10.3847/1538-4357/adce78},
archivePrefix = {arXiv},
       eprint = {2408.06153},
 primaryClass = {astro-ph.CO},
       adsurl = {https://ui.adsabs.harvard.edu/abs/2025ApJ...985..203F}
}

@ARTICLE{2020A&A...641A...5P,
       author = {{Planck Collaboration} and {Aghanim}, N. and {Akrami}, Y. and {Ashdown}, M. and {Aumont}, J. and {Baccigalupi}, C. and {Ballardini}, M. and {Banday}, A.~J. and {Barreiro}, R.~B. and {Bartolo}, N. and {Basak}, S. and {Benabed}, K. and {Bernard}, J. -P. and {Bersanelli}, M. and {Bielewicz}, P. and {Bock}, J.~J. and {Bond}, J.~R. and {Borrill}, J. and {Bouchet}, F.~R. and {Boulanger}, F. and {Bucher}, M. and {Burigana}, C. and {Butler}, R.~C. and {Calabrese}, E. and {Cardoso}, J. -F. and {Carron}, J. and {Casaponsa}, B. and {Challinor}, A. and {Chiang}, H.~C. and {Colombo}, L.~P.~L. and {Combet}, C. and {Crill}, B.~P. and {Cuttaia}, F. and {de Bernardis}, P. and {de Rosa}, A. and {de Zotti}, G. and {Delabrouille}, J. and {Delouis}, J. -M. and {Di Valentino}, E. and {Diego}, J.~M. and {Dor{\'e}}, O. and {Douspis}, M. and {Ducout}, A. and {Dupac}, X. and {Dusini}, S. and {Efstathiou}, G. and {Elsner}, F. and {En{\ss}lin}, T.~A. and {Eriksen}, H.~K. and {Fantaye}, Y. and {Fernandez-Cobos}, R. and {Finelli}, F. and {Frailis}, M. and {Fraisse}, A.~A. and {Franceschi}, E. and {Frolov}, A. and {Galeotta}, S. and {Galli}, S. and {Ganga}, K. and {G{\'e}nova-Santos}, R.~T. and {Gerbino}, M. and {Ghosh}, T. and {Giraud-H{\'e}raud}, Y. and {Gonz{\'a}lez-Nuevo}, J. and {G{\'o}rski}, K.~M. and {Gratton}, S. and {Gruppuso}, A. and {Gudmundsson}, J.~E. and {Hamann}, J. and {Handley}, W. and {Hansen}, F.~K. and {Herranz}, D. and {Hivon}, E. and {Huang}, Z. and {Jaffe}, A.~H. and {Jones}, W.~C. and {Keih{\"a}nen}, E. and {Keskitalo}, R. and {Kiiveri}, K. and {Kim}, J. and {Kisner}, T.~S. and {Krachmalnicoff}, N. and {Kunz}, M. and {Kurki-Suonio}, H. and {Lagache}, G. and {Lamarre}, J. -M. and {Lasenby}, A. and {Lattanzi}, M. and {Lawrence}, C.~R. and {Le Jeune}, M. and {Levrier}, F. and {Lewis}, A. and {Liguori}, M. and {Lilje}, P.~B. and {Lilley}, M. and {Lindholm}, V. and {L{\'o}pez-Caniego}, M. and {Lubin}, P.~M. and {Ma}, Y. -Z. and {Mac{\'\i}as-P{\'e}rez}, J.~F. and {Maggio}, G. and {Maino}, D. and {Mandolesi}, N. and {Mangilli}, A. and {Marcos-Caballero}, A. and {Maris}, M. and {Martin}, P.~G. and {Mart{\'\i}nez-Gonz{\'a}lez}, E. and {Matarrese}, S. and {Mauri}, N. and {McEwen}, J.~D. and {Meinhold}, P.~R. and {Melchiorri}, A. and {Mennella}, A. and {Migliaccio}, M. and {Millea}, M. and {Miville-Desch{\^e}nes}, M. -A. and {Molinari}, D. and {Moneti}, A. and {Montier}, L. and {Morgante}, G. and {Moss}, A. and {Natoli}, P. and {N{\o}rgaard-Nielsen}, H.~U. and {Pagano}, L. and {Paoletti}, D. and {Partridge}, B. and {Patanchon}, G. and {Peiris}, H.~V. and {Perrotta}, F. and {Pettorino}, V. and {Piacentini}, F. and {Polenta}, G. and {Puget}, J. -L. and {Rachen}, J.~P. and {Reinecke}, M. and {Remazeilles}, M. and {Renzi}, A. and {Rocha}, G. and {Rosset}, C. and {Roudier}, G. and {Rubi{\~n}o-Mart{\'\i}n}, J.~A. and {Ruiz-Granados}, B. and {Salvati}, L. and {Sandri}, M. and {Savelainen}, M. and {Scott}, D. and {Shellard}, E.~P.~S. and {Sirignano}, C. and {Sirri}, G. and {Spencer}, L.~D. and {Sunyaev}, R. and {Suur-Uski}, A. -S. and {Tauber}, J.~A. and {Tavagnacco}, D. and {Tenti}, M. and {Toffolatti}, L. and {Tomasi}, M. and {Trombetti}, T. and {Valiviita}, J. and {Van Tent}, B. and {Vielva}, P. and {Villa}, F. and {Vittorio}, N. and {Wandelt}, B.~D. and {Wehus}, I.~K. and {Zacchei}, A. and {Zonca}, A.},
        title = "{Planck 2018 results. V. CMB power spectra and likelihoods}",
      journal = {\aap},
         year = 2020,
        month = sep,
       volume = {641},
          eid = {A5},
        pages = {A5},
          doi = {10.1051/0004-6361/201936386},
archivePrefix = {arXiv},
       eprint = {1907.12875},
 primaryClass = {astro-ph.CO},
       adsurl = {https://ui.adsabs.harvard.edu/abs/2020A&A...641A...5P}
}

@ARTICLE{2020MNRAS.498.1420W,
       author = {{Wong}, Kenneth C. and {Suyu}, Sherry H. and {Chen}, Geoff C. -F. and {Rusu}, Cristian E. and {Millon}, Martin and {Sluse}, Dominique and {Bonvin}, Vivien and {Fassnacht}, Christopher D. and {Taubenberger}, Stefan and {Auger}, Matthew W. and {Birrer}, Simon and {Chan}, James H.~H. and {Courbin}, Frederic and {Hilbert}, Stefan and {Tihhonova}, Olga and {Treu}, Tommaso and {Agnello}, Adriano and {Ding}, Xuheng and {Jee}, Inh and {Komatsu}, Eiichiro and {Shajib}, Anowar J. and {Sonnenfeld}, Alessandro and {Blandford}, Roger D. and {Koopmans}, L{\'e}on V.~E. and {Marshall}, Philip J. and {Meylan}, Georges},
        title = "{H0LiCOW - XIII. A 2.4 per cent measurement of H$_{0}$ from lensed quasars: 5.3{\ensuremath{\sigma}} tension between early- and late-Universe probes}",
      journal = {\mnras},
         year = 2020,
        month = oct,
       volume = {498},
       number = {1},
        pages = {1420-1439},
          doi = {10.1093/mnras/stz3094},
archivePrefix = {arXiv},
       eprint = {1907.04869},
 primaryClass = {astro-ph.CO},
       adsurl = {https://ui.adsabs.harvard.edu/abs/2020MNRAS.498.1420W}
}

@ARTICLE{2021A&A...649A..88T,
       author = {{Tr{\"o}ster}, Tilman and {Asgari}, Marika and {Blake}, Chris and {Cataneo}, Matteo and {Heymans}, Catherine and {Hildebrandt}, Hendrik and {Joachimi}, Benjamin and {Lin}, Chieh-An and {S{\'a}nchez}, Ariel G. and {Wright}, Angus H. and {Bilicki}, Maciej and {Bose}, Benjamin and {Crocce}, Martin and {Dvornik}, Andrej and {Erben}, Thomas and {Giblin}, Benjamin and {Glazebrook}, Karl and {Hoekstra}, Henk and {Joudaki}, Shahab and {Kannawadi}, Arun and {K{\"o}hlinger}, Fabian and {Kuijken}, Konrad and {Lidman}, Chris and {Lombriser}, Lucas and {Mead}, Alexander and {Parkinson}, David and {Shan}, HuanYuan and {Wolf}, Christian and {Xia}, Qianli},
        title = "{KiDS-1000 Cosmology: Constraints beyond flat {\ensuremath{\Lambda}}CDM}",
      journal = {\aap},
         year = 2021,
        month = may,
       volume = {649},
          eid = {A88},
        pages = {A88},
          doi = {10.1051/0004-6361/202039805},
archivePrefix = {arXiv},
       eprint = {2010.16416},
 primaryClass = {astro-ph.CO},
       adsurl = {https://ui.adsabs.harvard.edu/abs/2021A&A...649A..88T}
}

@ARTICLE{2022PhR...984....1S,
       author = {{Sch{\"o}neberg}, Nils and {Abell{\'a}n}, Guillermo Franco and {S{\'a}nchez}, Andrea P{\'e}rez and {Witte}, Samuel J. and {Poulin}, Vivian and {Lesgourgues}, Julien},
        title = "{The H$_{0}$ Olympics: A fair ranking of proposed models}",
      journal = {\physrep},
         year = 2022,
        month = oct,
       volume = {984},
        pages = {1-55},
          doi = {10.1016/j.physrep.2022.07.001},
archivePrefix = {arXiv},
       eprint = {2107.10291},
 primaryClass = {astro-ph.CO},
       adsurl = {https://ui.adsabs.harvard.edu/abs/2022PhR...984....1S}
}

@ARTICLE{2023ARNPS..73..153K,
       author = {{Kamionkowski}, Marc and {Riess}, Adam G.},
        title = "{The Hubble Tension and Early Dark Energy}",
      journal = {Ann. Rev. Nucl. Part. Sci.},
         year = 2023,
        month = sep,
       volume = {73},
        pages = {153-180},
          doi = {10.1146/annurev-nucl-111422-024107},
archivePrefix = {arXiv},
       eprint = {2211.04492},
 primaryClass = {astro-ph.CO},
       adsurl = {https://ui.adsabs.harvard.edu/abs/2023ARNPS..73..153K}
}

@ARTICLE{2019ApJ...876...85R,
       author = {{Riess}, Adam G. and {Casertano}, Stefano and {Yuan}, Wenlong and {Macri}, Lucas M. and {Scolnic}, Dan},
        title = "{Large Magellanic Cloud Cepheid Standards Provide a 1\% Foundation for the Determination of the Hubble Constant and Stronger Evidence for Physics beyond {\ensuremath{\Lambda}}CDM}",
      journal = {\apj},
         year = 2019,
        month = may,
       volume = {876},
       number = {1},
          eid = {85},
        pages = {85},
          doi = {10.3847/1538-4357/ab1422},
archivePrefix = {arXiv},
       eprint = {1903.07603},
 primaryClass = {astro-ph.CO},
       adsurl = {https://ui.adsabs.harvard.edu/abs/2019ApJ...876...85R}
}

@ARTICLE{2021MNRAS.500.1201H,
       author = {{Hou}, Jiamin and {S{\'a}nchez}, Ariel G. and {Ross}, Ashley J. and {Smith}, Alex and {Neveux}, Richard and {Bautista}, Julian and {Burtin}, Etienne and {Zhao}, Cheng and {Scoccimarro}, Rom{\'a}n and {Dawson}, Kyle S. and {de Mattia}, Arnaud and {de la Macorra}, Axel and {du Mas des Bourboux}, H{\'e}lion and {Eisenstein}, Daniel J. and {Gil-Mar{\'\i}n}, H{\'e}ctor and {Lyke}, Brad W. and {Mohammad}, Faizan G. and {Mueller}, Eva-Maria and {Percival}, Will J. and {Rossi}, Graziano and {Vargas Maga{\~n}a}, Mariana and {Zarrouk}, Pauline and {Zhao}, Gong-Bo and {Brinkmann}, Jonathan and {Brownstein}, Joel R. and {Chuang}, Chia-Hsun and {Myers}, Adam D. and {Newman}, Jeffrey A. and {Schneider}, Donald P. and {Vivek}, M.},
        title = "{The completed SDSS-IV extended Baryon Oscillation Spectroscopic Survey: BAO and RSD measurements from anisotropic clustering analysis of the quasar sample in configuration space between redshift 0.8 and 2.2}",
      journal = {\mnras},
         year = 2021,
        month = jan,
       volume = {500},
       number = {1},
        pages = {1201-1221},
          doi = {10.1093/mnras/staa3234},
archivePrefix = {arXiv},
       eprint = {2007.08998},
 primaryClass = {astro-ph.CO},
       adsurl = {https://ui.adsabs.harvard.edu/abs/2021MNRAS.500.1201H}
}

@ARTICLE{2025arXiv250314738D,
       author = {{DESI Collaboration} and {Abdul-Karim}, M. and {Aguilar}, J. and {Ahlen}, S. and {Alam}, S. and {Allen}, L. and {Allende Prieto}, C. and {Alves}, O. and {Anand}, A. and {Andrade}, U. and {Armengaud}, E. and {Aviles}, A. and {Bailey}, S. and {Baltay}, C. and {Bansal}, P. and {Bault}, A. and {Behera}, J. and {BenZvi}, S. and {Bianchi}, D. and {Blake}, C. and {Brieden}, S. and {Brodzeller}, A. and {Brooks}, D. and {Buckley-Geer}, E. and {Burtin}, E. and {Calderon}, R. and {Canning}, R. and {Carnero Rosell}, A. and {Carrilho}, P. and {Casas}, L. and {Castander}, F.~J. and {Cereskaite}, R. and {Charles}, M. and {Chaussidon}, E. and {Chaves-Montero}, J. and {Chebat}, D. and {Chen}, X. and {Claybaugh}, T. and {Cole}, S. and {Cooper}, A.~P. and {Cuceu}, A. and {Dawson}, K.~S. and {de la Macorra}, A. and {de Mattia}, A. and {Deiosso}, N. and {Della Costa}, J. and {Demina}, R. and {Dey}, A. and {Dey}, B. and {Ding}, Z. and {Doel}, P. and {Edelstein}, J. and {Eisenstein}, D.~J. and {Elbers}, W. and {Fagrelius}, P. and {Fanning}, K.},
        title = "{DESI DR2 results. II. Measurements of baryon acoustic oscillations and cosmological constraints}",
      journal = {\prd},
         year = 2025,
        month = oct,
       volume = {112},
       number = {8},
          eid = {083515},
        pages = {083515},
          doi = {10.1103/tr6y-kpc6},
archivePrefix = {arXiv},
       eprint = {2503.14738},
 primaryClass = {astro-ph.CO},
       adsurl = {https://ui.adsabs.harvard.edu/abs/2025PhRvD.112h3515A}
}

@ARTICLE{2001IJMPD..10..213C,
       author = {{Chevallier}, Michel and {Polarski}, David},
        title = "{Accelerating Universes with Scaling Dark Matter}",
      journal = {Int. J. Mod. Phys. D},
         year = 2001,
        month = jan,
       volume = {10},
       number = {2},
        pages = {213-223},
          doi = {10.1142/S0218271801000822},
archivePrefix = {arXiv},
       eprint = {gr-qc/0009008},
 primaryClass = {gr-qc},
       adsurl = {https://ui.adsabs.harvard.edu/abs/2001IJMPD..10..213C}
}

@ARTICLE{2003PhRvL..90i1301L,
       author = {{Linder}, Eric V.},
        title = "{Exploring the Expansion History of the Universe}",
      journal = {\prl},
         year = 2003,
        month = mar,
       volume = {90},
       number = {9},
          eid = {091301},
        pages = {091301},
          doi = {10.1103/PhysRevLett.90.091301},
archivePrefix = {arXiv},
       eprint = {astro-ph/0208512},
 primaryClass = {astro-ph},
       adsurl = {https://ui.adsabs.harvard.edu/abs/2003PhRvL..90i1301L}
}

@ARTICLE{1998AJ....116.1009R,
       author = {{Riess}, Adam G. and {Filippenko}, Alexei V. and {Challis}, Peter and {Clocchiatti}, Alejandro and {Diercks}, Alan and {Garnavich}, Peter M. and {Gilliland}, Ron L. and {Hogan}, Craig J. and {Jha}, Saurabh and {Kirshner}, Robert P. and {Leibundgut}, B. and {Phillips}, M.~M. and {Reiss}, David and {Schmidt}, Brian P. and {Schommer}, Robert A. and {Smith}, R. Chris and {Spyromilio}, J. and {Stubbs}, Christopher and {Suntzeff}, Nicholas B. and {Tonry}, John},
        title = "{Observational Evidence from Supernovae for an Accelerating Universe and a Cosmological Constant}",
      journal = {\aj},
         year = 1998,
        month = sep,
       volume = {116},
       number = {3},
        pages = {1009-1038},
          doi = {10.1086/300499},
archivePrefix = {arXiv},
       eprint = {astro-ph/9805201},
 primaryClass = {astro-ph},
       adsurl = {https://ui.adsabs.harvard.edu/abs/1998AJ....116.1009R}
}

@ARTICLE{2025arXiv250319442S,
       author = {{St{\"o}lzner}, Benjamin and {Wright}, Angus H. and {Asgari}, Marika and {Heymans}, Catherine and {Hildebrandt}, Hendrik and {Hoekstra}, Henk and {Joachimi}, Benjamin and {Kuijken}, Konrad and {Li}, Shun-Sheng and {Mahony}, Constance and {Reischke}, Robert and {Yoon}, Mijin and {Bilicki}, Maciej and {Burger}, Pierre and {Chisari}, Nora Elisa and {Dvornik}, Andrej and {Georgiou}, Christos and {Giblin}, Benjamin and {Harnois-D{\'e}raps}, Joachim and {Jalan}, Priyanka and {William}, Anjitha John and {Joudaki}, Shahab and {Lesci}, Giorgio Francesco and {Linke}, Laila and {Loureiro}, Arthur and {Maturi}, Matteo and {Moscardini}, Lauro and {Napolitano}, Nicola R. and {Porth}, Lucas and {Radovich}, Mario and {Tr{\"o}ster}, Tilman and {von Wietersheim-Kramsta}, Maximilian and {Wittje}, Anna and {Yan}, Ziang and {Zhang}, Yun-Hao},
        title = "{KiDS-Legacy: Consistency of cosmic shear measurements and joint cosmological constraints with external probes}",
      journal = {\aap},
	url= "https://doi.org/10.1051/0004-6361/202554893",
	year = 2025,
	volume = 702,
	pages = "A169",
          doi = {10.1051/0004-6361/202554893},
archivePrefix = {arXiv},
       eprint = {2503.19442},
 primaryClass = {astro-ph.CO},
       adsurl = {https://ui.adsabs.harvard.edu/abs/2025arXiv250319442S}
}

@ARTICLE{2006PhR...429..307L,
       author = {{Lesgourgues}, Julien and {Pastor}, Sergio},
        title = "{Massive neutrinos and cosmology}",
      journal = {\physrep},
         year = 2006,
        month = jul,
       volume = {429},
       number = {6},
        pages = {307-379},
          doi = {10.1016/j.physrep.2006.04.001},
archivePrefix = {arXiv},
       eprint = {astro-ph/0603494},
 primaryClass = {astro-ph},
       adsurl = {https://ui.adsabs.harvard.edu/abs/2006PhR...429..307L}
}

@ARTICLE{gerbino_status_2017,
       author = {{Gerbino}, Martina and {Lattanzi}, Massimiliano},
        title = "{Status of neutrino properties and future prospects - Cosmological and astrophysical constraints}",
      journal = {Front. Phys.},
         year = 2017,
        month = dec,
       volume = {5},
          eid = {70},
        pages = {70},
          doi = {10.3389/fphy.2017.00070},
archivePrefix = {arXiv},
       eprint = {1712.07109},
 primaryClass = {astro-ph.CO},
       adsurl = {https://ui.adsabs.harvard.edu/abs/2017FrP.....5...70G}
}

@ARTICLE{abazajian_neutrinos_2015,
       author = {{Abazajian}, K.~N. and {Arnold}, K. and {Austermann}, J. and {Benson}, B.~A. and {Bischoff}, C. and {Bock}, J. and {Bond}, J.~R. and {Borrill}, J. and {Calabrese}, E. and {Carlstrom}, J.~E. and {Carvalho}, C.~S. and {Chang}, C.~L. and {Chiang}, H.~C. and {Church}, S. and {Cooray}, A. and {Crawford}, T.~M. and {Dawson}, K.~S. and {Das}, S. and {Devlin}, M.~J. and {Dobbs}, M. and {Dodelson}, S. and {Dor{\'e}}, O. and {Dunkley}, J. and {Errard}, J. and {Fraisse}, A. and {Gallicchio}, J. and {Halverson}, N.~W. and {Hanany}, S. and {Hildebrandt}, S.~R. and {Hincks}, A. and {Hlozek}, R. and {Holder}, G. and {Holzapfel}, W.~L. and {Honscheid}, K. and {Hu}, W. and {Hubmayr}, J. and {Irwin}, K. and {Jones}, W.~C. and {Kamionkowski}, M. and {Keating}, B. and {Keisler}, R. and {Knox}, L. and {Komatsu}, E. and {Kovac}, J. and {Kuo}, C. -L. and {Lawrence}, C. and {Lee}, A.~T. and {Leitch}, E. and {Linder}, E. and {Lubin}, P. and {McMahon}, J. and {Miller}, A. and {Newburgh}, L. and {Niemack}, M.~D. and {Nguyen}, H. and {Nguyen}, H.~T. and {Page}, L. and {Pryke}, C. and {Reichardt}, C.~L. and {Ruhl}, J.~E. and {Sehgal}, N. and {Seljak}, U. and {Sievers}, J. and {Silverstein}, E. and {Slosar}, A. and {Smith}, K.~M. and {Spergel}, D. and {Staggs}, S.~T. and {Stark}, A. and {Stompor}, R. and {Vieregg}, A.~G. and {Wang}, G. and {Watson}, S. and {Wollack}, E.~J. and {Wu}, W.~L.~K. and {Yoon}, K.~W. and {Zahn}, O.},
        title = "{Neutrino physics from the cosmic microwave background and large scale structure}",
      journal = {Astropart. Phys.},
         year = 2015,
        month = mar,
       volume = {63},
        pages = {66-80},
          doi = {10.1016/j.astropartphys.2014.05.014},
archivePrefix = {arXiv},
       eprint = {1309.5383},
 primaryClass = {astro-ph.CO},
       adsurl = {https://ui.adsabs.harvard.edu/abs/2015APh....63...66A}
}

@article{Esteban:2020cvm,
       author = {{Esteban}, Ivan and {Gonzalez-Garcia}, M.~C. and {Maltoni}, Michele and {Schwetz}, Thomas and {Zhou}, Albert},
        title = "{The fate of hints: updated global analysis of three-flavor neutrino oscillations}",
      journal = {JHEP},
         year = 2020,
        month = sep,
       volume = {2020},
       number = {9},
          eid = {178},
        pages = {178},
          doi = {10.1007/JHEP09(2020)178},
archivePrefix = {arXiv},
       eprint = {2007.14792},
 primaryClass = {hep-ph},
       adsurl = {https://ui.adsabs.harvard.edu/abs/2020JHEP...09..178E}
}

@ARTICLE{2025JCAP...05..065L,
       author = {{Lewis}, Antony and {Chamberlain}, Ewan},
        title = "{Understanding acoustic scale observations: the one-sided fight against {\ensuremath{\Lambda}}}",
      journal = {\jcap},
         year = 2025,
        month = may,
       volume = {2025},
       number = {5},
          eid = {065},
        pages = {065},
          doi = {10.1088/1475-7516/2025/05/065},
archivePrefix = {arXiv},
       eprint = {2412.13894},
 primaryClass = {astro-ph.CO},
       adsurl = {https://ui.adsabs.harvard.edu/abs/2025JCAP...05..065L}
}

@ARTICLE{2024JCAP...12..007C,
       author = {{Cort{\^e}s}, Marina and {Liddle}, Andrew R.},
        title = "{Interpreting DESI's evidence for evolving dark energy}",
      journal = {\jcap},
         year = 2024,
        month = dec,
       volume = {2024},
       number = {12},
          eid = {007},
        pages = {007},
          doi = {10.1088/1475-7516/2024/12/007},
archivePrefix = {arXiv},
       eprint = {2404.08056},
 primaryClass = {astro-ph.CO},
       adsurl = {https://ui.adsabs.harvard.edu/abs/2024JCAP...12..007C}
}

@ARTICLE{2025arXiv250612004H,
       author = {{Herold}, Laura and {Karwal}, Tanvi},
        title = "{Bayesian and frequentist perspectives agree on dynamical dark energy}",
      journal = {arXiv e-prints},
         year = 2025,
        month = jun,
          eid = {arXiv:2506.12004},
        pages = {arXiv:2506.12004},
          doi = {10.48550/arXiv.2506.12004},
archivePrefix = {arXiv},
       eprint = {2506.12004},
 primaryClass = {astro-ph.CO},
       adsurl = {https://ui.adsabs.harvard.edu/abs/2025arXiv250612004H}
}

@ARTICLE{2025JCAP...08..065B,
       author = {{Bayat}, Zahra and {Hertzberg}, Mark P.},
        title = "{Examining quintessence models with DESI data}",
      journal = {\jcap},
         year = 2025,
        month = aug,
       volume = {2025},
       number = {8},
          eid = {065},
        pages = {065},
          doi = {10.1088/1475-7516/2025/08/065},
archivePrefix = {arXiv},
       eprint = {2505.18937},
 primaryClass = {astro-ph.CO},
       adsurl = {https://ui.adsabs.harvard.edu/abs/2025JCAP...08..065B}
}

@ARTICLE{2025arXiv250403829S,
       author = {{Steinhardt}, Charles L. and {Phillips}, Preston and {Wojtak}, Radoslaw},
        title = "{Dark Energy Constraints and Joint Cosmological Inference from Mutually Inconsistent Observations}",
      journal = {arXiv e-prints},
         year = 2025,
        month = apr,
          eid = {arXiv:2504.03829},
        pages = {arXiv:2504.03829},
          doi = {10.48550/arXiv.2504.03829},
archivePrefix = {arXiv},
       eprint = {2504.03829},
 primaryClass = {astro-ph.CO},
       adsurl = {https://ui.adsabs.harvard.edu/abs/2025arXiv250403829S}
}

@ARTICLE{2025arXiv250913318T,
       author = {{Toomey}, Michael W. and {Montefalcone}, Gabriele and {McDonough}, Evan and {Freese}, Katherine},
        title = "{How Theory-Informed Priors Affect DESI Evidence for Evolving Dark Energy}",
      journal = {arXiv e-prints},
         year = 2025,
        month = sep,
          eid = {arXiv:2509.13318},
        pages = {arXiv:2509.13318},
          doi = {10.48550/arXiv.2509.13318},
archivePrefix = {arXiv},
       eprint = {2509.13318},
 primaryClass = {astro-ph.CO},
       adsurl = {https://ui.adsabs.harvard.edu/abs/2025arXiv250913318T}
}

@ARTICLE{2021MNRAS.502.1401M,
       author = {{Mead}, A.~J. and {Brieden}, S. and {Tr{\"o}ster}, T. and {Heymans}, C.},
        title = "{HMCODE-2020: improved modelling of non-linear cosmological power spectra with baryonic feedback}",
      journal = {\mnras},
         year = 2021,
        month = mar,
       volume = {502},
       number = {1},
        pages = {1401-1422},
          doi = {10.1093/mnras/stab082},
archivePrefix = {arXiv},
       eprint = {2009.01858},
 primaryClass = {astro-ph.CO},
       adsurl = {https://ui.adsabs.harvard.edu/abs/2021MNRAS.502.1401M}
}

@ARTICLE{2000ApJ...538..473L,
       author = {{Lewis}, Antony and {Challinor}, Anthony and {Lasenby}, Anthony},
        title = "{Efficient Computation of Cosmic Microwave Background Anisotropies in Closed Friedmann-Robertson-Walker Models}",
      journal = {\apj},
         year = 2000,
        month = aug,
       volume = {538},
       number = {2},
        pages = {473-476},
          doi = {10.1086/309179},
archivePrefix = {arXiv},
       eprint = {astro-ph/9911177},
 primaryClass = {astro-ph},
       adsurl = {https://ui.adsabs.harvard.edu/abs/2000ApJ...538..473L}
}

@ARTICLE{ivanov_cosmological_2020,
       author = {{Ivanov}, Mikhail M. and {Simonovi{\'c}}, Marko and {Zaldarriaga}, Matias},
        title = "{Cosmological parameters and neutrino masses from the final Planck and full-shape BOSS data}",
      journal = {\prd},
         year = 2020,
        month = apr,
       volume = {101},
       number = {8},
          eid = {083504},
        pages = {083504},
          doi = {10.1103/PhysRevD.101.083504},
archivePrefix = {arXiv},
       eprint = {1912.08208},
 primaryClass = {astro-ph.CO},
       adsurl = {https://ui.adsabs.harvard.edu/abs/2020PhRvD.101h3504I}
}

@article{deSalas:2020pgw,
       author = {{de Salas}, P.~F. and {Forero}, D.~V. and {Gariazzo}, S. and {Mart{\'\i}nez-Mirav{\'e}}, P. and {Mena}, O. and {Ternes}, C.~A. and {T{\'o}rtola}, M. and {Valle}, J.~W.~F.},
        title = "{2020 global reassessment of the neutrino oscillation picture}",
      journal = {JHEP},
         year = 2021,
        month = feb,
       volume = {2021},
       number = {2},
          eid = {71},
        pages = {71},
          doi = {10.1007/JHEP02(2021)071},
archivePrefix = {arXiv},
       eprint = {2006.11237},
 primaryClass = {hep-ph},
       adsurl = {https://ui.adsabs.harvard.edu/abs/2021JHEP...02..071D}
}

@ARTICLE{lange2023,
       author = {{Lange}, Johannes U.},
        title = "{NAUTILUS: boosting Bayesian importance nested sampling with deep learning}",
      journal = {\mnras},
         year = 2023,
        month = oct,
       volume = {525},
       number = {2},
        pages = {3181-3194},
          doi = {10.1093/mnras/stad2441},
archivePrefix = {arXiv},
       eprint = {2306.16923},
 primaryClass = {astro-ph.IM},
       adsurl = {https://ui.adsabs.harvard.edu/abs/2023MNRAS.525.3181L}
}

@ARTICLE{2023PhRvD.108l3519D,
       author = {{Dalal}, Roohi and {Li}, Xiangchong and {Nicola}, Andrina and {Zuntz}, Joe and {Strauss}, Michael A. and {Sugiyama}, Sunao and {Zhang}, Tianqing and {Rau}, Markus M. and {Mandelbaum}, Rachel and {Takada}, Masahiro and {More}, Surhud and {Miyatake}, Hironao and {Kannawadi}, Arun and {Shirasaki}, Masato and {Taniguchi}, Takanori and {Takahashi}, Ryuichi and {Osato}, Ken and {Hamana}, Takashi and {Oguri}, Masamune and {Nishizawa}, Atsushi J. and {Malag{\'o}n}, Andr{\'e}s A. Plazas and {Sunayama}, Tomomi and {Alonso}, David and {Slosar}, An{\v{z}}e and {Luo}, Wentao and {Armstrong}, Robert and {Bosch}, James and {Hsieh}, Bau-Ching and {Komiyama}, Yutaka and {Lupton}, Robert H. and {Lust}, Nate B. and {MacArthur}, Lauren A. and {Miyazaki}, Satoshi and {Murayama}, Hitoshi and {Nishimichi}, Takahiro and {Okura}, Yuki and {Price}, Paul A. and {Tait}, Philip J. and {Tanaka}, Masayuki and {Wang}, Shiang-Yu},
        title = "{Hyper Suprime-Cam Year 3 results: Cosmology from cosmic shear power spectra}",
      journal = {\prd},
         year = 2023,
        month = dec,
       volume = {108},
       number = {12},
          eid = {123519},
        pages = {123519},
          doi = {10.1103/PhysRevD.108.123519},
archivePrefix = {arXiv},
       eprint = {2304.00701},
 primaryClass = {astro-ph.CO},
       adsurl = {https://ui.adsabs.harvard.edu/abs/2023PhRvD.108l3519D}
}

@ARTICLE{2022A&A...664A.170V,
       author = {{van den Busch}, J.~L. and {Wright}, A.~H. and {Hildebrandt}, H. and {Bilicki}, M. and {Asgari}, M. and {Joudaki}, S. and {Blake}, C. and {Heymans}, C. and {Kannawadi}, A. and {Shan}, H.~Y. and {Tr{\"o}ster}, T.},
        title = "{KiDS-1000: Cosmic shear with enhanced redshift calibration}",
      journal = {\aap},
         year = 2022,
        month = aug,
       volume = {664},
          eid = {A170},
        pages = {A170},
          doi = {10.1051/0004-6361/202142083},
archivePrefix = {arXiv},
       eprint = {2204.02396},
 primaryClass = {astro-ph.CO},
       adsurl = {https://ui.adsabs.harvard.edu/abs/2022A&A...664A.170V}
}

@ARTICLE{2025PhRvL.134r1002Y,
       author = {{Ye}, Gen and {Martinelli}, Matteo and {Hu}, Bin and {Silvestri}, Alessandra},
        title = "{Hints of Nonminimally Coupled Gravity in DESI 2024 Baryon Acoustic Oscillation Measurements}",
      journal = {\prl},
         year = 2025,
        month = may,
       volume = {134},
       number = {18},
          eid = {181002},
        pages = {181002},
          doi = {10.1103/PhysRevLett.134.181002},
archivePrefix = {arXiv},
       eprint = {2407.15832},
 primaryClass = {astro-ph.CO},
       adsurl = {https://ui.adsabs.harvard.edu/abs/2025PhRvL.134r1002Y}
}

@ARTICLE{2025arXiv251107526C,
       author = {{Caldwell}, Robert R. and {Linder}, Eric V.},
        title = "{Null Impact of the Null Energy Condition in Current Cosmology}",
      journal = {arXiv e-prints},
         year = 2025,
        month = nov,
          eid = {arXiv:2511.07526},
        pages = {arXiv:2511.07526},
          doi = {10.48550/arXiv.2511.07526},
archivePrefix = {arXiv},
       eprint = {2511.07526},
 primaryClass = {astro-ph.CO},
       adsurl = {https://ui.adsabs.harvard.edu/abs/2025arXiv251107526C}
}

@ARTICLE{wright_kids_2025,
       author = {{Wright}, Angus H. and {St{\"o}lzner}, Benjamin and {Asgari}, Marika and {Bilicki}, Maciej and {Giblin}, Benjamin and {Heymans}, Catherine and {Hildebrandt}, Hendrik and {Hoekstra}, Henk and {Joachimi}, Benjamin and {Kuijken}, Konrad and {Li}, Shun-Sheng and {Reischke}, Robert and {von Wietersheim-Kramsta}, Maximilian and {Yoon}, Mijin and {Burger}, Pierre and {Chisari}, Nora Elisa and {de Jong}, Jelte and {Dvornik}, Andrej and {Georgiou}, Christos and {Harnois-D{\'e}raps}, Joachim and {Jalan}, Priyanka and {William}, Anjitha John and {Joudaki}, Shahab and {Lesci}, Giorgio Francesco and {Linke}, Laila and {Loureiro}, Arthur and {Mahony}, Constance and {Maturi}, Matteo and {Miller}, Lance and {Moscardini}, Lauro and {Napolitano}, Nicola R. and {Porth}, Lucas and {Radovich}, Mario and {Schneider}, Peter and {Tr{\"o}ster}, Tilman and {Wittje}, Anna and {Yan}, Ziang and {Zhang}, Yun-Hao},
        title = "{KiDS-Legacy: Cosmological constraints from cosmic shear with the complete Kilo-Degree Survey}",
      journal = {\aap},
         year = 2025,
        month = nov,
       volume = {703},
          eid = {A158},
        pages = {A158},
          doi = {10.1051/0004-6361/202554908},
archivePrefix = {arXiv},
       eprint = {2503.19441},
 primaryClass = {astro-ph.CO},
       adsurl = {https://ui.adsabs.harvard.edu/abs/2025arXiv250319441W}
}

@ARTICLE{Bigwood2024,
       author = {{Bigwood}, L. and {Amon}, A. and {Schneider}, A. and {Salcido}, J. and {McCarthy}, I.~G. and {Preston}, C. and {Sanchez}, D. and {Sijacki}, D. and {Schaan}, E. and {Ferraro}, S. and {Battaglia}, N. and {Chen}, A. and {Dodelson}, S. and {Roodman}, A. and {Pieres}, A. and {Fert{\'e}}, A. and {Alarcon}, A. and {Drlica-Wagner}, A. and {Choi}, A. and {Navarro-Alsina}, A. and {Campos}, A. and {Ross}, A.~J. and {Carnero Rosell}, A. and {Yin}, B. and {Yanny}, B. and {S{\'a}nchez}, C. and {Chang}, C. and {Davis}, C. and {Doux}, C. and {Gruen}, D. and {Rykoff}, E.~S. and {Huff}, E.~M. and {Sheldon}, E. and {Tarsitano}, F. and {Andrade-Oliveira}, F. and {Bernstein}, G.~M. and {Giannini}, G. and {Diehl}, H.~T. and {Huang}, H. and {Harrison}, I. and {Sevilla-Noarbe}, I. and {Tutusaus}, I. and {Elvin-Poole}, J. and {McCullough}, J. and {Zuntz}, J. and {Blazek}, J. and {DeRose}, J. and {Cordero}, J. and {Prat}, J. and {Myles}, J. and {Eckert}, K. and {Bechtol}, K. and {Herner}, K. and {Secco}, L.~F. and {Gatti}, M. and {Raveri}, M. and {Kind}, M. Carrasco and {Becker}, M.~R. and {Troxel}, M.~A. and {Jarvis}, M. and {MacCrann}, N. and {Friedrich}, O. and {Alves}, O. and {Leget}, P. -F. and {Chen}, R. and {Rollins}, R.~P. and {Wechsler}, R.~H. and {Gruendl}, R.~A. and {Cawthon}, R. and {Allam}, S. and {Bridle}, S.~L. and {Pandey}, S. and {Everett}, S. and {Shin}, T. and {Hartley}, W.~G. and {Fang}, X. and {Zhang}, Y. and {Aguena}, M. and {Annis}, J. and {Bacon}, D. and {Bertin}, E. and {Bocquet}, S. and {Brooks}, D. and {Carretero}, J. and {Castander}, F.~J. and {da Costa}, L.~N. and {Pereira}, M.~E.~S. and {De Vicente}, J. and {Desai}, S. and {Doel}, P. and {Ferrero}, I. and {Flaugher}, B. and {Frieman}, J. and {Garc{\'\i}a-Bellido}, J. and {Gaztanaga}, E. and {Gutierrez}, G. and {Hinton}, S.~R. and {Hollowood}, D.~L. and {Honscheid}, K. and {Huterer}, D. and {James}, D.~J. and {Kuehn}, K. and {Lahav}, O. and {Lee}, S. and {Marshall}, J.~L. and {Mena-Fern{\'a}ndez}, J. and {Miquel}, R. and {Muir}, J. and {Paterno}, M. and {Plazas Malag{\'o}n}, A.~A. and {Porredon}, A. and {Romer}, A.~K. and {Samuroff}, S. and {Sanchez}, E. and {Sanchez Cid}, D. and {Smith}, M. and {Soares-Santos}, M. and {Suchyta}, E. and {Swanson}, M.~E.~C. and {Tarle}, G. and {To}, C. and {Weaverdyck}, N. and {Weller}, J. and {Wiseman}, P. and {Yamamoto}, M.},
        title = "{Weak lensing combined with the kinetic Sunyaev-Zel'dovich effect: a study of baryonic feedback}",
      journal = {\mnras},
         year = 2024,
        month = oct,
       volume = {534},
       number = {1},
        pages = {655-682},
          doi = {10.1093/mnras/stae2100},
archivePrefix = {arXiv},
       eprint = {2404.06098},
 primaryClass = {astro-ph.CO},
       adsurl = {https://ui.adsabs.harvard.edu/abs/2024MNRAS.534..655B}
}

@ARTICLE{kaiser2000,
       author = {{Kaiser}, Nick and {Wilson}, Gillian and {Luppino}, Gerard A.},
        title = "{Large-Scale Cosmic Shear Measurements}",
      journal = {arXiv e-prints},
         year = 2000,
        month = mar,
          eid = {astro-ph/0003338},
        pages = {astro-ph/0003338},
archivePrefix = {arXiv},
       eprint = {astro-ph/0003338},
 primaryClass = {astro-ph},
       adsurl = {https://ui.adsabs.harvard.edu/abs/2000astro.ph..3338K}
}

@ARTICLE{vanWaerbeke2000,
       author = {{Van Waerbeke}, L. and {Mellier}, Y. and {Erben}, T. and {Cuillandre}, J.~C. and {Bernardeau}, F. and {Maoli}, R. and {Bertin}, E. and {McCracken}, H.~J. and {Le F{\`e}vre}, O. and {Fort}, B. and {Dantel-Fort}, M. and {Jain}, B. and {Schneider}, P.},
        title = "{Detection of correlated galaxy ellipticities from CFHT data: first evidence for gravitational lensing by large-scale structures}",
      journal = {\aap},
         year = 2000,
        month = jun,
       volume = {358},
        pages = {30-44},
archivePrefix = {arXiv},
       eprint = {astro-ph/0002500},
 primaryClass = {astro-ph},
       adsurl = {https://ui.adsabs.harvard.edu/abs/2000A&A...358...30V}
}

@ARTICLE{bacon2000,
       author = {{Bacon}, David J. and {Refregier}, Alexandre R. and {Ellis}, Richard S.},
        title = "{Detection of weak gravitational lensing by large-scale structure}",
      journal = {\mnras},
         year = 2000,
        month = oct,
       volume = {318},
       number = {2},
        pages = {625-640},
          doi = {10.1046/j.1365-8711.2000.03851.x},
archivePrefix = {arXiv},
       eprint = {astro-ph/0003008},
 primaryClass = {astro-ph},
       adsurl = {https://ui.adsabs.harvard.edu/abs/2000MNRAS.318..625B}
}

@ARTICLE{wittman2000,
       author = {{Wittman}, David M. and {Tyson}, J. Anthony and {Kirkman}, David and {Dell'Antonio}, Ian and {Bernstein}, Gary},
        title = "{Detection of weak gravitational lensing distortions of distant galaxies by cosmic dark matter at large scales}",
      journal = {\nat},
         year = 2000,
        month = may,
       volume = {405},
       number = {6783},
        pages = {143-148},
          doi = {10.1038/35012001},
archivePrefix = {arXiv},
       eprint = {astro-ph/0003014},
 primaryClass = {astro-ph},
       adsurl = {https://ui.adsabs.harvard.edu/abs/2000Natur.405..143W}
}

@ARTICLE{gilmarin_2020,
       author = {{Gil-Mar{\'\i}n}, H{\'e}ctor and {Bautista}, Juli{\'a}n E. and {Paviot}, Romain and {Vargas-Maga{\~n}a}, Mariana and {de la Torre}, Sylvain and {Fromenteau}, Sebastien and {Alam}, Shadab and {{\'A}vila}, Santiago and {Burtin}, Etienne and {Chuang}, Chia-Hsun and {Dawson}, Kyle S. and {Hou}, Jiamin and {de Mattia}, Arnaud and {Mohammad}, Faizan G. and {M{\"u}ller}, Eva-Maria and {Nadathur}, Seshadri and {Neveux}, Richard and {Percival}, Will J. and {Raichoor}, Anand and {Rezaie}, Mehdi and {Ross}, Ashley J. and {Rossi}, Graziano and {Ruhlmann-Kleider}, Vanina and {Smith}, Alex and {Tamone}, Am{\'e}lie and {Tinker}, Jeremy L. and {Tojeiro}, Rita and {Wang}, Yuting and {Zhao}, Gong-Bo and {Zhao}, Cheng and {Brinkmann}, Jonathan and {Brownstein}, Joel R. and {Choi}, Peter D. and {Escoffier}, Stephanie and {de la Macorra}, Axel and {Moon}, Jeongin and {Newman}, Jeffrey A. and {Schneider}, Donald P. and {Seo}, Hee-Jong and {Vivek}, Mariappan},
        title = "{The Completed SDSS-IV extended Baryon Oscillation Spectroscopic Survey: measurement of the BAO and growth rate of structure of the luminous red galaxy sample from the anisotropic power spectrum between redshifts 0.6 and 1.0}",
      journal = {\mnras},
         year = 2020,
        month = oct,
       volume = {498},
       number = {2},
        pages = {2492-2531},
          doi = {10.1093/mnras/staa2455},
archivePrefix = {arXiv},
       eprint = {2007.08994},
 primaryClass = {astro-ph.CO},
       adsurl = {https://ui.adsabs.harvard.edu/abs/2020MNRAS.498.2492G}
}

@ARTICLE{bourboux2020,
       author = {{du Mas des Bourboux}, H{\'e}lion and {Rich}, James and {Font-Ribera}, Andreu and {de Sainte Agathe}, Victoria and {Farr}, James and {Etourneau}, Thomas and {Le Goff}, Jean-Marc and {Cuceu}, Andrei and {Balland}, Christophe and {Bautista}, Julian E. and {Blomqvist}, Michael and {Brinkmann}, Jonathan and {Brownstein}, Joel R. and {Chabanier}, Sol{\`e}ne and {Chaussidon}, Edmond and {Dawson}, Kyle and {Gonz{\'a}lez-Morales}, Alma X. and {Guy}, Julien and {Lyke}, Brad W. and {de la Macorra}, Axel and {Mueller}, Eva-Maria and {Myers}, Adam D. and {Nitschelm}, Christian and {Mu{\~n}oz Guti{\'e}rrez}, Andrea and {Palanque-Delabrouille}, Nathalie and {Parker}, James and {Percival}, Will J. and {P{\'e}rez-R{\`a}fols}, Ignasi and {Petitjean}, Patrick and {Pieri}, Matthew M. and {Ravoux}, Corentin and {Rossi}, Graziano and {Schneider}, Donald P. and {Seo}, Hee-Jong and {Slosar}, An{\v{z}}e and {Stermer}, Julianna and {Vivek}, M. and {Y{\`e}che}, Christophe and {Youles}, Samantha},
        title = "{The Completed SDSS-IV Extended Baryon Oscillation Spectroscopic Survey: Baryon Acoustic Oscillations with Ly{\ensuremath{\alpha}} Forests}",
      journal = {\apj},
         year = 2020,
        month = oct,
       volume = {901},
       number = {2},
          eid = {153},
        pages = {153},
          doi = {10.3847/1538-4357/abb085},
archivePrefix = {arXiv},
       eprint = {2007.08995},
 primaryClass = {astro-ph.CO},
       adsurl = {https://ui.adsabs.harvard.edu/abs/2020ApJ...901..153D}
}

@ARTICLE{2022JCAP...09..039C,
       author = {{Carron}, Julien and {Mirmelstein}, Mark and {Lewis}, Antony},
        title = "{CMB lensing from Planck PR4 maps}",
      journal = {\jcap},
         year = 2022,
        month = sep,
       volume = {2022},
       number = {9},
          eid = {039},
        pages = {039},
          doi = {10.1088/1475-7516/2022/09/039},
archivePrefix = {arXiv},
       eprint = {2206.07773},
 primaryClass = {astro-ph.CO},
       adsurl = {https://ui.adsabs.harvard.edu/abs/2022JCAP...09..039C}
}

@ARTICLE{2024ApJ...962..112Q,
       author = {{Qu}, Frank J. and {Sherwin}, Blake D. and {Madhavacheril}, Mathew S. and {Han}, Dongwon and {Crowley}, Kevin T. and {Abril-Cabezas}, Irene and {Ade}, Peter A.~R. and {Aiola}, Simone and {Alford}, Tommy and {Amiri}, Mandana and {Amodeo}, Stefania and {An}, Rui and {Atkins}, Zachary and {Austermann}, Jason E. and {Battaglia}, Nicholas and {Battistelli}, Elia Stefano and {Beall}, James A. and {Bean}, Rachel and {Beringue}, Benjamin and {Bhandarkar}, Tanay and {Biermann}, Emily and {Bolliet}, Boris and {Bond}, J. Richard and {Cai}, Hongbo and {Calabrese}, Erminia and {Calafut}, Victoria and {Capalbo}, Valentina and {Carrero}, Felipe and {Carron}, Julien and {Challinor}, Anthony and {Chesmore}, Grace E. and {Cho}, Hsiao-mei and {Choi}, Steve K. and {Clark}, Susan E. and {C{\'o}rdova Rosado}, Rodrigo and {Cothard}, Nicholas F. and {Coughlin}, Kevin and {Coulton}, William and {Dalal}, Roohi and {Darwish}, Omar and {Devlin}, Mark J. and {Dicker}, Simon and {Doze}, Peter and {Duell}, Cody J. and {Duff}, Shannon M. and {Duivenvoorden}, Adriaan J. and {Dunkley}, Jo and {D{\"u}nner}, Rolando and {Fanfani}, Valentina and {Fankhanel}, Max and {Farren}, Gerrit and {Ferraro}, Simone and {Freundt}, Rodrigo and {Fuzia}, Brittany and {Gallardo}, Patricio A. and {Garrido}, Xavier and {Gluscevic}, Vera and {Golec}, Joseph E. and {Guan}, Yilun and {Halpern}, Mark and {Harrison}, Ian and {Hasselfield}, Matthew and {Healy}, Erin and {Henderson}, Shawn and {Hensley}, Brandon and {Herv{\'\i}as-Caimapo}, Carlos and {Hill}, J. Colin and {Hilton}, Gene C. and {Hilton}, Matt and {Hincks}, Adam D. and {Hlo{\v{z}}ek}, Ren{\'e}e and {Ho}, Shuay-Pwu Patty and {Huber}, Zachary B. and {Hubmayr}, Johannes and {Huffenberger}, Kevin M. and {Hughes}, John P. and {Irwin}, Kent and {Isopi}, Giovanni and {Jense}, Hidde T. and {Keller}, Ben and {Kim}, Joshua and {Knowles}, Kenda and {Koopman}, Brian J. and {Kosowsky}, Arthur and {Kramer}, Darby and {Kusiak}, Aleksandra and {La Posta}, Adrien and {Lague}, Alex and {Lakey}, Victoria and {Lee}, Eunseong and {Li}, Zack and {Li}, Yaqiong and {Limon}, Michele and {Lokken}, Martine and {Louis}, Thibaut and {Lungu}, Marius and {MacCrann}, Niall and {MacInnis}, Amanda and {Maldonado}, Diego and {Maldonado}, Felipe and {Mallaby-Kay}, Maya and {Marques}, Gabriela A. and {McMahon}, Jeff and {Mehta}, Yogesh and {Menanteau}, Felipe and {Moodley}, Kavilan and {Morris}, Thomas W. and {Mroczkowski}, Tony and {Naess}, Sigurd and {Namikawa}, Toshiya and {Nati}, Federico and {Newburgh}, Laura and {Nicola}, Andrina and {Niemack}, Michael D. and {Nolta}, Michael R. and {Orlowski-Scherer}, John and {Page}, Lyman A. and {Pandey}, Shivam and {Partridge}, Bruce and {Prince}, Heather and {Puddu}, Roberto and {Radiconi}, Federico and {Robertson}, Naomi and {Rojas}, Felipe and {Sakuma}, Tai and {Salatino}, Maria and {Schaan}, Emmanuel and {Schmitt}, Benjamin L. and {Sehgal}, Neelima and {Shaikh}, Shabbir and {Sierra}, Carlos and {Sievers}, Jon and {Sif{\'o}n}, Crist{\'o}bal and {Simon}, Sara and {Sonka}, Rita and {Spergel}, David N. and {Staggs}, Suzanne T. and {Storer}, Emilie and {Switzer}, Eric R. and {Tampier}, Niklas and {Thornton}, Robert and {Trac}, Hy and {Treu}, Jesse and {Tucker}, Carole and {Ullom}, Joel and {Vale}, Leila R. and {Van Engelen}, Alexander and {Van Lanen}, Jeff and {van Marrewijk}, Joshiwa and {Vargas}, Cristian and {Vavagiakis}, Eve M. and {Wagoner}, Kasey and {Wang}, Yuhan and {Wenzl}, Lukas and {Wollack}, Edward J. and {Xu}, Zhilei and {Zago}, Fernando and {Zheng}, Kaiwen},
        title = "{The Atacama Cosmology Telescope: A Measurement of the DR6 CMB Lensing Power Spectrum and Its Implications for Structure Growth}",
      journal = {\apj},
         year = 2024,
        month = feb,
       volume = {962},
       number = {2},
          eid = {112},
        pages = {112},
          doi = {10.3847/1538-4357/acfe06},
archivePrefix = {arXiv},
       eprint = {2304.05202},
 primaryClass = {astro-ph.CO},
       adsurl = {https://ui.adsabs.harvard.edu/abs/2024ApJ...962..112Q}
}

@ARTICLE{2025arXiv250314451N,
       author = {{Naess}, Sigurd and {Guan}, Yilun and {Duivenvoorden}, Adriaan J. and {Hasselfield}, Matthew and {Wang}, Yuhan and {Abril-Cabezas}, Irene and {Addison}, Graeme E. and {Ade}, Peter A.~R. and {Aiola}, Simone and {Alford}, Tommy and {Alonso}, David and {Amiri}, Mandana and {An}, Rui and {Atkins}, Zachary and {Austermann}, Jason E. and {Barbavara}, Eleonora and {Battaglia}, Nicholas and {Battistelli}, Elia Stefano and {Beall}, James A. and {Bean}, Rachel and {Beheshti}, Ali and {Beringue}, Benjamin and {Bhandarkar}, Tanay and {Biermann}, Emily and {Bolliet}, Boris and {Bond}, J. Richard and {Calabrese}, Erminia and {Capalbo}, Valentina and {Carrero}, Felipe and {Chen}, Stephen and {Chesmore}, Grace and {Cho}, Hsiao-mei and {Choi}, Steve K. and {Clark}, Susan E. and {Rosado}, Rodrigo Cordova and {Cothard}, Nicholas F. and {Coughlin}, Kevin and {Coulton}, William and {Crichton}, Devin and {Crowley}, Kevin T. and {Devlin}, Mark J. and {Dicker}, Simon and {Duell}, Cody J. and {Duff}, Shannon M. and {Dunkley}, Jo and {Dunner}, Rolando and {Embil Villagra}, Carmen and {Fankhanel}, Max and {Farren}, Gerrit S. and {Ferraro}, Simone and {Foster}, Allen and {Freundt}, Rodrigo and {Fuzia}, Brittany and {Gallardo}, Patricio A. and {Garrido}, Xavier and {Giardiello}, Serena and {Gill}, Ajay and {Givans}, Jahmour and {Gluscevic}, Vera and {Golec}, Joseph E. and {Gong}, Yulin and {Halpern}, Mark and {Harrison}, Ian and {Healy}, Erin and {Henderson}, Shawn and {Hensley}, Brandon and {Herv{\'\i}as-Caimapo}, Carlos and {Hill}, J. Colin and {Hilton}, Gene C. and {Hilton}, Matt and {Hincks}, Adam D. and {Hlo{\v{z}}ek}, Ren{\'e}e and {Ho}, Shuay-Pwu Patty and {Hood}, John and {Hornecker}, Erika and {Huber}, Zachary B. and {Hubmayr}, Johannes and {Huffenberger}, Kevin M. and {Hughes}, John P. and {Ikape}, Margaret and {Irwin}, Kent and {Isopi}, Giovanni and {Jense}, Hidde T. and {Joshi}, Neha and {Keller}, Ben and {Kim}, Joshua and {Knowles}, Kenda and {Koopman}, Brian J. and {Kosowsky}, Arthur and {Kramer}, Darby and {Kusiak}, Aleksandra and {La Posta}, Adrien and {Lagu{\"e}}, Alex and {Lakey}, Victoria and {Lee}, Eunseong and {Li}, Yaqiong and {Li}, Zack and {Limon}, Michele and {Lokken}, Martine and {Louis}, Thibaut and {Lungu}, Marius and {MacCrann}, Niall and {MacInnis}, Amanda and {Madhavacheril}, Mathew S. and {Maldonado}, Diego and {Maldonado}, Felipe and {Mallaby-Kay}, Maya and {Marques}, Gabriela A. and {van Marrewijk}, Joshiwa and {McCarthy}, Fiona and {McMahon}, Jeff and {Mehta}, Yogesh and {Menanteau}, Felipe and {Moodley}, Kavilan and {Morris}, Thomas W. and {Mroczkowski}, Tony and {Namikawa}, Toshiya and {Nati}, Federico and {Nerval}, Simran K. and {Newburgh}, Laura and {Nicola}, Andrina and {Niemack}, Michael D. and {Nolta}, Michael R. and {Orlowski-Scherer}, John and {Page}, Lyman A. and {Pandey}, Shivam and {Partridge}, Bruce and {Perez Sarmiento}, Karen and {Prince}, Heather and {Puddu}, Roberto and {Qu}, Frank J. and {Ragavan}, Damien C. and {Ried Guachalla}, Bernardita and {Rogers}, Keir K. and {Rojas}, Felipe and {Sakuma}, Tai and {Schaan}, Emmanuel and {Schmitt}, Benjamin L. and {Sehgal}, Neelima and {Shaikh}, Shabbir and {Sherwin}, Blake D. and {Sierra}, Carlos and {Sievers}, Jon and {Sif{\'o}n}, Crist{\'o}bal and {Simon}, Sara and {Sonka}, Rita and {London}, Alexander Spencer and {Spergel}, David N. and {Staggs}, Suzanne T. and {Storer}, Emilie and {Surrao}, Kristen and {Switzer}, Eric R. and {Tampier}, Niklas and {Thornton}, Robert and {Trac}, Hy and {Tucker}, Carole and {Ullom}, Joel and {Vale}, Leila R. and {Van Engelen}, Alexander and {Van Lanen}, Jeff and {Vargas}, Cristian and {Vavagiakis}, Eve M. and {Wagoner}, Kasey and {Wenzl}, Lukas and {Wollack}, Edward J. and {Zheng}, Kaiwen and {The Atacama Cosmology Telescope collaboration}},
        title = "{The Atacama Cosmology Telescope: DR6 maps}",
      journal = {\jcap},
         year = 2025,
        month = nov,
       volume = {2025},
       number = {11},
          eid = {061},
        pages = {061},
          doi = {10.1088/1475-7516/2025/11/061},
archivePrefix = {arXiv},
       eprint = {2503.14451},
 primaryClass = {astro-ph.CO},
       adsurl = {https://ui.adsabs.harvard.edu/abs/2025JCAP...11..061N}
}

@ARTICLE{2024A&A...686A..10B,
       author = {{Balkenhol}, L. and {Trendafilova}, C. and {Benabed}, K. and {Galli}, S.},
        title = "{candl: cosmic microwave background analysis with a differentiable likelihood}",
      journal = {\aap},
         year = 2024,
        month = jun,
       volume = {686},
          eid = {A10},
        pages = {A10},
          doi = {10.1051/0004-6361/202449432},
archivePrefix = {arXiv},
       eprint = {2401.13433},
 primaryClass = {astro-ph.CO},
       adsurl = {https://ui.adsabs.harvard.edu/abs/2024A&A...686A..10B}
}

@ARTICLE{2023PhRvD.107b3529O,
       author = {{Omori}, Y. and {Baxter}, E.~J. and {Chang}, C. and {Friedrich}, O. and {Alarcon}, A. and {Alves}, O. and {Amon}, A. and {Andrade-Oliveira}, F. and {Bechtol}, K. and {Becker}, M.~R. and {Bernstein}, G.~M. and {Blazek}, J. and {Bleem}, L.~E. and {Camacho}, H. and {Campos}, A. and {Carnero Rosell}, A. and {Carrasco Kind}, M. and {Cawthon}, R. and {Chen}, R. and {Choi}, A. and {Cordero}, J. and {Crawford}, T.~M. and {Crocce}, M. and {Davis}, C. and {DeRose}, J. and {Dodelson}, S. and {Doux}, C. and {Drlica-Wagner}, A. and {Eckert}, K. and {Eifler}, T.~F. and {Elsner}, F. and {Elvin-Poole}, J. and {Everett}, S. and {Fang}, X. and {Fert{\'e}}, A. and {Fosalba}, P. and {Gatti}, M. and {Giannini}, G. and {Gruen}, D. and {Gruendl}, R.~A. and {Harrison}, I. and {Herner}, K. and {Huang}, H. and {Huff}, E.~M. and {Huterer}, D. and {Jarvis}, M. and {Krause}, E. and {Kuropatkin}, N. and {Leget}, P. -F. and {Lemos}, P. and {Liddle}, A.~R. and {MacCrann}, N. and {McCullough}, J. and {Muir}, J. and {Myles}, J. and {Navarro-Alsina}, A. and {Pandey}, S. and {Park}, Y. and {Porredon}, A. and {Prat}, J. and {Raveri}, M. and {Rollins}, R.~P. and {Roodman}, A. and {Rosenfeld}, R. and {Ross}, A.~J. and {Rykoff}, E.~S. and {S{\'a}nchez}, C. and {Sanchez}, J. and {Secco}, L.~F. and {Sevilla-Noarbe}, I. and {Sheldon}, E. and {Shin}, T. and {Troxel}, M.~A. and {Tutusaus}, I. and {Varga}, T.~N. and {Weaverdyck}, N. and {Wechsler}, R.~H. and {Wu}, W.~L.~K. and {Yanny}, B. and {Yin}, B. and {Zhang}, Y. and {Zuntz}, J. and {Abbott}, T.~M.~C. and {Aguena}, M. and {Allam}, S. and {Annis}, J. and {Bacon}, D. and {Benson}, B.~A. and {Bertin}, E. and {Bocquet}, S. and {Brooks}, D. and {Burke}, D.~L. and {Carlstrom}, J.~E. and {Carretero}, J. and {Chang}, C.~L. and {Chown}, R. and {Costanzi}, M. and {da Costa}, L.~N. and {Crites}, A.~T. and {Pereira}, M.~E.~S. and {de Haan}, T. and {De Vicente}, J. and {Desai}, S. and {Diehl}, H.~T. and {Dobbs}, M.~A. and {Doel}, P. and {Everett}, W. and {Ferrero}, I. and {Flaugher}, B. and {Friedel}, D. and {Frieman}, J. and {Garc{\'\i}a-Bellido}, J. and {Gaztanaga}, E. and {George}, E.~M. and {Giannantonio}, T. and {Halverson}, N.~W. and {Hinton}, S.~R. and {Holder}, G.~P. and {Hollowood}, D.~L. and {Holzapfel}, W.~L. and {Honscheid}, K. and {Hrubes}, J.~D. and {James}, D.~J. and {Knox}, L. and {Kuehn}, K. and {Lahav}, O. and {Lee}, A.~T. and {Lima}, M. and {Luong-Van}, D. and {March}, M. and {McMahon}, J.~J. and {Melchior}, P. and {Menanteau}, F. and {Meyer}, S.~S. and {Miquel}, R. and {Mocanu}, L. and {Mohr}, J.~J. and {Morgan}, R. and {Natoli}, T. and {Padin}, S. and {Palmese}, A. and {Paz-Chinch{\'o}n}, F. and {Pieres}, A. and {Plazas Malag{\'o}n}, A.~A. and {Pryke}, C. and {Reichardt}, C.~L. and {Romer}, A.~K. and {Ruhl}, J.~E. and {Sanchez}, E. and {Schaffer}, K.~K. and {Schubnell}, M. and {Serrano}, S. and {Shirokoff}, E. and {Smith}, M. and {Staniszewski}, Z. and {Stark}, A.~A. and {Suchyta}, E. and {Tarle}, G. and {Thomas}, D. and {To}, C. and {Vieira}, J.~D. and {Weller}, J. and {Williamson}, R. and {DES} and {SPT Collaborations}},
        title = "{Joint analysis of Dark Energy Survey Year 3 data and CMB lensing from SPT and Planck. I. Construction of CMB lensing maps and modeling choices}",
      journal = {\prd},
         year = 2023,
        month = jan,
       volume = {107},
       number = {2},
          eid = {023529},
        pages = {023529},
          doi = {10.1103/PhysRevD.107.023529},
archivePrefix = {arXiv},
       eprint = {2203.12439},
 primaryClass = {astro-ph.CO},
       adsurl = {https://ui.adsabs.harvard.edu/abs/2023PhRvD.107b3529O}
}

@ARTICLE{2023A&A...673A.111Y,
       author = {{Yao}, Ji and {Shan}, Huanyuan and {Zhang}, Pengjie and {Liu}, Xiangkun and {Heymans}, Catherine and {Joachimi}, Benjamin and {Asgari}, Marika and {Bilicki}, Maciej and {Hildebrandt}, Hendrik and {Kuijken}, Konrad and {Tr{\"o}ster}, Tilman and {van den Busch}, Jan Luca and {Wright}, Angus and {Yan}, Ziang},
        title = "{KiDS-1000: Cross-correlation with Planck cosmic microwave background lensing and intrinsic alignment removal with self-calibration}",
      journal = {\aap},
         year = 2023,
        month = may,
       volume = {673},
          eid = {A111},
        pages = {A111},
          doi = {10.1051/0004-6361/202346020},
archivePrefix = {arXiv},
       eprint = {2301.13437},
 primaryClass = {astro-ph.CO},
       adsurl = {https://ui.adsabs.harvard.edu/abs/2023A&A...673A.111Y}
}

@ARTICLE{2025arXiv250420038Q,
       author = {{Qu}, Frank J. and {Ge}, Fei and {Wu}, W.~L. Kimmy and {Abril-Cabezas}, Irene and {Madhavacheril}, Mathew S. and {Millea}, Marius and {Ahmed}, Zeeshan and {Anderes}, Ethan and {Anderson}, Adam J. and {Ansarinejad}, Behzad and {Archipley}, Melanie and {Atkins}, Zachary and {Balkenhol}, Lennart and {Battaglia}, Nicholas and {Benabed}, Karim and {Bender}, Amy N. and {Benson}, Bradford A. and {Bianchini}, Federico and {Bleem}, Lindsey. E. and {Bolliet}, Boris and {Bond}, J. Richard and {Bouchet}, Fran{\c{c}}ois. R. and {Bryant}, Lincoln and {Calabrese}, Erminia and {Camphuis}, Etienne and {Carlstrom}, John E. and {Carron}, Julien and {Challinor}, Anthony and {Chang}, Clarence L. and {Chaubal}, Prakrut and {Chen}, Geoff and {Chichura}, Paul M. and {Choi}, Steve K. and {Chokshi}, Aman and {Chou}, Ti-Lin and {Coerver}, Anna and {Coulton}, William and {Crawford}, Thomas M. and {Daley}, Cail and {Darwish}, Omar and {de Haan}, Tijmen and {Devlin}, Mark J. and {Dibert}, Karia R. and {Dobbs}, Matthew A. and {Doohan}, Michael and {Doussot}, Aristide and {Duivenvoorden}, Adriaan J. and {Dunkley}, Jo and {Dunner}, Rolando and {Dutcher}, Daniel and {Villagra}, Carmen Embil and {Everett}, Wendy and {Farren}, Gerrit S. and {Feng}, Chang and {Ferraro}, Simone and {Ferguson}, Kyle R. and {Fichman}, Kyra and {Finson}, Emily and {Foster}, Allen and {Gallardo}, Patricio A. and {Galli}, Silvia and {Gambrel}, Anne E. and {Gardner}, Rob W. and {Goeckner-Wald}, Neil and {Gualtieri}, Riccardo and {Guidi}, Federica and {Guns}, Sam and {Halpern}, Mark and {Halverson}, Nils W. and {Hill}, J. Colin and {Hilton}, Matt and {Hivon}, Eric and {Holder}, Gilbert P. and {Holzapfel}, William L. and {Hood}, John C. and {Howe}, Doug and {Hryciuk}, Alec and {Huang}, Nicholas and {Hubmayr}, Johannes and {K{\'e}ruzor{\'e}}, Florian and {Khalife}, Ali R. and {Kim}, Joshua and {Knox}, Lloyd and {Korman}, Milo and {Kornoelje}, Kayla and {Kosowsky}, Arthur and {Kuo}, Chao-Lin and {Jense}, Hidde T. and {La Posta}, Adrien and {Levy}, Kevin and {Lowitz}, Amy E. and {Louis}, Thibaut and {Lu}, Chunyu and {Lynch}, Gabriel P. and {MacCrann}, Niall and {Maniyar}, Abhishek and {Martsen}, Emily S. and {McMahon}, Jeff and {Menanteau}, Felipe and {Montgomery}, Joshua and {Nakato}, Yuka and {Moodley}, Kavilan and {Namikawa}, Toshiya and {Natoli}, Tyler and {Niemack}, Michael D. and {Noble}, Gavin I. and {Omori}, Yuuki and {Ouellette}, Aaron and {Page}, Lyman A. and {Pan}, Zhaodi and {Paschos}, Pascal and {Phadke}, Kedar A. and {Pollak}, Alexander W. and {Prabhu}, Karthik and {Quan}, Wei and {Raghunathan}, Srinivasan and {Rahimi}, Mahsa and {Rahlin}, Alexandra and {Reichardt}, Christian L. and {Riebel}, Dave and {Rouble}, Maclean and {Ruhl}, John E. and {Schaan}, Emmanuel and {Schiappucci}, Eduardo and {Sehgal}, Neelima and {Sierra}, Carlos E. and {Simpson}, Aidan and {Sherwin}, Blake D. and {Sif{\'o}n}, Crist{\'o}bal and {Spergel}, David N. and {Staggs}, Suzanne T. and {Sobrin}, Joshua A. and {Stark}, Antony A. and {Stephen}, Judith and {Tandoi}, Chris and {Thorne}, Ben and {Trendafilova}, Cynthia and {Umilta}, Caterina and {Van Engelen}, Alexander and {Vieira}, Joaquin D. and {Vitrier}, Aline and {Wan}, Yujie and {Whitehorn}, Nathan and {Wollack}, Edward J. and {Young}, Matthew R. and {Zebrowski}, Jessica A. and {(ACT + SPT-3G Collaborations)}},
        title = "{Unified and Consistent Structure Growth Measurements from Joint ACT, SPT, and Planck CMB Lensing}",
      journal = {\prl},
         year = 2026,
        month = jan,
       volume = {136},
       number = {2},
          eid = {021001},
        pages = {021001},
          doi = {10.1103/k5yr-3h6d},
archivePrefix = {arXiv},
       eprint = {2504.20038},
 primaryClass = {astro-ph.CO},
       adsurl = {https://ui.adsabs.harvard.edu/abs/2026PhRvL.136b1001Q}
}

@ARTICLE{2023PhRvD.108l2005P,
       author = {{Pan}, Z. and {Bianchini}, F. and {Wu}, W.~L.~K. and {Ade}, P.~A.~R. and {Ahmed}, Z. and {Anderes}, E. and {Anderson}, A.~J. and {Ansarinejad}, B. and {Archipley}, M. and {Aylor}, K. and {Balkenhol}, L. and {Barry}, P.~S. and {Basu Thakur}, R. and {Benabed}, K. and {Bender}, A.~N. and {Benson}, B.~A. and {Bleem}, L.~E. and {Bouchet}, F.~R. and {Bryant}, L. and {Byrum}, K. and {Camphuis}, E. and {Carlstrom}, J.~E. and {Carter}, F.~W. and {Cecil}, T.~W. and {Chang}, C.~L. and {Chaubal}, P. and {Chen}, G. and {Chichura}, P.~M. and {Cho}, H. -M. and {Chou}, T. -L. and {Cliche}, J. -F. and {Coerver}, A. and {Crawford}, T.~M. and {Cukierman}, A. and {Daley}, C. and {de Haan}, T. and {Denison}, E.~V. and {Dibert}, K.~R. and {Ding}, J. and {Dobbs}, M.~A. and {Doussot}, A. and {Dutcher}, D. and {Everett}, W. and {Feng}, C. and {Ferguson}, K.~R. and {Fichman}, K. and {Foster}, A. and {Fu}, J. and {Galli}, S. and {Gambrel}, A.~E. and {Gardner}, R.~W. and {Ge}, F. and {Goeckner-Wald}, N. and {Gualtieri}, R. and {Guidi}, F. and {Guns}, S. and {Gupta}, N. and {Halverson}, N.~W. and {Harke-Hosemann}, A.~H. and {Harrington}, N.~L. and {Henning}, J.~W. and {Hilton}, G.~C. and {Hivon}, E. and {Holder}, G.~P. and {Holzapfel}, W.~L. and {Hood}, J.~C. and {Howe}, D. and {Huang}, N. and {Irwin}, K.~D. and {Jeong}, O. and {Jonas}, M. and {Jones}, A. and {K{\'e}ruzor{\'e}}, F. and {Khaire}, T.~S. and {Knox}, L. and {Kofman}, A.~M. and {Korman}, M. and {Kubik}, D.~L. and {Kuhlmann}, S. and {Kuo}, C. -L. and {Lee}, A.~T. and {Leitch}, E.~M. and {Levy}, K. and {Lowitz}, A.~E. and {Lu}, C. and {Maniyar}, A. and {Menanteau}, F. and {Meyer}, S.~S. and {Michalik}, D. and {Millea}, M. and {Montgomery}, J. and {Nadolski}, A. and {Nakato}, Y. and {Natoli}, T. and {Nguyen}, H. and {Noble}, G.~I. and {Novosad}, V. and {Omori}, Y. and {Padin}, S. and {Paschos}, P. and {Pearson}, J. and {Posada}, C.~M. and {Prabhu}, K. and {Quan}, W. and {Raghunathan}, S. and {Rahimi}, M. and {Rahlin}, A. and {Reichardt}, C.~L. and {Riebel}, D. and {Riedel}, B. and {Ruhl}, J.~E. and {Sayre}, J.~T. and {Schiappucci}, E. and {Shirokoff}, E. and {Smecher}, G. and {Sobrin}, J.~A. and {Stark}, A.~A. and {Stephen}, J. and {Story}, K.~T. and {Suzuki}, A. and {Takakura}, S. and {Tandoi}, C. and {Thompson}, K.~L. and {Thorne}, B. and {Trendafilova}, C. and {Tucker}, C. and {Umilta}, C. and {Vale}, L.~R. and {Vanderlinde}, K. and {Vieira}, J.~D. and {Wang}, G. and {Whitehorn}, N. and {Yefremenko}, V. and {Yoon}, K.~W. and {Young}, M.~R. and {Zebrowski}, J.~A.},
        title = "{Measurement of gravitational lensing of the cosmic microwave background using SPT-3G 2018 data}",
      journal = {\prd},
         year = 2023,
        month = dec,
       volume = {108},
       number = {12},
          eid = {122005},
        pages = {122005},
          doi = {10.1103/PhysRevD.108.122005},
archivePrefix = {arXiv},
       eprint = {2308.11608},
 primaryClass = {astro-ph.CO},
       adsurl = {https://ui.adsabs.harvard.edu/abs/2023PhRvD.108l2005P}
}

@article{Skilling_Nested_2007,
author = {John Skilling},
title = {{Nested sampling for general Bayesian computation}},
volume = {1},
journal = {Bayesian Analysis},
number = {4},
publisher = {International Society for Bayesian Analysis},
pages = {833 -- 859},
year = {2006},
doi = {10.1214/06-BA127},
URL = {https://doi.org/10.1214/06-BA127},
adsurl = {https://ui.adsabs.harvard.edu/abs/2004AIPC..735..395S},

}

\begin{appendix}

\section{Feedback prior}
\label{app:1}

\begin{figure}
    \centering
    \includegraphics[width=0.95\linewidth]{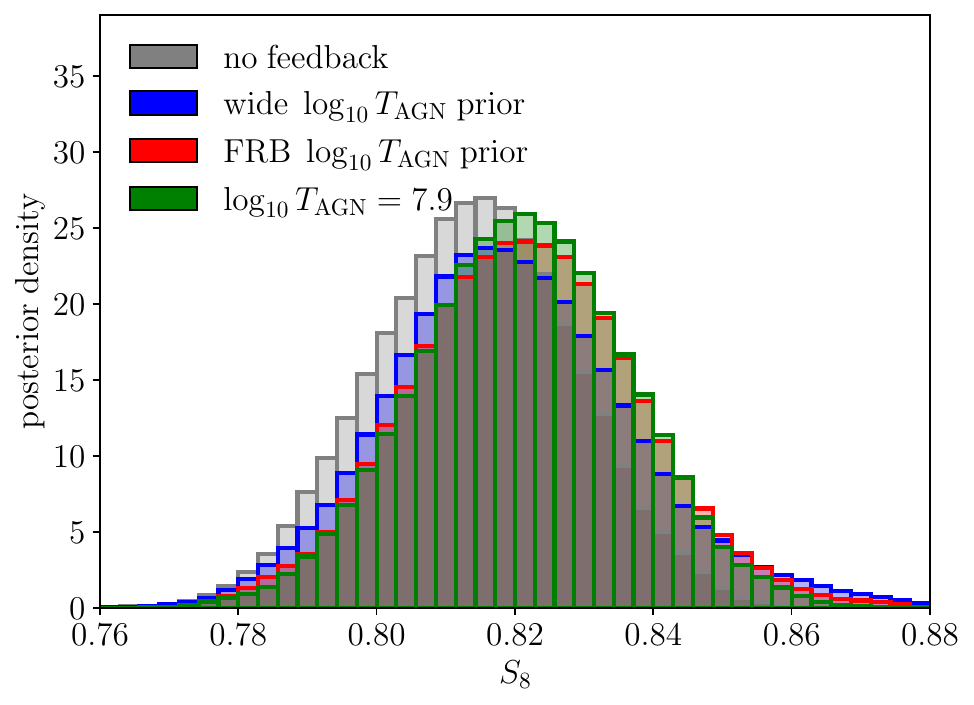}
    \caption{Marginal posterior distribution of $S_8$ for different prior choices for the feedback parameter $\log_{10}T_\mathrm{AGN}$, only using KiDS-Legacy cosmic shear data.}
    \label{fig:s8feedback}
\end{figure}

 \revint{The feedback parameter, $\log_{10}T_\mathrm{AGN}$, is barely constrained (compare \Cref{tab:constraints}). If at all, one can find an upper limit for $\log_{10}T_\mathrm{AGN}$ for some probe combinations. Since large $\log_{10}T_\mathrm{AGN}$ corresponds to stronger feedback, this is indicative of a ceiling to the amount of feedback. This is, of course, only true for the shape of the power spectrum suppression facilitated by $\log_{10}T_\mathrm{AGN}$ and for the summary statistic used. \citet{2025arXiv250908365B} for example, reconstruct the matter power spectrum using a different summary statistic, $\xi_\pm$, and find a somewhat larger suppression than indicated by the posterior for $\log_{10}T_\mathrm{AGN}$.}

Since the influence of $S_8$ on the extensions studied here was already established in the main text, the main question we can still address is whether different prior choices on $\log_{10}T_\mathrm{AGN}$ influence $S_8$. To this end, we run \revint{four} additional chains using only the KiDS-Legacy data, changing the prior range of $\log_{10}T_\mathrm{AGN}$ to a wide prior: $\log_{10}T_\mathrm{AGN}\in [6.8,9.6]$, (\revint{compare to the fiducial prior adopted in \Cref{tab:parameters}: $\log_{10}T_\mathrm{AGN}\in [7.3,8.3]$}). This prior range allows the suppression of the matter power spectrum to exceed unity; that is, the power will be enhanced. \revint{This happens if $\log_{10}T_\mathrm{AGN} < 7.3$ and affects only scales with $k > 1\,h\,\mathrm{Mpc}^{-1}$. However, the effect can reach $10\,\%$ at $k =10\,h\,\mathrm{Mpc}^{-1}$.} Furthermore, this broad prior is much wider than what {\sc HMCODE2020} has been calibrated on, so extrapolation might not give the most realistic results. However, we only want to show that we are not sensitive to that choice. 
Additionally, we use recent constraints on $\log_{10}T_\mathrm{AGN}$ from Fast Radio Bursts \citep[FRBs,][]{2025arXiv250717742R} to provide a somewhat tighter prior, in particular favouring stronger feedback models over the KiDS data. These constraints are also consistent with previous analyses using KiDS-1000 and the thermal Sunyaev-Zel'dovich effect \citep{2022A&A...660A..27T}.

The result of this exercise is shown in \Cref{fig:s8feedback}, illustrating that $S_8$ is very stable with respect to the prior choices. The tighter prior on $\log_{10}T_\mathrm{AGN}$ increases $S_8$ slightly and improves the constraints marginally. This is due to the modest preference for stronger feedback among the FRBs, which requires a larger $S_8$. Lastly, we check the posterior of $S_8$ for two other cases: firstly, for a fixed value of $\log_{10} T_\mathrm{AGN}= 7.9$, which corresponds to the value found in \citet{2025arXiv250717742R} and is consistent with the suppression measured in \citet{Bigwood2024,2025arXiv250707991K} and the $f_\mathrm{gas}-8\sigma$ {\sc{Flamingo}} simulation \citep{Schaye2023_flamingo,Schaller_pkFlamingo_2024}, that is fairly strong feedback. Secondly, we assume no feedback and infer $S_8$ from a pure CDM power spectrum. The results are shown as a green and grey histogram, respectively. \revint{In \citet{des/kids:2023} it was already shown that COSEBIs within the angular scales used here are expected to be rather insensitive to feedback.}
As before, we observe that larger feedback shifts $S_8$ slightly higher. 
The fact that we do not have to marginalise over the feedback results in marginally tighter constraints on $S_8$. 

In conclusion, this plot shows that the fiducial analysis of KiDS-Legacy and, in combination with the results shown here, also the extended cosmological analysis are not influenced by the prior on the feedback parameter when using COSEBIs as a summary statistic, in line with \citet{wright_kids_2025}.

\section{Full cosmological posteriors}
\label{app:2}

\begin{figure*}
    \centering
    \begin{subfigure}[b]{0.49\textwidth}
        \includegraphics[width=\textwidth]{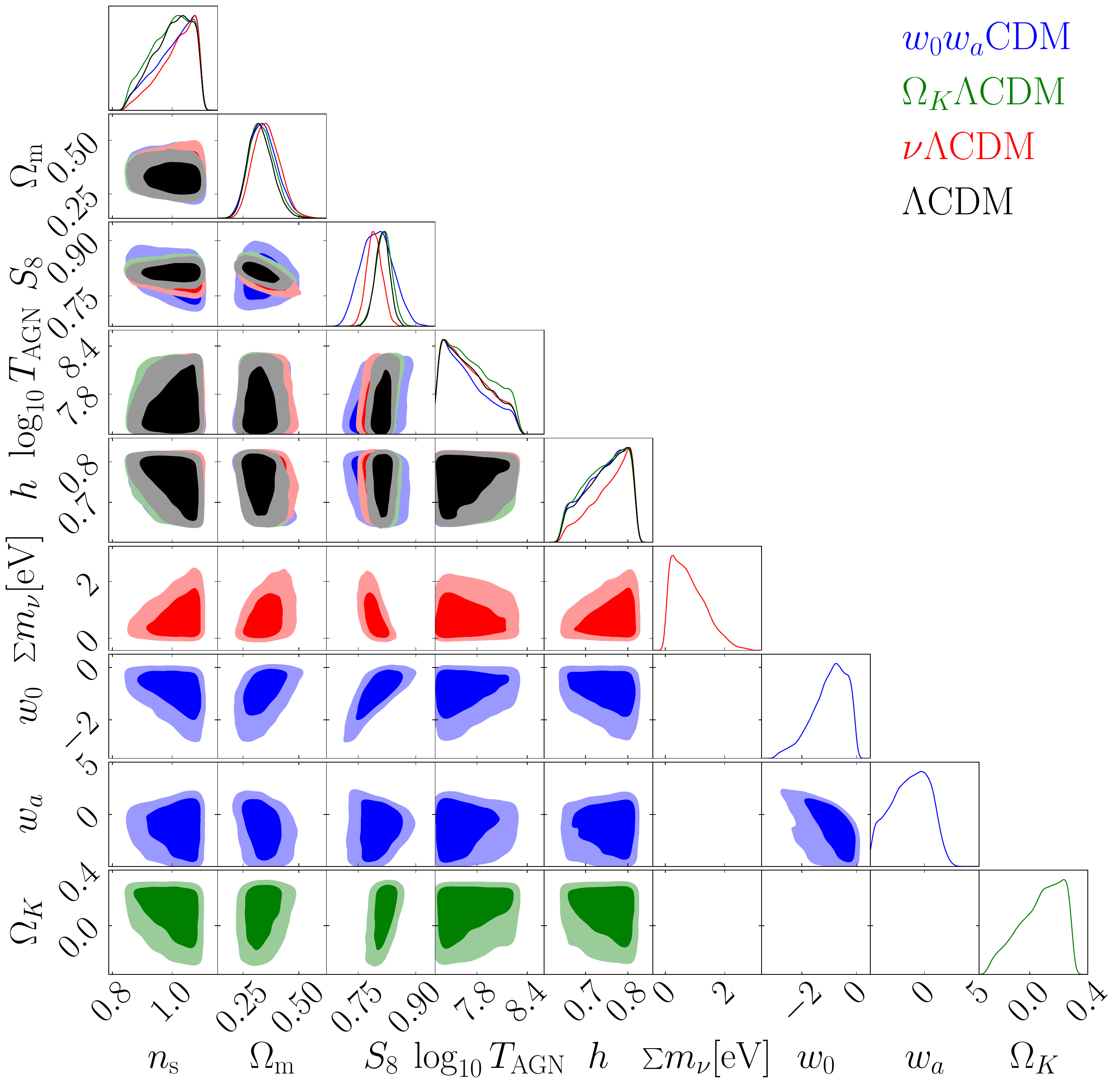}
        \caption{KiDS-Legacy}
        \label{fig:kids_full_post}
    \end{subfigure}
    \begin{subfigure}[b]{0.49\textwidth}
        \includegraphics[width=\textwidth]{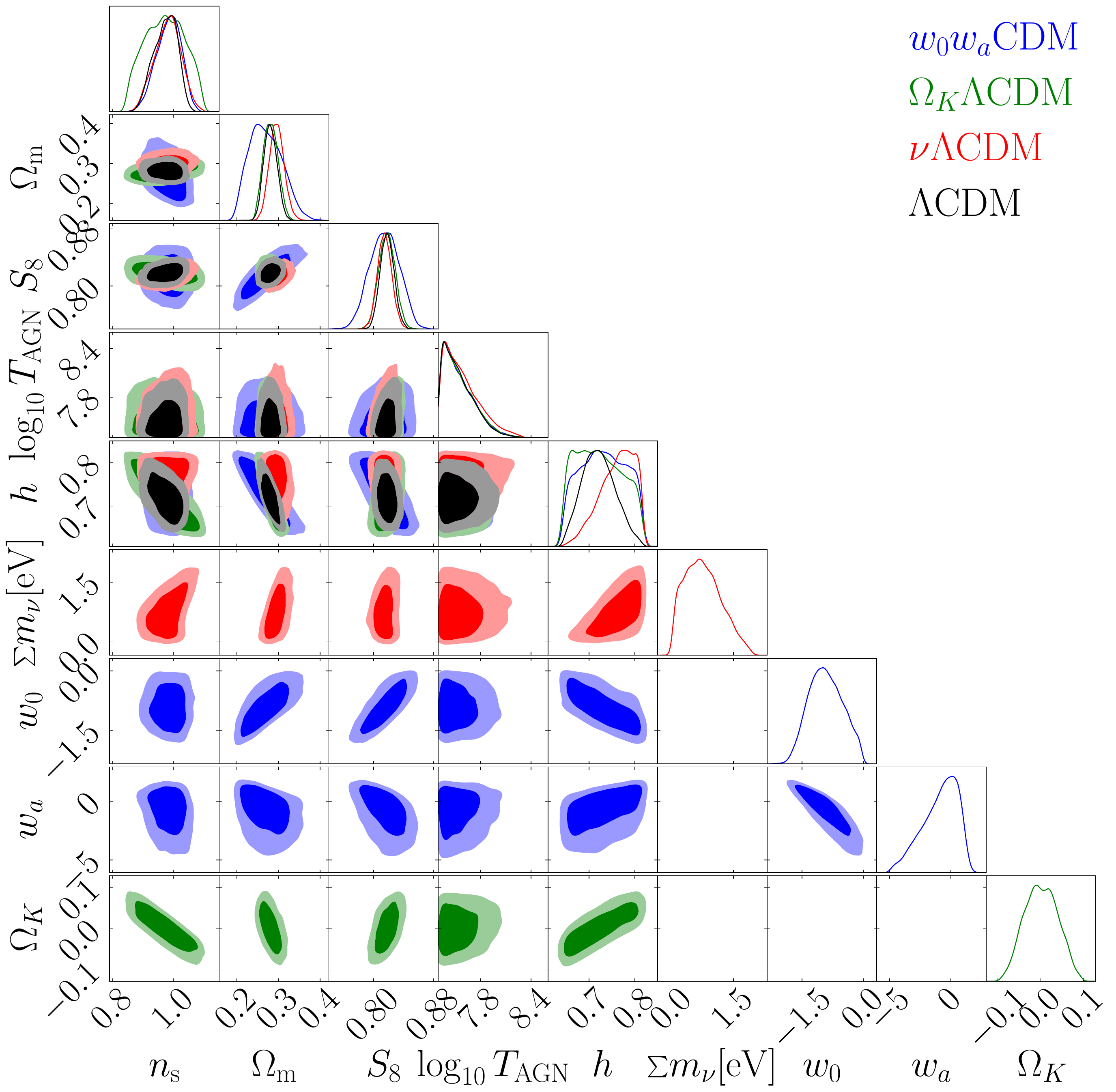}
        \caption{KiDS+DES Y3+CMB(lensing)}
        \label{fig:lensing_full_post}
    \end{subfigure}
    \vspace{.7cm}\\
    \begin{subfigure}[b]{0.49\textwidth}
        \includegraphics[width=\textwidth]{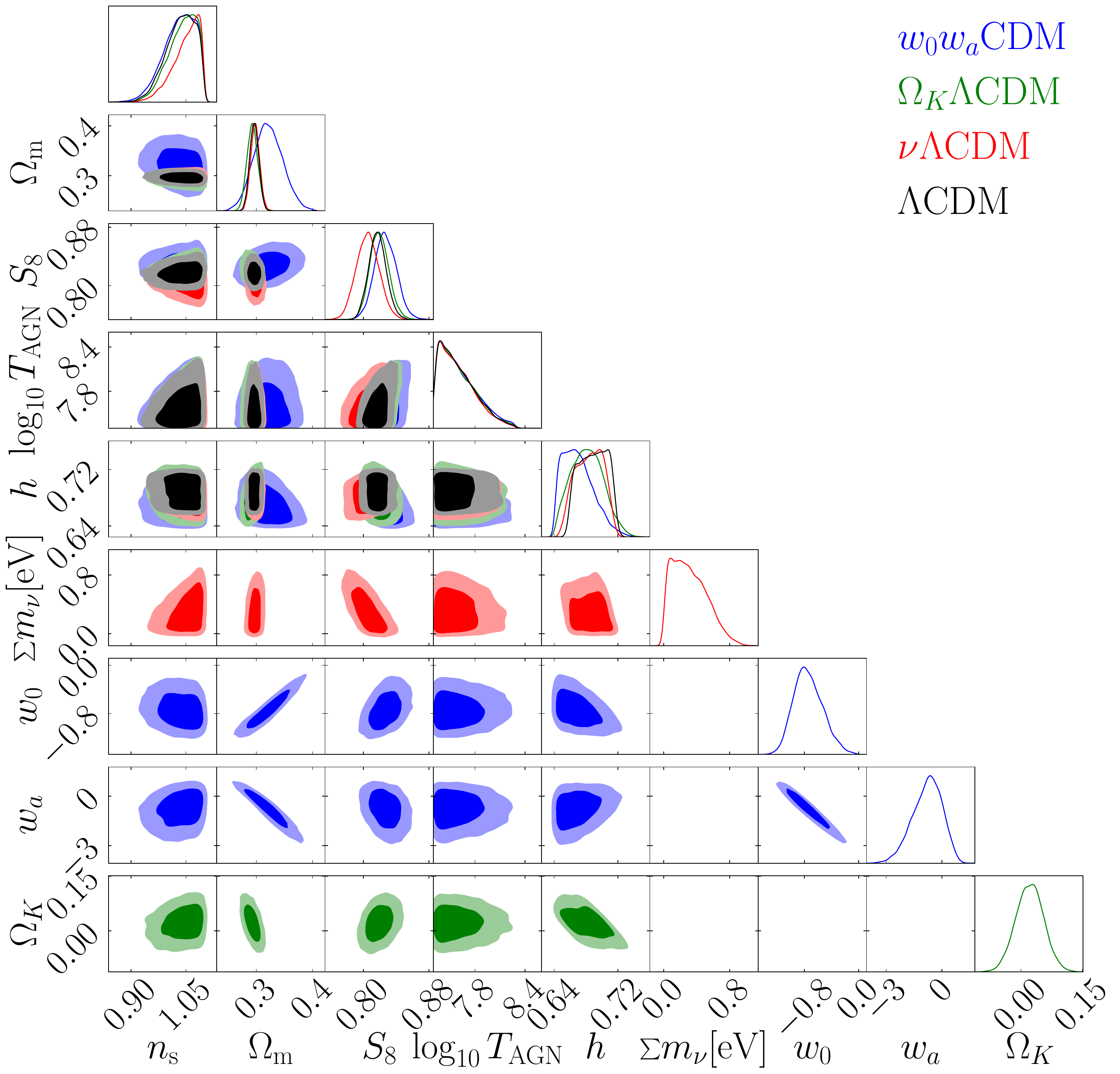}
        \caption{KiDS+DES Y3+DESI BAO+eBOSS RSD (low-$z$)}
        \label{fig:lowz_full_post}
    \end{subfigure}
    \begin{subfigure}[b]{0.49\textwidth}
        \includegraphics[width=\textwidth]{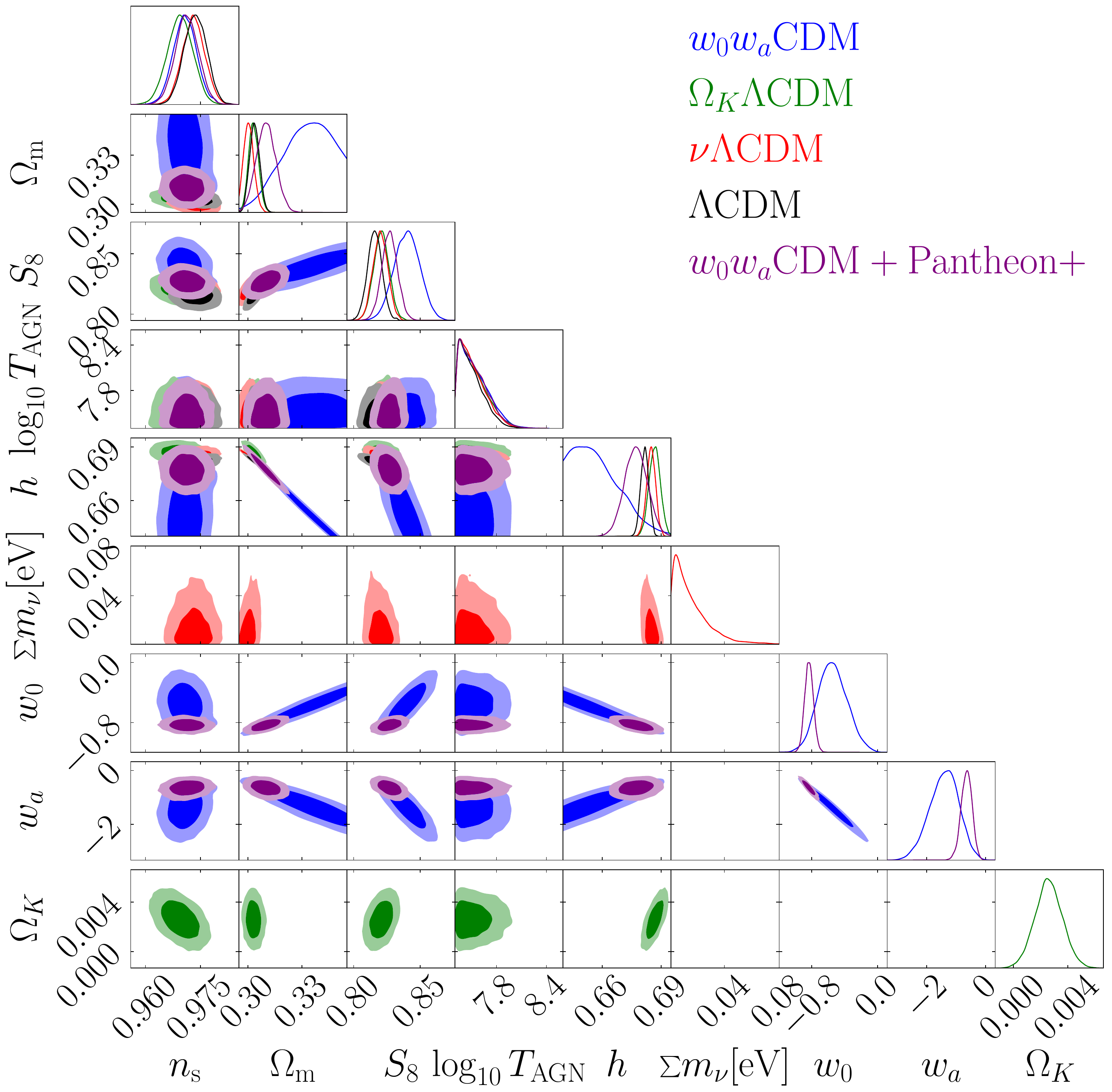}
        \caption{low-$z$+CMB(+Pantheon+, purple)}
        \label{fig:all_full_post}
    \end{subfigure}
    
    \caption{Marginal constraints for the full set of cosmological parameters and the feedback parameter for all probe combinations and model extensions discussed in this work.}
    \label{fig:figa1}
\end{figure*}

In this Appendix, we show the posterior distributions of the cosmological parameters and the feedback parameter $\log_{10}T_\mathrm{AGN}$. \Cref{fig:kids_full_post,fig:lensing_full_post,fig:lowz_full_post,fig:all_full_post} show the credible regions for the three extensions for KiDS-Legacy, lensing, low-$z$ and low-$z+$CMB (compare \Cref{tab:data_combiantions}), respectively.

Here, one can see a few trends discussed in the main text. \Cref{fig:kids_full_post,fig:lensing_full_post} e.g. show that allowing the neutrino mass to vary leads to more probability mass at lower $n_\mathrm{s}$ in \Cref{fig:kids_full_post}, thus lowering $S_8$ slightly. However, this degeneracy is broken by introducing any of the external probes studied here; large values of $n_\mathrm{s}$ become excluded. In \Cref{fig:lensing_full_post}, one can see that in $\Lambda$CDM, gravitational lensing can constrain the Hubble constant slightly (black contour). This is possible because the range of scales tested by CMB lensing and cosmic shear is large enough to resolve the peak of the power spectrum and the transition from linear to non-linear scales, both of which depend on $h$.

\Cref{fig:all_full_post} shows the effect of opening up the parameter space to dynamical dark energy when using the CMB, as the strong degeneracy between $\Omega_\mathrm{m}$ is exposed in this case. In $\Lambda$CDM, this is not visible since the expansion history is fixed by the combination of the CMB and BAO, limiting the range of $\Omega_\mathrm{m}$ values. This allows the exploration of low values of $h$ that are, in principle, rejected by our prior choices (see \Cref{tab:parameters}). For chains including the CMB and dynamical dark energy, however, we allow for a wider $h$ prior to discover the full shape of the posterior as driven by the likelihood. \citet{2021A&A...649A..88T} found a similar situation for $\Omega_{K}$; however, in this work, this is remedied by the inclusion of additional data, thereby removing the slight skewness towards negative $\Omega_K$ in the CMB. In fact, we find consistent results with the analysis presented in \citet{2025arXiv250620707C}, with $\Omega_K> 0$ at $2.6\sigma$. \review{This, however, does not warrant a model preference as we show in \Cref{tab:bayes_factor}.}
It is noteworthy that the addition of SNe from Pantheon+ blocks the high $\Omega_\mathrm{m}$ and low $h$ part of the parameter space again.

   \end{appendix} 
\end{document}